\documentclass{article}
\usepackage[utf8]{inputenc}
\usepackage{amssymb}
\usepackage{amsmath}
\usepackage{authblk}
\usepackage[colorlinks=true,linkcolor=blue,citecolor=blue,urlcolor=blue,unicode=true]{hyperref}

\usepackage{lineno}

\usepackage{mathtools}
\usepackage{amsfonts}
\usepackage{mathrsfs}
\usepackage{bbm}
\usepackage{slashed}

\usepackage{graphicx}
\usepackage{color}
\usepackage{array}

\usepackage{placeins}
\usepackage{booktabs}
\usepackage{makecell}
\usepackage{subcaption}
\usepackage{floatrow}
\usepackage{siunitx}
\usepackage{rotating}

\usepackage{bm}
\usepackage{todonotes}

\def\Fig#1{Fig.~\ref{#1}}
\newcommand{\ubr}[2]{\underbrace{#1}_{\text{#2}}}
\newcommand{\ave}[1]{\left\langle #1 \right\rangle}
\newcommand{\lpt}{l_\mathrm{patch}}
\newcommand{\mfp}{l_\mathrm{mfp}}
\newcommand{\vv}[1]{{\bf VK: #1}}

\newcommand{\pt}{p_{\mathrm{T}}}

\newcommand{\pp}{$p$+$p$}
\newcommand{\pPb}{$p$+Pb}
\newcommand{\PbPb}{Pb+Pb}
\newcommand{\AuAu}{Au+Au}
\newcommand{\BeBe}{Be+Be}
\newcommand{\ArSc}{Ar+Sc}
\newcommand{\XeLa}{Xe+La}
\newcommand{\AgAg}{Ag+Ag}
\newcommand{\CC}{C+C}
\newcommand{\SiSi}{Si+Si}

\title{Dynamics of critical fluctuations:\\ Theory -- phenomenology -- heavy-ion
collisions}

\author[1,2]{Marcus	Bluhm (organiser, editor)}	
\affil[1]{SUBATECH UMR 6457 (IMT Atlantique, Universit\'e de Nantes,
IN2P3/CNRS), 
4 rue Alfred Kastler, 44307 Nantes, France}
\affil[2]{ExtreMe Matter Institute EMMI, GSI, Planckstr. 1, 64291 Darmstadt,
Germany}
\author[1,2]{Marlene Nahrgang (organiser, editor)} 
\author[3a]{Alexander Kalweit (organiser, editor)}
\affil[3a]{Experimental Physics Department, CERN, CH-1211 Geneva 23,
Switzerland}
\affil[3b]{Theoretical Physics Department, CERN, CH-1211 Geneva 23, 
Switzerland}
\author[4a]{Mesut Arslandok}	
\affil[4a]{Physikalisches Institut, Universit\"at Heidelberg, Im Neuenheimer
Feld 226, D-69120 Heidelberg, Germany}
\affil[4b]{Institut f\"ur Theoretische Physik, Universit\"at Heidelberg,
Philosophenweg 16, D-69120, Heidelberg, Germany}
\author[2,4a,5]{Peter Braun-Munzinger}
\affil[5]{GSI Helmholtzzentrum f\"{u}r Schwerionenforschung GmbH, 64291 Darmstadt,
Germany}
\author[4b]{Stefan Floerchinger (editor)}
\author[6]{Eduardo S. Fraga}
\affil[6]{Instituto de F\'\i sica, Universidade Federal do Rio de Janeiro,
Caixa Postal 68528, 21941-972, Rio de Janeiro, RJ, Brazil}
\author[7a,8]{Marek	Gazdzicki}	
\affil[7a]{Institut f\"ur Kernphysik,
Goethe Universit\"at Frankfurt, Max-von-Laue-Str. 1, D-60438 Frankfurt am Main,
Germany}
\affil[7b]{Institut f\"ur Theoretische Physik,
Goethe Universit\"at Frankfurt, Max-von-Laue-Str. 1, D-60438 Frankfurt am Main,
Germany}
\affil[8]{Division of Nuclear Physics, Jan Kochanowski University, 25-406 Kielce, Poland}
\author[1]{Christoph Hartnack}	
\author[9]{Christoph Herold}	
\affil[9]{School of Physics and Center of Excellence in High Energy Physics \&
Astrophysics, Suranaree University of Technology, Nakhon Ratchasima 30000,
Thailand}
\author[5]{Romain Holzmann}
\author[1,10]{Iurii	Karpenko}
\affil[10]{Czech Technical University in Prague, FNSPE, B\v{r}ehov\'a 7, 
Prague 115 19,Czech Republic}
\author[11,12]{Masakiyo Kitazawa}	
\affil[11]{Department of Physics, Osaka University, Toyonaka, Osaka 560-0043,
Japan}
\affil[12]{J-PARC Branch, KEK Theory Center, Institute of Particle and Nuclear
Studies, KEK, 203-1, Shirakata, Tokai, Ibaraki, 319-1106, Japan}
\author[13]{Volker Koch (editor)}
\affil[13]{Nuclear Science Division, Lawrence Berkeley National Laboratory, 
1 Cyclotron Road, Berkeley, CA 94720, U.S.A.}
\author[14]{Stefan Leupold}
\affil[14]{Institutionen f\"or fysik och astronomi, Uppsala universitet, Box 516, S-75120 Uppsala, Sweden}
\author[3b,4b]{Aleksas Mazeliauskas (editor)}
\author[3a,15]{Bedangadas Mohanty}
\affil[15]{School of Physical Sciences, National Institute of Science Education
and Research, HBNI, Jatni 752050, India}
\author[4a,16a]{Alice Ohlson (editor)}
\affil[16a]{Lund University Department of Physics, Division of Particle Physics, Box 118, S-22100 Lund, Sweden}
\affil[16b]{Theoretical Particle Physics, Department of Astronomy and Theoretical Physics,
Lund University, S\"olvegatan 14A, SE-22362 Lund, Sweden}
\author[13]{Dmytro Oliinychenko}
\author[2,4b]{Jan M. Pawlowski}
\author[16b]{Christopher Plumberg}
\author[17]{Gregory	W. Ridgway}
\affil[17]{Center for Theoretical Physics, Massachusetts Institute of
Technology, Cambridge, MA 02139, U.S.A.}
\author[18]{Thomas Sch{\"a}fer (editor)}
\affil[18]{Department of Physics, North Carolina State University, Raleigh, 
NC 27695, U.S.A.}
\author[5,19]{Ilya	Selyuzhenkov}
\affil[19]{National Research Nuclear University MEPhI (Moscow Engineering Physics Institute), Kashirskoe highway 31, 115409, Moscow, Russia}
\author[4a]{Johanna	Stachel}
\author[20]{Mikhail	Stephanov}
\affil[20]{Department of Physics, University of Illinois, Chicago, IL 60607, U.S.A.}
\author[21]{Derek Teaney}
\affil[21]{Department of Physics and Astronomy, Stony Brook University, Stony Brook,
NY 11794, U.S.A.}
\author[1]{Nathan Touroux}
\author[7b,22]{Volodymyr Vovchenko}
\affil[22]{Frankfurt Institute for Advanced Studies, Giersch Science Center,
Ruth-Moufang-Str. 1, D-60438 Frankfurt am Main, Germany}
\author[4b]{Nicolas	Wink}

\begin{document}

\maketitle

\begin{abstract}
    This report summarizes the presentations and discussions during the 
    Rapid Reaction Task Force "Dynamics of critical fluctuations: Theory 
    -- phenomenology -- heavy-ion collisions", which was organized by the ExtreMe Matter Institute EMMI and held at GSI, Darmstadt,
    Germany in April 2019. We address the current understanding of the dynamics 
    of critical fluctuations in QCD and their measurement in heavy-ion collision
    experiments. In addition, we outline what might be learned from studying
    correlations in other physical systems, such as cold atomic gases.
\end{abstract}
\newpage

\tableofcontents
\section{Introduction}

 Ultra-relativistic heavy-ion collisions create small droplets of 
deconfined QCD matter -- the Quark Gluon Plasma (QGP). As the system expands, 
it cools and eventually hadronizes. As a function of beam energy, system size,
and rapidity the collision explores different regions of temperature $T$ and 
baryo-chemical potential $\mu_B$ in the QCD phase 
diagram \cite{Odyniec:2013aaa,Gazdzicki:995681,Luo:2015doi,Bzdak:2019pkr},
possibly including  a conjectured QCD critical point \cite{Stephanov:1998dy}.
This critical point is the endpoint of a line of first order QCD phase
transitions, analogous to the critical endpoint in the phase diagram of water. 

The main tool that connects the evolution of the matter produced in a
relativistic heavy-ion collision to bulk properties of QCD is viscous
relativistic fluid dynamics
\cite{Heinz:2013th,Teaney:2009qa,deSouza:2015ena,Schenke:2010nt}. 
Fluid dynamics can be understood as the effective theory of the long-time and
long-wavelength behavior of a classical or quantum many-body system. In 
this limit the system approaches approximate local thermal equilibrium,
and the dynamics is governed by the evolution of conserved charges. The 
system produced in relativistic heavy-ion collisions is not truly macroscopic
-- the number of produced hadrons ranges from about a hundred to several
tens of thousands -- and the question just how far the hydrodynamic paradigm 
can be pushed  towards smaller systems, lower energies, and more rare probes 
is an active area of study~\cite{Nagle:2018nvi,Schlichting:2019abc}. 

 Researchers are also investigating why the fluid dynamic description is 
so effective, even in systems that are very small and very rapidly
evolving~\cite{Florkowski:2017olj,Romatschke:2017ejr}. 
While no complete consensus has been achieved, a number of important factors 
have been identified. The first is the fact that the QGP behaves as a nearly
perfect fluid \cite{Hirano:2005wx,Schafer:2009dj}. In particular, the mean 
free path is short and transport coefficients such as the shear viscosity to
entropy density ratio $\eta/s$, are small. The second is rapid 
"hydrodynamization" \cite{Heller:2013oxa,Heller:2015dha}. There are 
indications, based on weak coupling kinetic models as well as strong coupling
holographic approaches, that the fluid dynamic description is valid even in a
regime where the system is still far from local thermal equilibrium. 

 The main observables that helped to establish the hydrodynamic paradigm
are the spectra of identified particles, flow observables, and the spectra
of certain rare probes, such as photons and dileptons  \cite{Stock:2010hoa,Braun-Munzinger:2015hba}. In this report
we will focus on fluctuation observables. There are several sources of 
fluctuations in relativistic heavy-ion collisions. The first is quantum 
fluctuations, in particular fluctuations in the initial multiplicity or energy
deposition. The second is thermal fluctuations. In heavy-ion collisions the 
volume that is locally equilibrated is quite small, and fluctuations due 
to the finite size of the system are sizeable. These fluctuations are
controlled by susceptibilities and related to the equation of state of 
the system. It is this connection that motivates a program of using 
fluctuation observables to investigate the phase structure of QCD. In 
particular, fluctuation observables may reflect the nature of the 
quasi-particles -- quarks or hadrons -- that carry the conserved charges,
baryon number, electric charge, and flavor \cite{Jeon:2000wg,Asakawa:2000wh}.
Furthermore, fluctuations probe the critical scaling of susceptibilities 
near a possible endpoint of a first order phase transition line in the QCD 
phase diagram \cite{Stephanov:1998dy}. 

  A quantitative effort motivated by these ideas has to incorporate all
sources of fluctuations in a heavy-ion collision. The central theme of 
this report is that such an analysis also requires a fully dynamical 
framework for the evolution of fluctuations. On a purely theoretical 
level, fluctuation-dissipation relations require that any dissipative 
theory of the evolution of a QGP has to include fluctuations, 
and any theory of fluctuations must incorporate dissipative effects. In 
thermal equilibrium, both effects balance, and a thermal spectrum of 
fluctuations emerges. Both dissipative effects as well as fluctuations
are relatively more important in small systems. 

  At a practical level, the relative size of different sources of 
fluctuations depends on the evolution of the system, and a careful 
modeling of fluctuations in relativistic heavy-ion collisions requires 
a framework for the dynamical evolution: 

\begin{itemize}
\item Initial state fluctuations: Fluctuations of the initial state 
are related to quantum mechanical fluctuations in the distribution of
initial sources in the transverse plane (``wounded nucleons"), and to
large multiplicity fluctuations in individual proton-proton collisions. 
The presence of large initial state fluctuations is experimentally well
established, based on the observation of odd Fourier moments of 
azimuthal flow \cite{Alver:2010gr}. 

 Initial fluctuations have to be propagated through the event using viscous 
fluid dynamics, combined with kinetic theory for the final stages. The rate 
at which the amplitude of a fluctuation is damped, as well as the rate at 
which fluctuations diffuse, depends on the value of transport coefficients 
and on the precise spatial structure of the initial state. In order to 
analyze data from the beam energy scan we also need to understand how initial
state fluctuations depend on beam energy and rapidity. 

\item Thermal fluctuations: As discussed above, local thermal fluctuations
arise from the finite size of the volume that thermodynamic variables are 
coarse grained over, and their magnitude is governed by equilibrium 
susceptibilities, which are derivatives of the equation of state. At RHIC
fluctuations of the net-proton number and charge have been observed
\cite{Adamczyk:2013dal,Adamczyk:2014fia,Rustamov:2017lio,Adam:2020unf}, and in principle
they can be related to lattice QCD calculations of the
susceptibilities \cite{Bazavov:2012vg,Borsanyi:2014ewa} provided one corrects
for baryon-number conservation \cite{Jeon:2003gk,Schuster:2009jv,Bzdak:2012an} as well as for
the fact that the experiment only measures protons 
\cite{Kitazawa:2011wh,Kitazawa:2012at}. 

  In an expanding system the growth, decay, and diffusion of fluctuations
depends on the history of the system, the length scale of the fluctuation 
and the transport coefficients. This is of particular importance
for critical fluctuations, because dynamical scaling implies that 
long-wavelength fluctuations evolve very slowly near a critical point. Furthermore,
different moments of fluctuation observables evolve at different rates
\cite{Kitazawa:2013bta,Mukherjee:2015swa}, 
making a naive comparison between a dynamical transit of a critical point
and an equilibrium estimate at the freeze-out surface impossible. In this
report we will discuss several implementations of the dynamical theory of 
fluctuations, based either on stochastic equations, or on deterministic
equations for higher order correlation functions. We will also discuss the 
problem of backreaction, the degree to which large fluctuations may 
affect the equation of state or the transport properties of the QGP. 

\item Hadronization: The formation of hadrons from a QGP is
an intrinsically quantum mechanical process and involves fluctuations. This 
is evident from  measurements of hadron production in $pp$ collisions, 
which clearly show non-thermal tails in multiplicity and momentum 
distributions. This feature is also present in most models of hadronization, 
which involve stochastic processes such as string fragmentation or 
coalescence. In fluid dynamics hadronization is typically implemented 
using the Cooper-Frye formula~\cite{Cooper:1974mv}. This particlization method is based on matching the 
conserved quantum numbers between fluid dynamical densities and kinetic
distribution functions across the freeze-out surface. If the kinetic framework
is based on particles, as in molecular dynamics, then this process also
involves a stochastic element, because we have to sample particles from
a distribution function~\cite{Oliinychenko:2019zfk}. 

  Any dynamical scheme for the evolution of fluctuation observables has to 
include not only a hadronization mechanism, but also a kinetic scheme for 
propagating fluctuations in the hadronic phase. Given that hadronization 
is a stochastic process, there is a question to what degree hadronization 
may wash out existing fluctuations, or create additional sources of 
fluctuations and correlations. 

\item Detection: Detectors have finite acceptance and imperfect detection
efficiency. Finite acceptance, coupled with global charge conservation 
leads to corrections to the measured fluctuation observables. Imperfect
efficiency also leads to additional sources of fluctuations not present in 
the underlying event \cite{Bzdak:2012ab,Bzdak:2016qdc,Nonaka:2018mgw}.

 Quantifying the magnitude of these corrections not only requires a
detailed understanding of the detector, but also detailed modeling of 
the evolution of initial state or dynamically created fluctuations
in rapidity and transverse momentum. 

\end{itemize}

This report provides a summary of the discussions and presentations at the Rapid Reaction Task Force (RRTF) "Dynamics of critical fluctuations: 
Theory -- phenomenology -- heavy-ion collisions" organized by the ExtreMe Matter Institute EMMI. It describes ideas in 
an active and ongoing research effort, and the discussions at the workshop
represented many different points of view. As a result, not all statements
in this report necessarily reflect the opinion of every single author. 

The document is organized as follows: In Section \ref{sec:dyn-fluc} we
discuss dynamical approaches to fluctuations in fluid dynamics. There are
two main frameworks, based on either stochastic equations for fluid dynamical variables (stochastic fluid dynamics), or on deterministic equations for 
correlation functions (hydro-kinetics). We also discuss the problem of 
hadronization and the issue of backreaction of fluctuations on the fluid
dynamical evolution. In Section \ref{sec:Exp} we discuss experimental challenges. In Section \ref{sec:Other} we discuss intersections and experimental opportunities related to fluctuation probes in other systems, in particular ultra-cold atomic gases. Additional details regarding a number of dynamical approaches are provided in an Appendix. 


\section{Theory of dynamical fluctuations}
\label{sec:dyn-fluc}

The study of physical effects arising from the presence of fluid dynamical
fluctuations in the context of relativistic heavy-ion collisions was for a long
time restricted to idealised systems with a large number of
symmetries~\cite{Kovtun:2003vj,Gavin:2006xd}. However, in recent years
significant theoretical and phenomenological 
effort has been made to bring the simulations of fluctuating fluid dynamics
closer to realistic scenarios. To this end two main avenues of simulating
fluid dynamics with noise have emerged: \emph{stochastic fluid dynamics} and
\emph{hydro-kinetics}, which are addressed in Sections \ref{sec:stochastic} and \ref{sec:hydrokin}\footnote{There is also a top-down approach of formulating the effective action for stochastic fluid dynamics~\cite{Kovtun:2014hpa, Crossley:2015evo,Haehl:2018lcu}, which we will not
discuss here.}.
Stochastic fluid dynamics refers to numerical implementations of viscous relativistic fluid dynamics
with a stochastic conservation law~\cite{LandauStatPart1,LandauStatPart2}
\begin{align}
  \label{eq:stochasticeom}
  \partial_\mu T^{\mu\nu} & = 0,\quad T^{\mu\nu}=T^{\mu\nu}_\text{ideal}+T^{\mu\nu}_\text{viscous}+S_\text{noise}^{\mu\nu} \,, \\
  \label{eq:stochasticcc}
  \partial_\mu J^{\mu} & = 0,\quad J^{\mu}=J^{\mu}_\text{ideal}+J^{\mu}_\text{viscous}+I_\text{noise}^{\mu} \,.
\end{align}
In this approach discretized  noise is sampled event-by-event and
the final observables are calculated after statistical averaging.
The other approach, called hydro-kinetics, corresponds to a set of deterministic kinetic equations for the two-point 
functions of fluid dynamical fields, which are derived from the linearisation of  stochastic
fluid dynamics around a background flow. 
In this approach the statistical average of noise is performed analytically 
in the derivation of the deterministic equations. 

We note that for the study of critical fluctuations, notably in form of higher-order cumulants, the inclusion of non-linearities is essential. In such studies, fluctuating fluid dynamics needs to be supplemented by 
a model containing critical fluctuations, which may be done by using existing fluid dynamical fields, as done in Sections \ref{sec:stodiff} and \ref{sec:transits}, or by introducing new, non-fluid dynamical degrees of freedom (see Sections~\ref{sec:chiral} and \ref{sec:Hydro+}) depending on which quantity one considers to be the critical slow mode. Similarly, one can choose to solve stochastic or deterministic equations of motion.
Finally, the experimental observables are given in terms of correlations of produced particles, therefore the conversion from fluid fields to particle degrees of freedom, i.e. particlization, is a necessary step, which we discuss in Section \ref{sec:conversion}.
During the RRTF meeting the current status, advantages and challenges of these approaches were discussed.

Before discussing the details of possible implementations, it is important to recognize the multiple scales
in the problem.  In one limiting case, the largest wavelength perturbations will be dominated by
the initial conditions. These perturbations of size $l_{\text{hydro}} \sim R_{\text{nucleus}}$ are not completely damped by dissipative
processes and will survive until the end of the expansion. However the evolution of such modes
can be affected by the influence of smaller scale $l_{\text{noise}}$ fluctuations, e.g., by 
the renormalization of effective transport coefficients and the equation of state. In the other limit,
the smaller scale structure of initial conditions will be damped or mixed with the stochastic noise
produced during the evolution.
Although these two scales are often well separated at each 
point $\tau,x$
\begin{equation}
   l_{\text{noise}}(\tau,x)
 \ll l_{\text{hydro}}(\tau,x)
,\end{equation}
in an expanding system propagating perturbations can move from one domain to another, e.g., 
even small thermal fluctuations  at initial time can be stretched
to long wavelengths at later times. Similarly, the 
divergence of the correlation length close to the critical point, will be capped by the dynamics of the system.
Therefore both spatial and temporal evolutions of
fluctuations have to be understood to identify the relevant physical observables to be measured in the experiments.
In the following subsections we discuss different implementations of dynamical fluctuations and the relevant scales in the problem.

\subsection{Implementation of stochastic fluid dynamics\label{sec:stochastic}}

The modeling of viscous relativistic fluid dynamics for heavy-ion collisions has made
significant conceptional and technological advances~\cite{Romatschke:2017ejr}, which goes beyond the relativistic
Navier-Stokes equations~\cite{landau2013fluid}. Numerical implementations of 3 + 1 dimensional fluid dynamics
using relaxation type equations exist and are publicly available (e.g. vHLLE~\cite{Karpenko:2013wva}, MUSIC~\cite{Schenke:2010nt}, ECHO-QGP~\cite{DelZanna:2013eua}).
However stochastic fluid dynamics, although rather advanced in non-relativistic settings~\cite{bell2007numerical,donev2011diffusive,balboa2012staggered},
has been a challenge to implement for the modeling of heavy-ion collisions. 
The stochastic energy-momentum tensor includes a thermal noise term, whose correlator is given by~\cite{LandauStatPart1,LandauStatPart2,Kovtun:2012rj,Kapusta:2011gt}
\begin{align}
&\langle S^{\mu\nu}(x_1)S^{\alpha\beta}(x_2)\rangle 
= 2T\left[
\begin{aligned}
&\eta \left(\Delta^{\mu\alpha}\Delta^{\nu\beta} + \Delta^{\mu\beta}\Delta^{\nu\alpha}\right) \\
&+\left(\zeta - \frac{2}{3}\eta\right)\Delta^{\mu\nu}\Delta^{\alpha\beta}
\end{aligned}
\right] \delta^{(4)}(x_1-x_2) .\label{eq:noisecor}
\end{align}
Similarly, the stochastic current contains a noise term which satisfies
\begin{equation}
\label{eq:currentnoisecorr}
  \langle I^\mu(x_1) I^\nu(x_2)\rangle = 2T\sigma \Delta^{\mu\nu} \delta^{(4)}(x_1-x_2)\,.
\end{equation}
Here, $\eta$, $\zeta$ and $\sigma$ denote the relevant transport coefficients shear viscosity, bulk viscosity and charge conductivity for the conserved charge, respectively. 
The local approximation of white noise, given by the Dirac $\delta$-function is an approximation of more complicated noise kernels, that can be obtained from microscopic calculations \cite{Xu:1999aq,Murase:2013tma} or causality arguments \cite{Kapusta:2017hfi}. The discretization of this Dirac-$\delta$ function leads to stochastic terms, which diverge $\delta \sim \frac{1}{\Delta t \Delta V}$ with decreasing grid spacing. There are several issues connected to it:
\begin{enumerate}
    \item Stochastic noise introduces a lattice spacing dependence,
    \item Correction terms due to renormalization become large for small lattice spacings,
    \item Large noise contributions can locally lead to negative densities,
    \item Large gradients introduced by the uncorrelated noise is a problem for partial differential equation (PDE) solvers.
\end{enumerate}
The currently available implementations of fluctuating fluid dynamics have shown that for more than one spatial dimension one is forced to limit the resolution scale of the stochastic terms, i.e.\ only attempt to simulate noise down to a particular filter length scale, which is larger
than the numerical grid spacing applied for the discretization of the deterministic fluid dynamical fields
 \begin{equation}
  l_\text{grid}< l_\text{filter} \lesssim l_{\text{noise}} \ll l_\text{hydro} \,.
\end{equation}
This can be justified physically, since at the shortest scale fluctuations decay
almost instantaneously to equilibrium and from the point of view of measurable observables, there is no need to simulate them dynamically.
A similar issue has been observed in nonequilibrium chiral fluid dynamics discussed in Section~\ref{sec:chiral}, where the noise field was effectively coarse-grained over the spatial extension of the equilibrium correlation length. Here, we discuss the various possibilities applied in stochastic hydrodynamical approaches.

\textbf{Murase et al.}: In \cite{Murase:2016rhl,Hirano:2018diu} the noise term is smeared by a Gauss distribution in rapidity and transverse direction. The widths of these Gaussians are chosen to be $\sigma_\eta=\sigma_\perp/\textrm{fm}=1-1.5$. The dependence on this choice is not discussed. A large enhancement of the flow coefficients $v_n$ is observed when noise is included. 

\textbf{Nahrgang et al.}: In \cite{Nahrgang:2017oqp, Bluhm:2018plm} the noise term is either propagated on a second grid with larger spacings $\Delta x=1\,\text{fm}$  than typically used for the deterministic hydrodynamical fields or coarse-grained over the same scale. Both the energy density and the variance of the energy density fluctuations show a strong linear dependence on $1/\Delta V$. It is therefore mandatory to introduce correction terms on the level of the equation of state and the transport coefficients.

\textbf{Singh et al.}: In \cite{Singh:2018dpk} a high-mode filter is applied. Locally a cut-off of $p_{\rm cut}=0.6/\tau_\pi$ is determined in each fluid cell. Then the noise field is Fourier transformed and all modes with $k>p_{\rm cut}$ are set to zero. After an inverse Fourier transform the noise field is smoothed. It is reported that energy conservation is verified and that the $v_n(2)$ are within statistical errors independent of $p_{\rm cut}$. In addition, it is shown that charged hadron multiplicities are little affected by the inclusion of fluctuations at this cut-off scale. One sees, however, that the coarse-graining scale that is introduced is quite large $>1$~fm in the transverse plane.

The renormalization of the equation of state and the transport coefficients in stochastic fluid dynamics codes in the presence of fluctuations is a challenging task. The nonlinearities which are introduced by the full fluid dynamical equations lead to corrections \cite{Kovtun:2011np,Chafin:2012eq}, as one can for example observe in the retarded shear-shear correlator 
\begin{equation}
 G^{xyxy}_{R,{\rm shear-shear}}(\omega,\mathbf{0})=-\frac{7T}{90\pi^2}\Lambda^3-i\omega\frac{7T}{60\pi^2}\frac{\Lambda}{\gamma_\eta}+{(i+1)\omega^{3/2}\frac{7T}{90\pi^2}\frac{1}{\gamma_\eta^{3/2}}}\, .
\label{eq:contr}
\end{equation}
One can identify the first term in Eq.~\eqref{eq:contr} as a cutoff-dependent contribution to the equilibrium pressure, while the second term is a cutoff-dependent contribution to the shear viscosity $\eta$. How this renormalization can be performed on the level of the numerical implementations represents an ongoing effort.

In summary, the clear advantage to implement the full $3+1$ dimensional event-by-event stochastic  fluid dynamics is obvious: it allows us to evaluate all the relevant observables like the $n$-point correlation functions within the existing frameworks for simulations of heavy-ion collisions. It is therefore straightforward to include the kinematic cuts as applied in the experiment as well as taking initial and final state fluctuations into account. In return, stochastic fluid dynamics can easily incorporate the study of e.g.\ heavy and hard probes in order to investigate the impact of fluctuations on other observables in heavy-ion collisions beyond criticality.

However, the numerical challenges of implementing stochastic noise, validation of the effective equation of state and the statistical averaging over a sufficient number of events is a significant computational task requiring large ressources.

\subsection{Implementation of deterministic hydro-kinetics \label{sec:hydrokin}}

As we have just discussed, solving stochastic fluid dynamics brings multiple new challenges compared to ordinary fluid dynamics. 
The Dirac $\delta$-function correlation of the noise in Eqs.~\eqref{eq:noisecor} and~\eqref{eq:currentnoisecorr} has to be regularized
in any numerical implementation and the stochastic terms make it difficult to apply standard PDE solvers.
More subtly, the non-linearities of fluid dynamical equations lead to noise induced corrections to the effective
equation of state and transport coefficients with divergent terms depending on the noise regularization cut-off. Therefore to simulate the cut-off independent physics the properties of fluid dynamical models have to be chosen in a non-trivial cut-off dependent way. 
Reproducing and understanding these subtle effects on a discrete grid is a considerable challenge
and an alternative way of solving stochastic fluid dynamical equations, known as the hydro-kinetic approach, was developed recently~\cite{Akamatsu:2016llw,An:2019osr}, although similar ideas in the non-relativistic setting have
been discussed earlier~\cite{andreev1971two, andreev1978corrections}.
The advantage of this approach is that
the divergent cut-off dependent terms are  absorbed in the renormalization of background fields and
the evolution equations for the two-point correlation functions can be formulated in terms of deterministic kinetic equations. 
In applications for heavy-ion collisions this approach was studied in the case of Bjorken boost-invariant expansion~\cite{Akamatsu:2016llw, Akamatsu:2017rdu, Martinez:2018wia} and recently generalized to arbitrary backgrounds in Ref.~\cite{An:2019osr}.

Hydro-kinetics depends on the separation of scales 
between long-wavelength fluid dynamical modes and short wavelength fluctuations, which stay
in equilibrium despite the expansion (see discussions in \cite{Akamatsu:2016llw, An:2019osr} and  also Appendix~\ref{sec:hydro-kinetics}).
Denoting the characteristic length-scale
$ l_{\text{noise}}$ marking the boundary between the expansion and dissipation dominated fluctuations we have
\begin{equation}
 l_{\text{micro}} \ll
   l_{\text{noise}} \ll l_{\text{hydro}}\,,
\end{equation}
where $ l_{\text{micro}}$ is the microscopic scale, e.g.\ the mean free path or inverse temperature $1/T$.
The length scale at which the diffusive processes begin to over-come the macroscopic gradients driving the system out of equilibrium is given by
\begin{equation}
    l_\text{noise} \sim ( \gamma l_\text{hydro}/c_s)^{1/2},
\end{equation}
where $\gamma$ is the corresponding diffusion constant, e.g.\ $\gamma_\eta \sim \eta/(e+p)$ for shear dissipation.
Then the equal time correlation function of fluid dynamical fields $\phi_A(t,{\bf x})$ represented by 
\begin{equation}
G_{AB}(t, {\bf x}, { \bf y}) = \left<\phi_A(t,{\bf x}), \phi_B(t,{\bf y})\right>
\end{equation}
will satisfy the equilibrium fluctuation-dissipation relation at length scales  $|{\bf x}-{\bf y}|\ll l_\text{noise}$, but will be driven away from equilibrium by long wavelength gradients over distances $|{\bf x}-{\bf y}|\gtrsim l_\text{noise}$.
The deviation of $G_{AB}(t, {\bf x}, { \bf y})$ from equilibrium gives the non-trivial corrections to the constitutive equations, which can be estimated to be of characteristic size $\sim (c_s/(\gamma l_\text{hydro}))^{3/2}$ and are known in the literature as ``long time tails" of fluid dynamical response~\cite{Kovtun:2003vj, Kovtun:2011np,andreev1971two,andreev1978corrections}.
It is important to note that such corrections are non-analytic indicating their non-local nature. In addition, in the fluid dynamical gradient expansion of constituent equations they come formally before the second order gradient terms, which are often included in relativistic fluid dynamical codes for stability and causality~\cite{Israel:1979wp}.

It is convenient to study the Wigner transform of the correlation function 
\begin{equation}
W_{AB}(t,{\bf x}, {\bf q}) = \int d^3 y\, G_{AB}(t, {\bf x}+{\bf y}/2, {\bf x}-{\bf y}/2) e^{-i{\bf qy}},
\end{equation}
as the separation of scales allows us to write  hydro-kinetic equations local in ${\bf x}$ for the relaxation of $W_{AB}(t,{\bf x}, {\bf q})$ to equilibrium. For the non-trivial relativistic case the notion of equal time correlation  functions has to be revised, which was recently accomplished in ref.~\cite{An:2019osr}.
Linearizing the equations of motion, Eq.~\eqref{eq:stochasticeom}, one derives
the evolution equations for the perturbation fields $\phi^A=(c_s \delta e, w \delta u^\mu)$, which in turn can be used to calculate the time dependence of the two-point correlation functions. After lengthy calculations~\cite{Akamatsu:2016llw,An:2019osr} one arrives at hydro-kinetic equations for two propagating sound modes ($\pm$) and three diffusive modes for a fluid with no conserved charges. For example, for a sound mode one has
\begin{equation}
    \left[(u+v)\cdot \bar{\nabla} +f\cdot \frac{\partial}{\partial q}\right] W_+ =
    - \gamma_L q^2 (W_+ - W^{(0)}) + K'' W_+\label{eq:hkinetics}\,,
\end{equation}
where the left hand side is equivalent to the Liouville operator for a phonon with space-time dependent dispersion relation. On the right hand side one gets the relaxation term to equilibrium and the forcing term $K''$ proportional to fluid gradients.
Once the $W_{AB}(x,q)$ is determined, the contribution to the energy momentum tensor at a point is given by the momentum integral of the Wigner distribution. The analysis of such contributions reveals the
divergent universal corrections to the background equation of state and transport coefficients, which can be absorbed or renormalized. The remaining finite term (long-time tails) is particular to the given background expansion and has to be evolved dynamically.

The outstanding challenge of deterministic hydro-kinetics is the
application to a realistic QGP expansion in nuclear collisions.
Formally the hydro-kinetic equation, Eq.~\eqref{eq:hkinetics}, requires solving 3+3+1 dimensional equations, i.e.\ 3-dimensional momentum space equations
for each space-time point, to find out the equal-time correlation functions of the fluid dynamical fields. This is obviously numerically demanding in general, but the hydro-kinetic equations are linear and smooth, therefore one does not need fine momentum-space
discretization to accurately solve the equations. In addition,  hydro-kinetic equations
could be solved using fictitious test particles which move on top of a 
fluid dynamical background solved using traditional approaches.
One should note here that 
deterministic fluid dynamical simulations do not need to be repeated to obtain the statistical averages over thermal fluctuations. 
However, the currently derived hydro-kinetic equations are limited to two-point functions.
Interesting higher order correlation functions therefore require the generalization of this scheme, which is currently not done even for
simple backgrounds.

\subsection{Implementation of stochastic diffusion
\label{sec:stodiff}}

Numerical simulations of the dynamics of fluctuations in the conserved net-baryon number $N_B$ both on the crossover and first-order phase transition sides near the conjectured QCD critical point have recently been performed for one spatial dimension without~\cite{Sakaida:2017rtj,Nahrgang:2017hkh,Bluhm:2019yfb} and with non-linearities~\cite{Nahrgang:2018afz}. Considering the net-baryon density $n_B$ as the slow critical mode~\cite{Hohenberg:1977ym,Son:2004iv,Fujii:2004jt}, the dynamics of critical fluctuations may be studied by means of a stochastic diffusion equation in the form 
\begin{equation}
 \partial_t n_B = \Gamma \nabla^2\bigg(\frac{\delta {\cal F}[n_B]}{\delta n_B}\bigg) + \vec{\nabla}\cdot\vec{{\cal J}} \,.
 \label{eq:diffeq1}
\end{equation}
This equation describes the non-relativistic evolution of the current $J_B^\mu$ in Eqs.~\eqref{eq:stochasticcc} with~\eqref{eq:currentnoisecorr}, which is decoupled from the evolution of energy and momentum densities, under the assumption of a spatially homogeneous temperature and a space-time independent fluid velocity field. The fluctuation dynamics is governed by the minimization of the free energy ${\cal F}$ in the system. The particular form of the free energy studied in the numerical simulations together with a discussion of the parameters and how criticality is embedded can be found in Appendix~\ref{app:stochasticdiff}. For a stochastic current $\vec{{\cal J}}$ of the form
\begin{equation}
 \vec{{\cal J}} = \sqrt{2 T \Gamma} \vec{\zeta}
\end{equation}
and mobility coefficient $\Gamma=Dn_c/T$ Eq.~\eqref{eq:diffeq1} becomes
\begin{multline}
 \partial_t n_B(x,t) = \frac{D}{n_c} \left(m^2 \nabla_x^2 n_B - K\nabla_x^4 n_B\right) + \sqrt{2Dn_c/A}\,\nabla_x\zeta_x(x,t) 
 \\ 
 + D\nabla_x^2\left(\frac{\lambda_3}{n_c^2}\, (\Delta n_B)^2 + \frac{\lambda_4}{n_c^3}\, (\Delta n_B)^3 + \frac{\lambda_6}{n_c^5}\, (\Delta n_B)^5\right) \,.
\label{eq:stochdiff}
\end{multline}
Here, $D$ is the diffusion coefficient and $\zeta_x$ is the white noise $x$-component with zero mean and covariance $\langle\zeta_x(x,t),\zeta_x(x',t')\rangle=\delta(x-x')\delta(t-t')$. This ensures that the fluctuation-dissipation balance is guaranteed. 

The stochastic diffusion equation is solved numerically with a semi-implicit predictor-corrector scheme in which the non-linear terms in $\Delta n_B$ are treated explicitly. Equation~\eqref{eq:stochdiff} is valid for the propagation of fluctuations in one spatial dimension where the physics in the transverse area $A$ has been scaled out. A static box of finite length $L$ is considered with a resolution $\Delta x=L/N_x$ for $N_x$ lattice sites. Charge conservation is exactly realized by imposing periodic boundary conditions. The numerical framework has been tested extensively in both limits of a Gaussian ($K=\lambda_i=0$) and Gauss$+$surface ($\lambda_i=0$) model as discussed in~\cite{Nahrgang:2017hkh} and~\cite{Bluhm:2019yfb}, respectively. For these models analytic results both for the continuum and discretized space-time are available that the numerics can be confronted with. One notes that for a meaningful comparison charge conservation in a finite-size system must be included in the analytic results. It is found that the numerics can accurately reproduce the analytic expectations for the static and dynamic structure factor, the correlation function and the local variance for a given $\Delta x$. This implies that the lattice spacing dependence of physical observables is well under control. Moreover, the continuum expectations are approached with $\Delta x\to 0$ which highlights that there is neither the need for renormalization nor a coarse-graining or filtering of the noise and the algorithm can well handle white noise on a finite grid of $\Delta x$ and $\Delta t$.

\begin{figure}[t]
 \centering
 \includegraphics[width=0.34\textwidth]{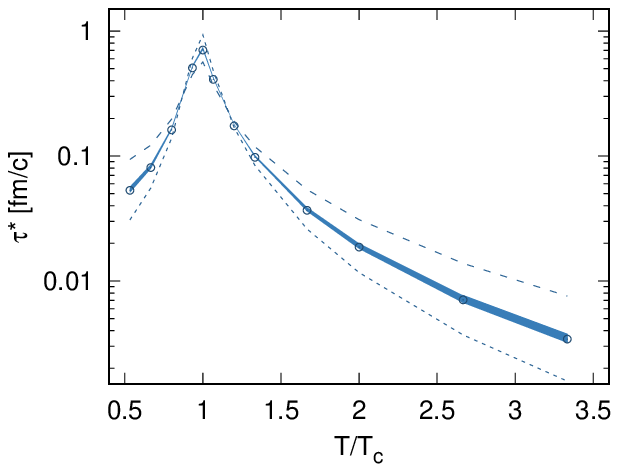}
 \hspace{-3mm}
 \includegraphics[width=0.34\textwidth]{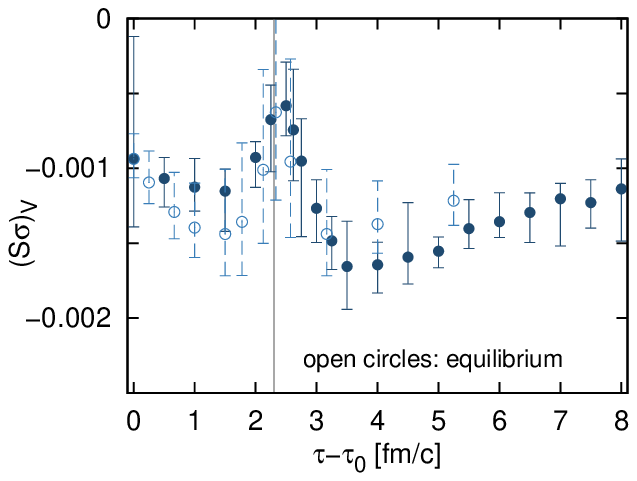}
 \hspace{-4mm}
 \includegraphics[width=0.34\textwidth]{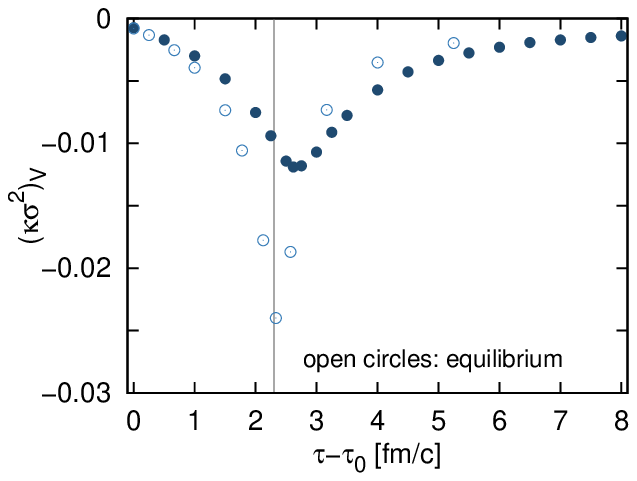}
 \caption{Left panel: scaling behavior of the relaxation time $\tau^*$ (circles) with $\xi$ for modes with $k^*=1/\xi$ as a function of $T/T_c$. The filled band shows the scaling $\propto\xi^z$ with $z=4\pm 0.1$. In comparison, $z=3$ (dashed line) and $z=5$ (dotted line) can be excluded. Figure taken from~\cite{Nahrgang:2018afz}. Middle and right panels: dynamical evolution (full circles) of volume-integrated skewness $(S\sigma)_V$ and kurtosis $(\kappa\sigma^2)_V$ for a system in which $T$ varies as a function of time $\tau-\tau_0$ in comparison with corresponding equilibrium results (open circles). $T_c$ is reached at $\tau-\tau_0=2.3$~fm$/$c. Figures modified from~\cite{Nahrgang:2018afz}.  \label{fig:stochdifffigures}}
\end{figure}
In Fig.~\ref{fig:stochdifffigures} some highlight results of this framework are shown. The employed parameters read $n_c=1/(3\,$fm$^3)$, $T_c=0.15$~GeV, $\xi_0=0.479$~fm, $\tilde K=1$, $\tilde\lambda_3=1$, $\tilde\lambda_4=10$ and $\tilde\lambda_6=3$, see Appendix~\ref{app:stochasticdiff}. In the left panel of Fig.~\ref{fig:stochdifffigures} the relaxation time $\tau^*$ (circles) of the critical mode with $k^*=1/\xi$ for a given fixed $T$ is contrasted with a scaling function proportional to $\xi^z$. It is found that the numerics is best described with $z\simeq 4$ (filled band) which shows that the expected dynamic critical scaling of model B is realized numerically. For this plot the correlation length $\xi$ is deduced from the behavior of the equal-time correlation function $\langle\Delta n_B(r)\Delta n_B(0)\rangle$. Moreover, the relaxation time $\tau_k$ is obtained from the exponential decay $\propto e^{-t/\tau_k}$ of the dynamic structure factor $\langle\Delta n_B(k,t_0+t)\Delta n_B(-k,t_0)\rangle$ with time. For fixed wave-number, $\tau_k$ is larger for temperatures near $T_c$ than further away, and it decreases with increasing $k$ for fixed $T$. In the middle and right panels of  Fig.~\ref{fig:stochdifffigures} the volume-integrated skewness $(S\sigma)_V$ and kurtosis $(\kappa\sigma^2)_V$ are shown. These are obtained for a subregion of observation $V\simeq 2$~fm smaller than $L$ for a dynamically evolving system (full circles) and compared to the static equilibrium limit (open circles). The evolution takes place in form of a time dependence of the background temperature via $T(\tau)=T_0(\tau_0/\tau)$ starting in equilibrium at $\tau_0=1$~fm with $T_0=0.5$~GeV and $D(\tau_0)=1$~fm which then decreases as $D(\tau)=D(\tau_0) T(\tau)/T_0$. The non-linear terms in Eq.~\eqref{eq:stochdiff} are essential for skewness and kurtosis to develop from purely white noise. One observes that the non-Gaussian fluctuations behave non-monotonically, and that in particular $(\kappa\sigma^2)_V$ increases significantly near $T_c$ compared to its value at $T_0$ or $\tau_0$. Nonetheless, even in equilibrium (open circles) finite-size effects can modify the infinite-volume expectations~\cite{Stephanov:2008qz} of the scaling behavior with $\xi$ dramatically~\cite{Nouhou:2019nhe}. This can, in particular, be seen in the structure of $(S\sigma)_V$ which is a consequence of the competition of different scalings, see~\cite{Nahrgang:2018afz}. The evolution of $T$ (full circles) results in dynamical, non-equilibrium effects notably a reduction of the fluctuation signals. Moreover, as a consequence of the finite relaxation times, the observables in the dynamical setting lag behind their equilibrium values. Both effects, which can also be seen in the variance~\cite{Nahrgang:2017hkh,Nahrgang:2018afz,Bluhm:2019yfb}, become more pronounced with decreasing $D(\tau_0)$. 

For a realistic modeling of the physics in a heavy-ion collision the current framework still needs to be extended. In particular, a realistic spatio-temporal evolution of the fireball must be embedded. A first step into this direction is to consider a sytem undergoing a Bjorken-type expansion. Corresponding works are currently underway. With this the coupling of the dynamics of critical fluctuations to the evolution of other fluctuating fluid dynamical fields becomes feasible. This will allow one to quantify, for example, the impact of the critical fluctuations on the medium and vice versa or to study the role of advection. Eventually, the framework must be extended to three spatial dimensions. Only then one may study to what extent the dynamics of the fluctuations in the longitudinal direction is decoupled from the dynamics in the transverse direction as was assumed so far. This will necessitate, however, a careful analysis and understanding of renormalization effects. Nonetheless, the coupling to the evolution of the transverse velocity field will allow one for the first time to study numerically the physics of model H as the assumed dynamical universality class of QCD. Further future developments range from including realistic fluctuating initial conditions, to study the interplay and competition of different fluctuation sources, to embedding the conversion to measurable particles at chemical freeze-out by explicit charge conservation on an event-by-event basis, see section~\ref{sec:conversion}. 

\subsection{Implementation of nonequilibrium chiral fluid dynamics (N$\chi$FD) \label{sec:chiral}}

In order to study the dynamics of critical fluctuations properly we need to include their evolution equations coupled to a fluid dynamical evolution. Within the framework of nonequilibrium chiral fluid dynamics, the chiral condensate $\sigma=\langle \bar{q}q\rangle$, which is considered as the critical mode, is propagated via a relaxation equation of the following form,
\begin{equation}
\label{eq:eomsigma1}
 \partial_\mu\partial^\mu\sigma+\eta\partial_t \sigma+\frac{\delta \Omega}{\delta\sigma}=\xi~.
\end{equation}
The damping coefficient $\eta$, the noise $\xi$, and the potential terms $\Omega$ can be obtained from an effective model of QCD, like the quark-meson (QM) or Polyakov-quark-meson (PQM) model. In the works \cite{Nahrgang:2011mg, Nahrgang:2011mv, Nahrgang:2011vn,Nahrgang:2013jx,Herold:2013bi,Herold:2014zoa, Herold:2016uvv,Herold:2017day,Herold:2018ptm} the mean-field approximation of the (P)QM model was applied. In a recent QCD assisted transport model \cite{Bluhm:2018qkf} the equilibrium input is provided by FRG calculations.

It is assumed that the fluid consisting of the fermionic degrees of freedom and the fast modes of the sigma field are the heat bath in which the chiral order parameter $\sigma$ evolves. Due to the mutual coupling the fluid equilibrates locally under the condition of the actual value of $\sigma$. The fluid dynamical pressure is therefore not determined at the mean-field value of $\sigma$ but includes the backreaction of $\sigma$ on the fluid. It depends explicitly on the fluctuations of the order parameter
\begin{equation}
\label{eq:pressure1}
 p(T,\mu; \sigma) = -\Omega_{\rm q\bar q}(T,\mu; \sigma)~.
\end{equation}

Contrary to standard Langevin-simulations the heat bath is not static, but evolves according to the equations of fluid dynamics, and describes the bulk evolution of a heavy-ion collision. Therefore, the total energy and momentum of the coupled system of the fluid and the order parameter need to be conserved. This is achieved by adding a source term to the standard fluid dynamical equations,
\begin{align}
\label{eq:fluidT1}
\partial_\mu T^{\mu\nu}&=-\partial_\mu T_\sigma^{\mu\nu}~,\\
\label{eq:fluidN1}
\partial_\mu N^{\mu}&=0~.
\end{align}
The stochastic nature of the source term on the right hand side of Eq.\ \eqref{eq:fluidT1} leads to a stochastic evolution for the fluid dynamical fields. Eqs.~\eqref{eq:eomsigma1} - \eqref{eq:fluidN1} are coupled and as a result of Eq.~\eqref{eq:pressure1}, the evolution of the fluid and the order parameter feed back on each of the other. More details on N$\chi$FD can be found in the Appendix \ref{app:nchifd}. It has been applied to calculating various observables in heavy-ion collisions, notably the critical enhancement of net-proton fluctuations \cite{Herold:2016uvv}.

\begin{figure}
    \centering
    \includegraphics[width=0.21\textwidth, angle=-90]{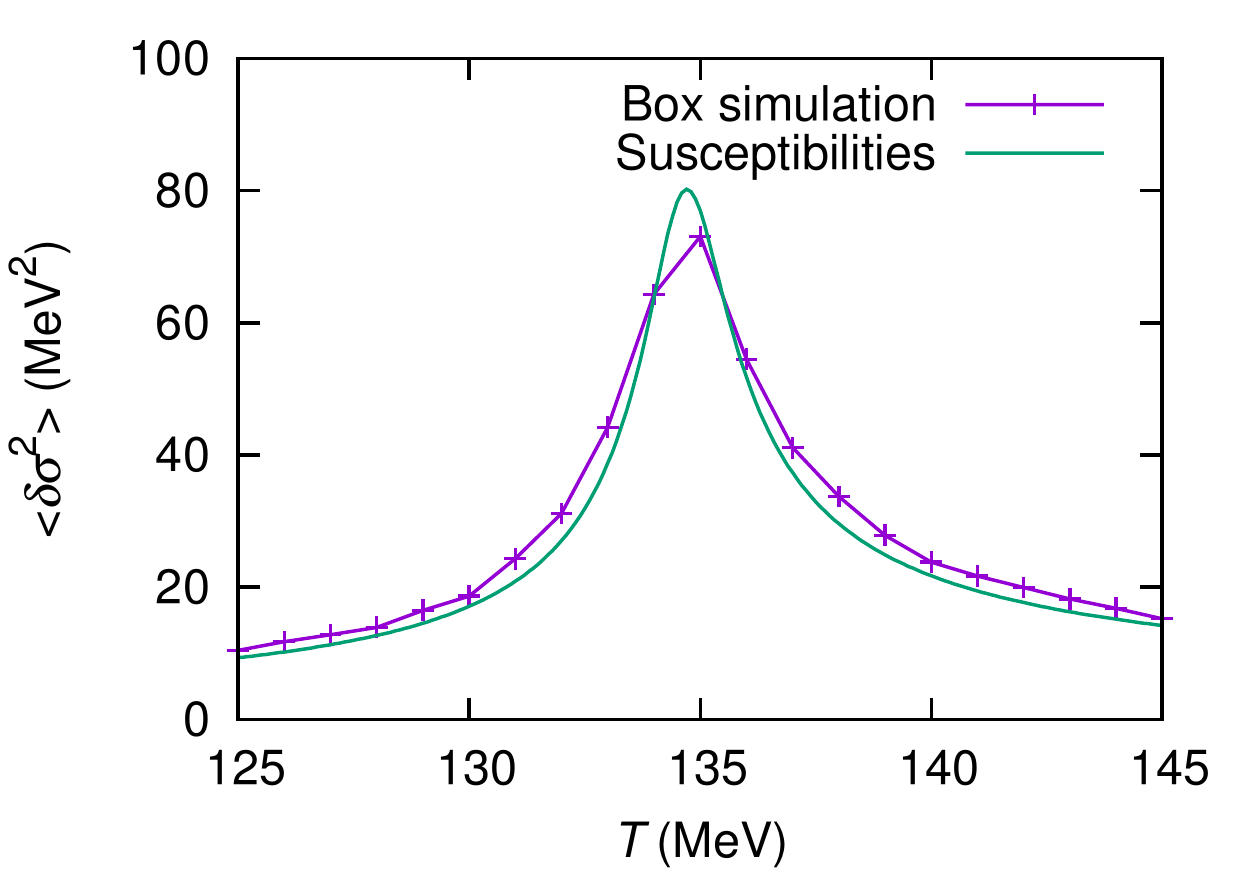}\includegraphics[width=0.21\textwidth, angle=-90]{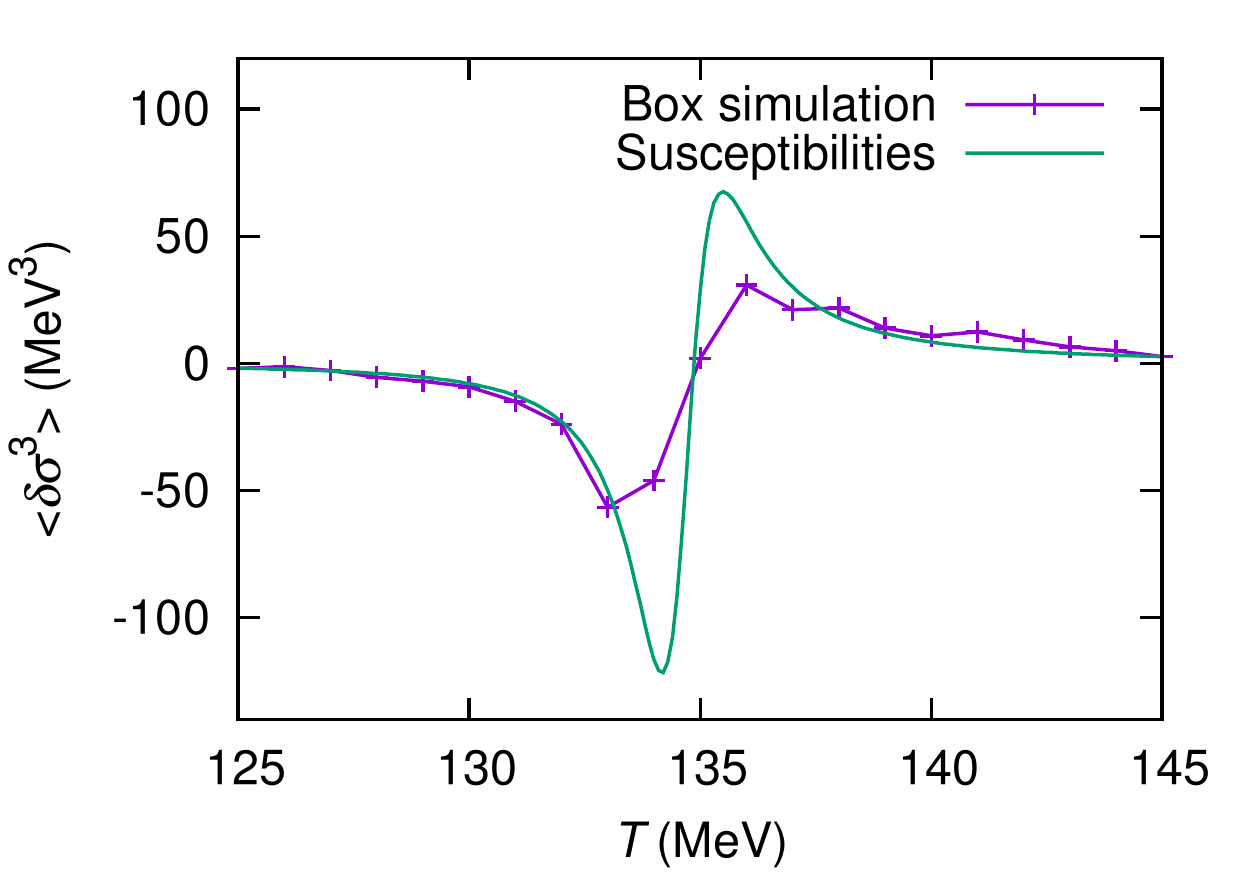}\includegraphics[width=0.21\textwidth, angle=-90]{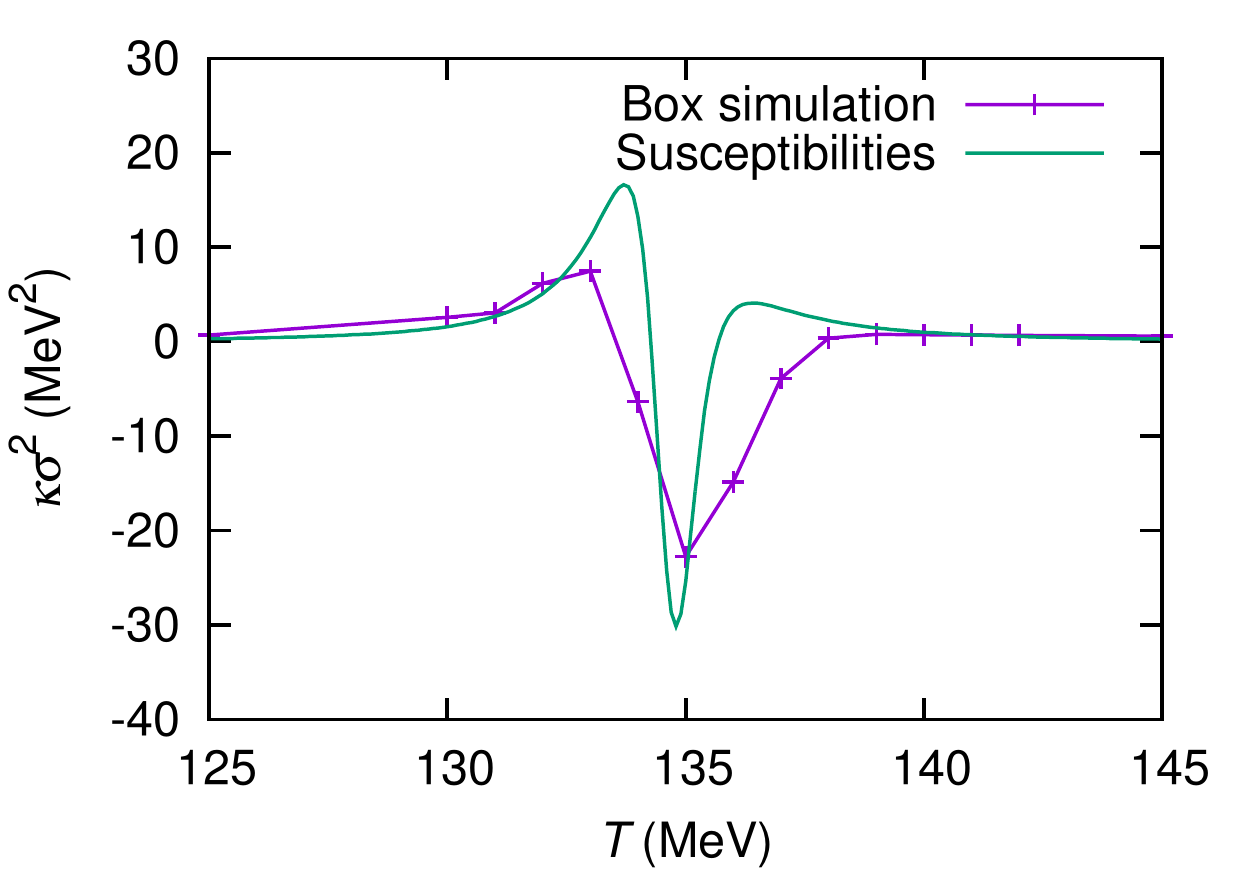}
    \caption{Comparison of susceptibilities obtained from nonequilibrium chiral fluid dynamics in a box compared to the thermodynamical expectation (left figure taken from~\cite{Herold:2017omo}).}
    \label{fig:nchifd}
\end{figure}

In order to avoid an unphysical dependence on the lattice spacing, we model a spatial correlation of the noise field over a correlation length of $1/m_\sigma$, where $m_\sigma$ is the local equilibrium screening mass. This procedure is a regularization method of the otherwise white noise correlator, as discussed previously.
The full solution of Eqs.\ \eqref{eq:eomsigma1},  \eqref{eq:fluidT1}, \eqref{eq:fluidN1} is obtained in $3+1$ dimensions. It can be expected that the input equation of state is modified due to the cutoff (either $\Delta x$ or the spatial correlation of the noise field). This could explain the quantitative differences of the susceptibilities, which are obtained in static box simulations, compared to the thermodynamic expectations, see Fig.~\ref{fig:nchifd}. 
One should therefore check the equation of state in these box simulations to see if modifications to the original $P_0(T,\mu_B)$ can be observed. This is rather complicated as many calculations need to be performed at various temperatures and baryo-chemical potentials all over the phase diagram. It is assumed to be easier to derive an analytic formula for the correction (see Section~\ref{sec:hydro-kinetics}) and fix the coefficients with a couple of test calculations. The boundary conditions must be fixed coherently and the finite piece of the correction needs to be treated separately. The corresponding calculations and tests are currently ongoing. 

To perform calculations in the entire phase diagram it is important to have a reliable equation of state, which correctly describes the hadronic phase at high baryon densities but also retains the non-equilibrium fluctuations of the order parameter. First calculations have been performed for the equation of state of a hadronic SU(3) non-linear sigma model with quarks \cite{Dexheimer:2009hi,Nahrgang:2016eou}.

In QCD-assisted transport \cite{Bluhm:2018qkf} a similar equation of motion for the chiral condensate as in Eq.~\eqref{eq:eomsigma1} is solved. It contains a kinetic term related to the real part of the effective action $\Gamma^{\, (2)}_{\sigma\sigma}$, a diffusion term sensitive to the imaginary part of $\Gamma^{\, (2)}_{\sigma\sigma}$, and an effective potential, which can be obtained in FRG calculations. This description provides a systematic approach to the dynamics of the chiral order parameter, which is valid beyond mean field and beyond the scaling region around the critical point, which might be very small. A detailed description can be found in the Appendix \ref{app:QCDassisted}.

\begin{figure}
	\begin{subfigure}[t]{0.47\textwidth}
		\includegraphics[width=\textwidth]{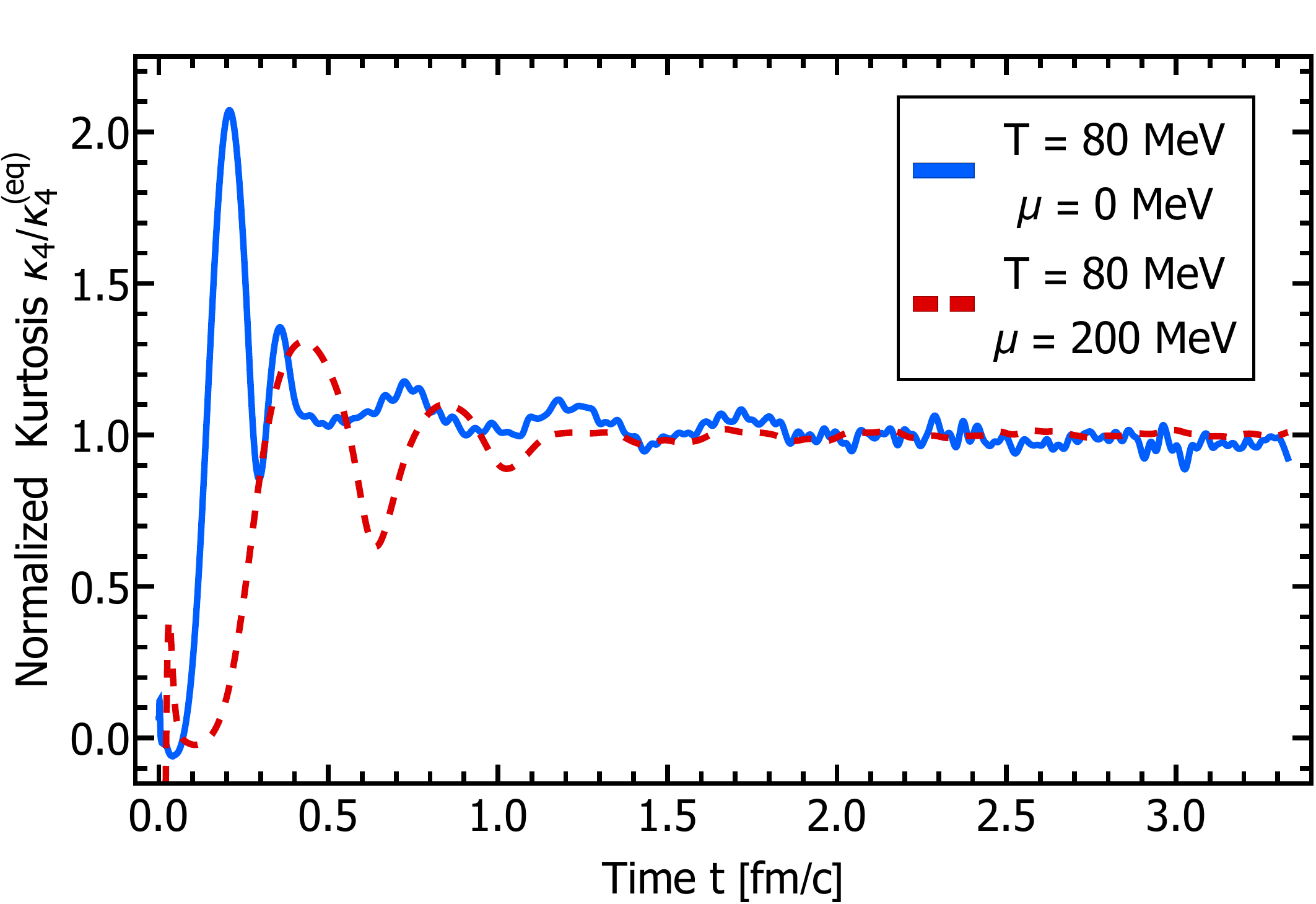}
	\end{subfigure}
	\hfill
	\begin{subfigure}[t]{0.47\textwidth}
		\includegraphics[width=\textwidth]{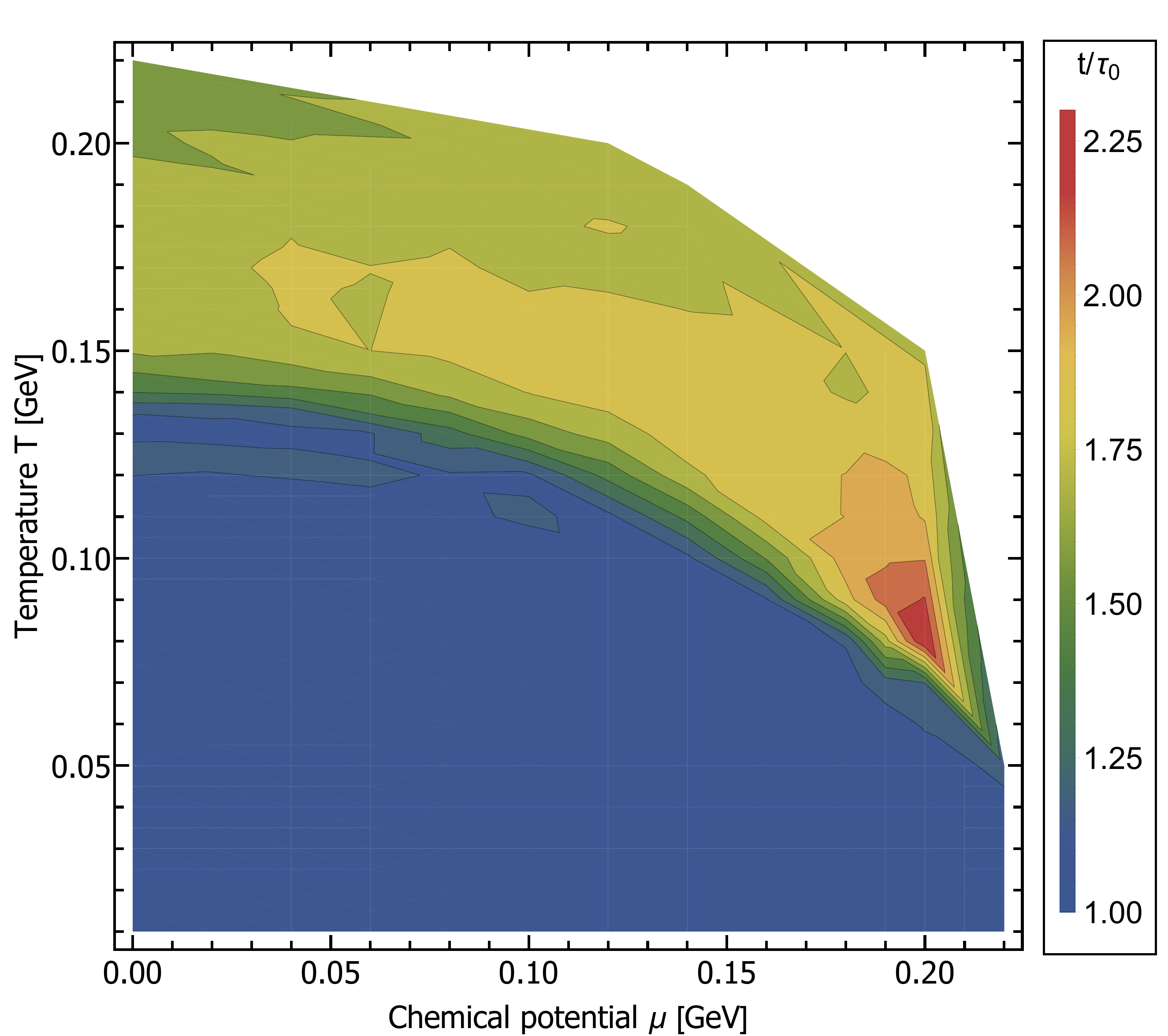}
	\end{subfigure}
	\caption{Left: Scaled kurtosis as a function of time for a
		quench from high $T$ to two different points in the phase
		diagram. Within statistical deviations, the equilibration
		time is found to be significantly increased near the
		critical endpoint (red, dashed curve) compared to a quench
		far away from it (blue, solid curve). Right: Equilibration
		time $t$ in units of $\tau_0\simeq 0.4$~fm/c in the QCD
		phase diagram based on the analysis of the scaled kurtosis
		in the quench scenario (see left panel). 
		Figures taken from~\cite{Bluhm:2018qkf}.}
	\label{fig:equilibration}
\end{figure}

As an example result of QCD assisted transport we show in \autoref{fig:equilibration} (left panel) the time-evolution of the kurtosis scaled by its late-time equilibrium limit for the quench from high temperatures to two different points in the QCD phase diagram. Far away from the critical endpoint the scaled kurtosis exhibits a rather quick equilibration while close to it the corresponding time scale is clearly increased. For the quench through the phase boundary one furthermore observes that the equilibrium value is approached from above as the equilibrium kurtosis is larger near the phase boundary than in the low-temperature phase.

Based on these results for the scaled kurtosis in the quench scenario, one may estimate the equilibration time of the critical fluctuations within the QCD phase diagram. This is shown in \autoref{fig:equilibration} (right panel). One can clearly identify both the phase boundary and the region near the critical endpoint and observe the expected increase of the equilibration time in that region. Nevertheless, this increase is found to be rather moderate suggesting that phenomena associated with critical slowing down are only moderately pronounced. This hints towards equilibrium dominated measurements and, thus, to the feasibility of studying the QCD phase diagram by means of heavy-ion collisions. For quantitative statements, however, the dynamical modeling of the fluctuations remains necessary. 

\subsection{Implementation of Hydro+ \label{sec:Hydro+}}

%
%
{

\def\half{{\textstyle\frac{1}{2}}}
\def\scl{{\rm cr}}
\def\crs{{\rm cr}}
\def\Cp{{C_p}}
\def\third{{\textstyle\frac{1}{3}}}
\def\quarter{{\textstyle\frac{1}{4}}}
\def\fifth{{\textstyle\frac{1}{5}}}
\newcommand{\punch}[1]{{\color{red} \emph{#1}}}

\newcommand*\bbar[1]{%
  \vbox{%
    \hrule height 0.5pt
    \kern-0.4ex
    \hbox{%
      \kern-0.2em
      \ifmmode#1\else\ensuremath{#1}\fi
      \kern-0.1em
    }
  }
}

%
%
\def \na {\nabla}
\def \pd {\partial}

%

\def\p{{\bf p}}

\def\st{\begin{equation}}
\def\stp{\end{equation}}
\def\bg{\begin{eqnarray}}
\def\nd{\end{eqnarray}}
\def\Eq#1{Eq.~(\ref{#1})}
\def\Eqs#1{Eqs.~(\ref{#1})}
\def\eq#1{(\ref{#1})}
\def\app#1{Appendix~\ref{#1}}
\def\Fig#1{Fig.~\ref{#1}}
\def\Figs#1{Figs.~\ref{#1}}
\def\Sect#1{Sect.~\ref{#1}}
\def\Ref#1{Ref.~\cite{#1}}
\def\snn{{\sqrt{s_{\scriptscriptstyle NN}}}}
\def\llangle{\left\langle}
\def\rrangle{\right\rangle}
\def \bes {\begin{subequations}}
\def \ees {\end{subequations}}

\def \S{{\mathcal S}}
\def\L{{\mathcal L}}
\def\N{{\mathcal N}}
\def\nh{{\hat n}}
\def\E{{\mathcal E}}

%
%
\def \a {\alpha}
\def \b {\beta}
\def \c {\chi}
\def \d {\delta}
\def \e {\epsilon}
\def \g {\gamma}
\def \k {{\bm k}}
\def \o {\omega}
\def \l {\lambda}
\def \m {\mu}
\def \n {\nu}
\def \s {\sigma}
\def \t {\tau}
\def \th {\theta}
\def \x {\xi}
\def \D {\Delta}
\def \G {\Gamma}
\def \O {\Omega}
\def \P {\psi}

%
%
\def \vp {\bm{p}}
\def \vq {\bm{q}}
\def \vx {\bm{x}}
\def \vy {\bm{y}}
\def \vs {\bm{s}}
\def \vo {\bm{\o}}
\def \vr {\bm {r}}
\def \va {\bm{a}}
\def \vb {\bm{b}}
\def \vk {\bm{k}}
\def \vj {\bm{j}}
\def \vA {\bm{A}}
\def \vG {\bm{G}}
\def \vL {\bm{L}}
\def \vB {\bm{B}}
\def \vH {\bm{H}}
\def \vR{\bm{R}}
\def \vV{\bm{V}}
\def \vE {\bm{E}}
\def \vQ {\bm{Q}}
\def \vW{\bm{W}}
\def \vT {\bm{T}}
\def \vP {\bm{P}}
\def \vv {\bm{v}}
\def \vY {\bm {Y}}
\def \vX {\bm{X}}

%
%
\def \<{\langle}
\def \>{\rangle}
\def \+{\dagger}
\def \({\left(}
\def \){\right)}
\def \[{\left[}
\def \]{\right]}
%
%
\def\no{\nonumber}

%
%

%
%
\def\crit{\textrm{crit}}
\def\Ising {\textrm{Is}}
\def\reg{\textrm{reg}}
\def\QGP{\textrm{QGP}}
\def\CO{\textrm{C.O.}}
\def\noCP{\textrm{no C.P.}}
\def\CP{\textrm{C.P.}}
\def\adia{\textrm{adia}}
\def\eff{\textrm{eff}}
\def \fm {\textrm{fm}}
\def \MeV {\textrm{MeV}}
\def \GeV {\textrm{GeV}}
\def\BR{\textrm{B.R.}}
\def\noBR{\textrm{no B.R.}}

%
%
\def\plus{(+)}
\def\phieq{\overline{\phi}}
\def\Gammaeq{\overline{\G}}
\def\CVS{{\cal C}_{V}}
\def\sec{Sect.}
\def \QKZ {Q^{*}}

\def \Qscaled{\widetilde{Q}}


\def \eq{Eq.~}

\newcommand{\YY}[1]{{\blue{(YY: #1 )}}}
\newcommand{\KR}[1]{{\red{  (KR: #1 )}}}
\newcommand\gr[1]{{\textbf{{\color{purple}(GR: #1)}}}}
\newcommand{\RW}[1]{{\dgreen{(RW: #1 )}}}

\newcommand{\blue}[1]{\textcolor{blue}{#1}}
\newcommand{\redout}[1]{\textcolor{red}{\sout{#1}}}
\newcommand{\red}[1]{\textcolor{red}{#1}}
\newcommand{\green}[1]{\textcolor{green}{#1}}
\definecolor{dgreen}{rgb}{0.0, 0.5, 0.0}
\newcommand{\dgreen}[1]{\textcolor{dgreen}{#1}}
\newcommand{\greenbox}[1]{\colorbox{green}{#1}}

In the spirit of hydro-kinetics, the recently developed Hydro+ formalism allows for a consistent, deterministic description of both the dynamics of a fluid -- 
which are described by the standard fluid dynamical variables $\varepsilon$ (the energy density), $u^\mu$ (the fluid four-velocity) and $n_B$ (the baryon number density) --
and the out-of-equilibrium critical fluctuations induced by a critical point, 
including the feedback between the fluid dynamical variables and critical fluctuations.  The formulation of Hydro+ can be found in Ref.~\cite{Stephanov:2017ghc} and its numerical implementation for a heavy-ion motivated model can be found in Ref.~\cite{Rajagopal:2019xwg}. Many details omitted in this section can be found in these two references. 

In Hydro+, the critical fluctuations are encoded in the Wigner transform of the equal-time two-point function of the fluctuation of an order parameter field $M(t,\vx)$:
\begin{eqnarray}
\label{phi-def}
    \phi_{\vQ}\(t,\vx\)
    \equiv \int d^3{\vy}\, \<\d M\(t,\vx-\vy/2\)\,\d M\(t,\vx+\vy/2\)\>\, e^{-i\vy\cdot\vQ}\, ,
\end{eqnarray}
where $\delta M\(t,\vx\) \equiv M\(t,\vx\) -\<M\(t,\vx\)\>$, with $\<\ldots\>$ denoting the ensemble average.
If we consider the dynamics of a cooling droplet of QGP with $\mu_B=0$, namely undoped QGP with zero net baryon number, allowing us to drop baryon density $n_{B}$,  
and if we set the bulk viscosity to zero (although the relaxation of $\phi_{\vQ}$ still leads to an effective bulk viscosity),
the Hydro+ equations become
\bes
\label{hydro-eq-we-use}
    \begin{eqnarray}
    D\, \varepsilon &=& -\(\varepsilon+p_{\plus}\)\theta + \frac{1}{2}\Pi^{\mu\nu}\, \sigma_{\nu\mu}\, , 
    \\ \label{acceleration-fluid}
    \(\varepsilon+p_{\plus}\)\, D \, u^{\mu}
    &=&\nabla^{\mu}p_{\plus} - \Delta^{\mu}_{\nu}\nabla_{\sigma}\Pi^{\nu\sigma}
    + \Pi^{\mu\nu}D\, u_{\nu}\, ,
    \\
    \label{Pi-eq}
    \tau_{\Pi}\, \Delta^{\mu}_{\alpha}\,\Delta^{\nu}_{\beta}\, D\, \Pi^{\a\b}&=&
    -\Pi^{\mu\nu}+\eta_{\plus}\,\sigma^{\mu\nu} - \tau_{\Pi}\, 
    \( \Pi^{\alpha\mu}\,\omega^{\nu}_{\alpha}+\Pi^{\alpha\nu}\,\omega^{\mu}_{\alpha} \)
    \\
    D\,\phi_{\vQ}(t,x)
    &=& -
    \Gamma_{\vQ}
    \left( \phi_{\vQ} - \phieq_{Q} \right),
\label{eqn:phi_relaxation}
\end{eqnarray}
\ees
where we have followed the Muller-Israel-Stewart formalism and have introduced
terms involving the shear tensor $\Pi^{\mu\nu}$ to maintain causality of our equations. We have defined $D = u^\mu \partial_\mu$ and $\phieq_{Q}$ as the equilibrium value of $\phi_{\vQ}$, with all other quantities defined in Ref.~\cite{Rajagopal:2019xwg}.
These equations are very similar to standard fluid dynamical equations~\cite{Baier:2006gy}, except now $\phi_{\vQ}\(t,x\)$ is treated as a dynamical variable in Eqn.~\eqref{eqn:phi_relaxation} and obeys a relaxation equation, and standard fluid dynamical variables like pressure $p$, shear viscosity $\eta$, and bulk viscosity $\zeta$ have been replaced
by generalized fluid dynamical variables $p_{\plus}$, $\eta_{\plus}$, and $\zeta_{\plus}$. These generalized fluid dynamical variables are dependent on $\phi_{\vQ}$ and are different than their standard counterparts when the $\phi_{\vQ}$ modes are out of equilibrium.
For example, the difference between the generalized entropy, which determines $p_{(+)}$, 
and the entropy is given by 
\begin{eqnarray}
    s_{(+)} - s = \frac{1}{2}\,\int \bbar{d}^{\,3}\vQ\, \[\log\(\frac{\phi_{\vQ}}{\phieq_{\vQ}}\)-\frac{\phi_{\vQ}}{\phieq_{\vQ}}+1\], 
\end{eqnarray}
which vanishes when $\phi_{\vQ}$ = $\phieq_{\vQ}$.
It is through these generalized variables that the evolution of the standard fluid dynamical variables experience feedback from the out-of-equilibrium dynamics of critical fluctuations, an effect we call ``backreaction,'' and it's through the explicit factor of $u^\mu$ and the implicit dependence of $\phieq_{Q}$ on $\varepsilon$ in Eqn.~\eqref{eqn:phi_relaxation} that the evolution of the critical fluctuations depends on the bulk evolution of the fluid. 
In Fig.~\ref{fig:phi_evolution} we show a numerical solution of Eqns.~\ref{hydro-eq-we-use} for a highly simplified, though heavy-ion collision inspired model, which includes a critical point~\cite{Rajagopal:2019xwg}. 
These plots demonstrate key non-equilibrium effects coming from the Hydro+ equations, namely the finite relaxation rate of the critical fluctuations, which causes $\phi_{\vQ}$ to lag behind its equilibrium value, the advection of the fluctuations, which causes $\phi(\vQ)$ to flow outward as the QGP droplet expands, and the existence of memory effects, resulting in the radially outflowing peaks in the right two plots of the figure. 

\begin{figure}
\center
\includegraphics[width=0.32\textwidth]{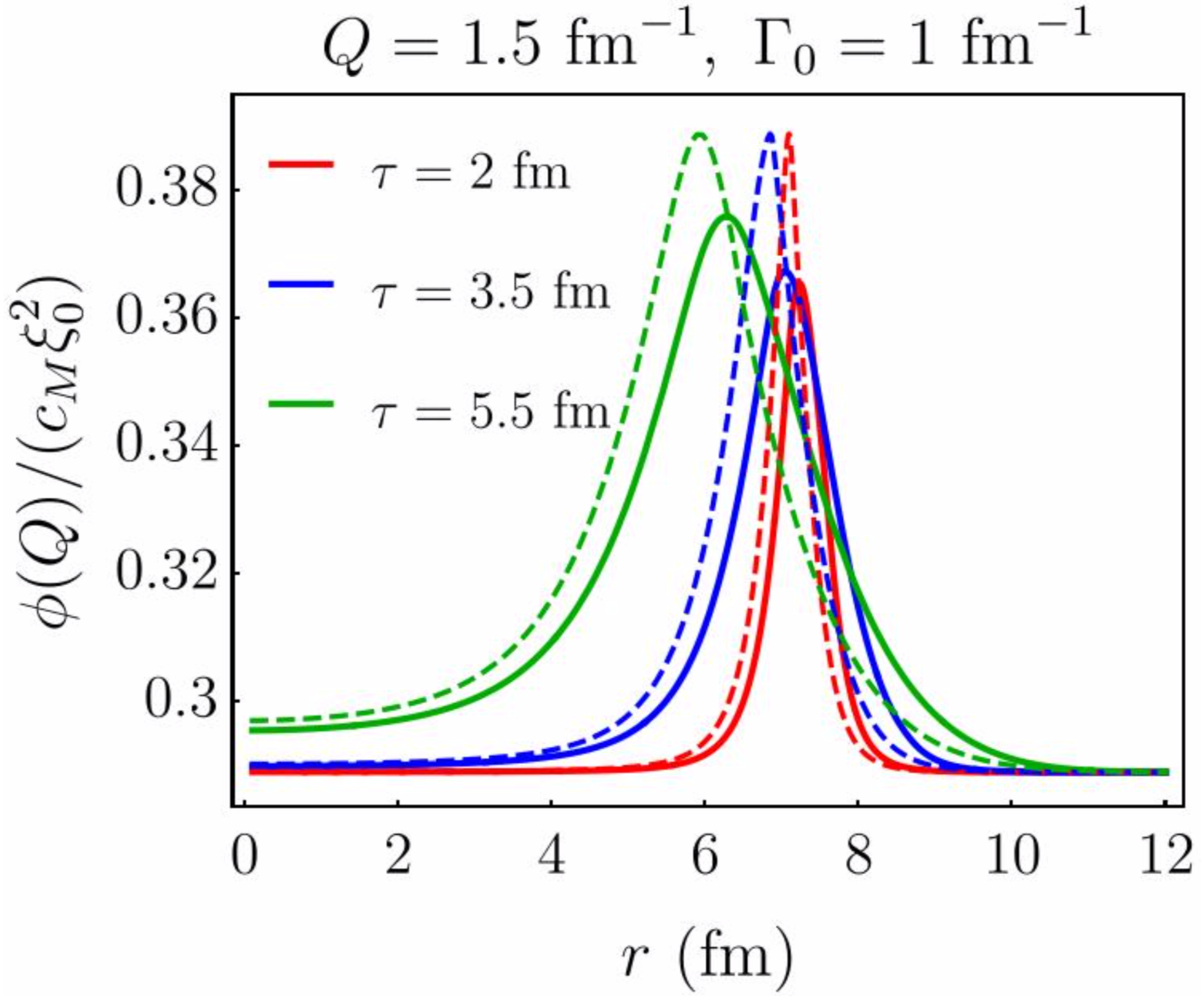}
\includegraphics[width=0.32\textwidth]{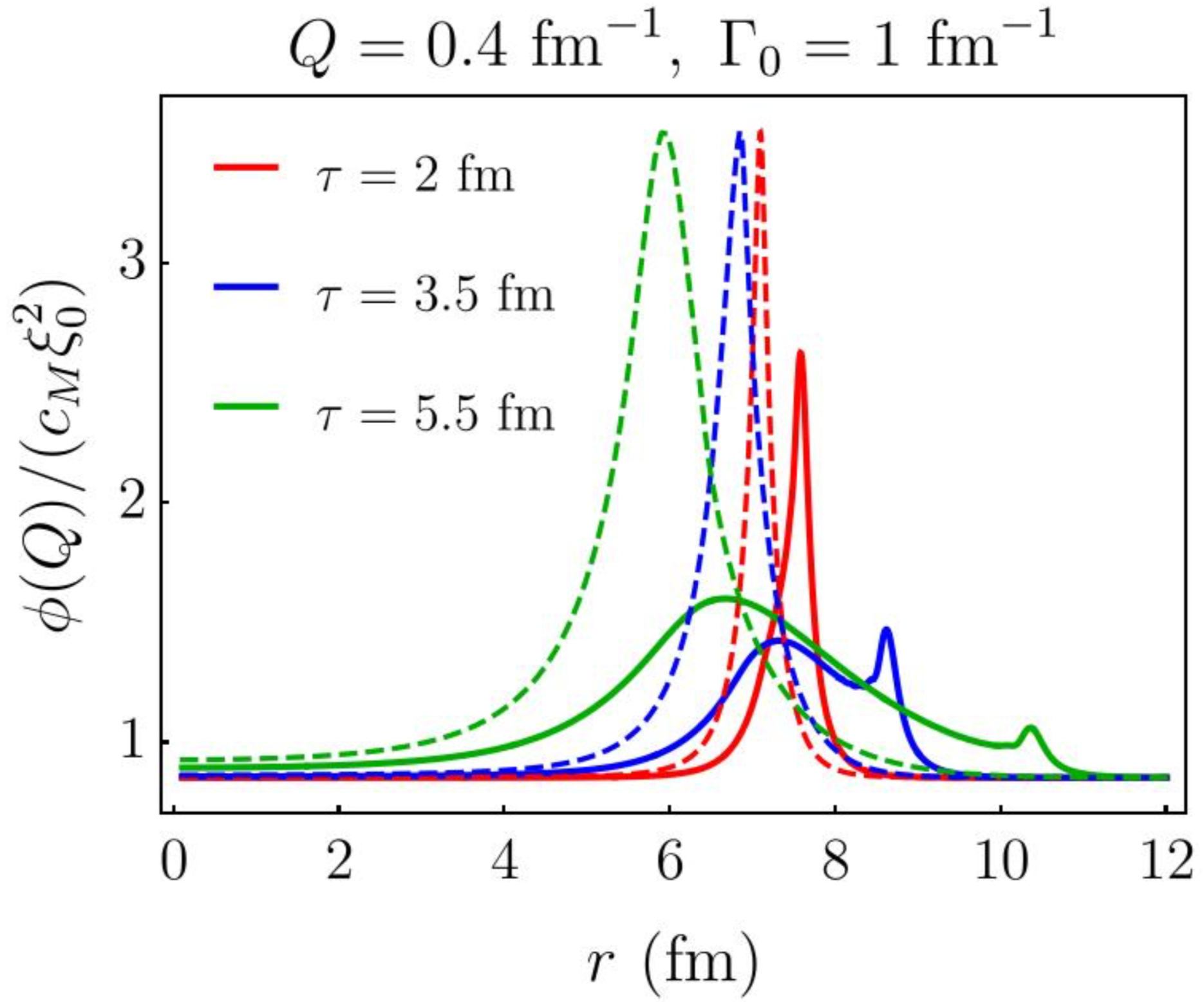}
\includegraphics[width=0.32\textwidth]{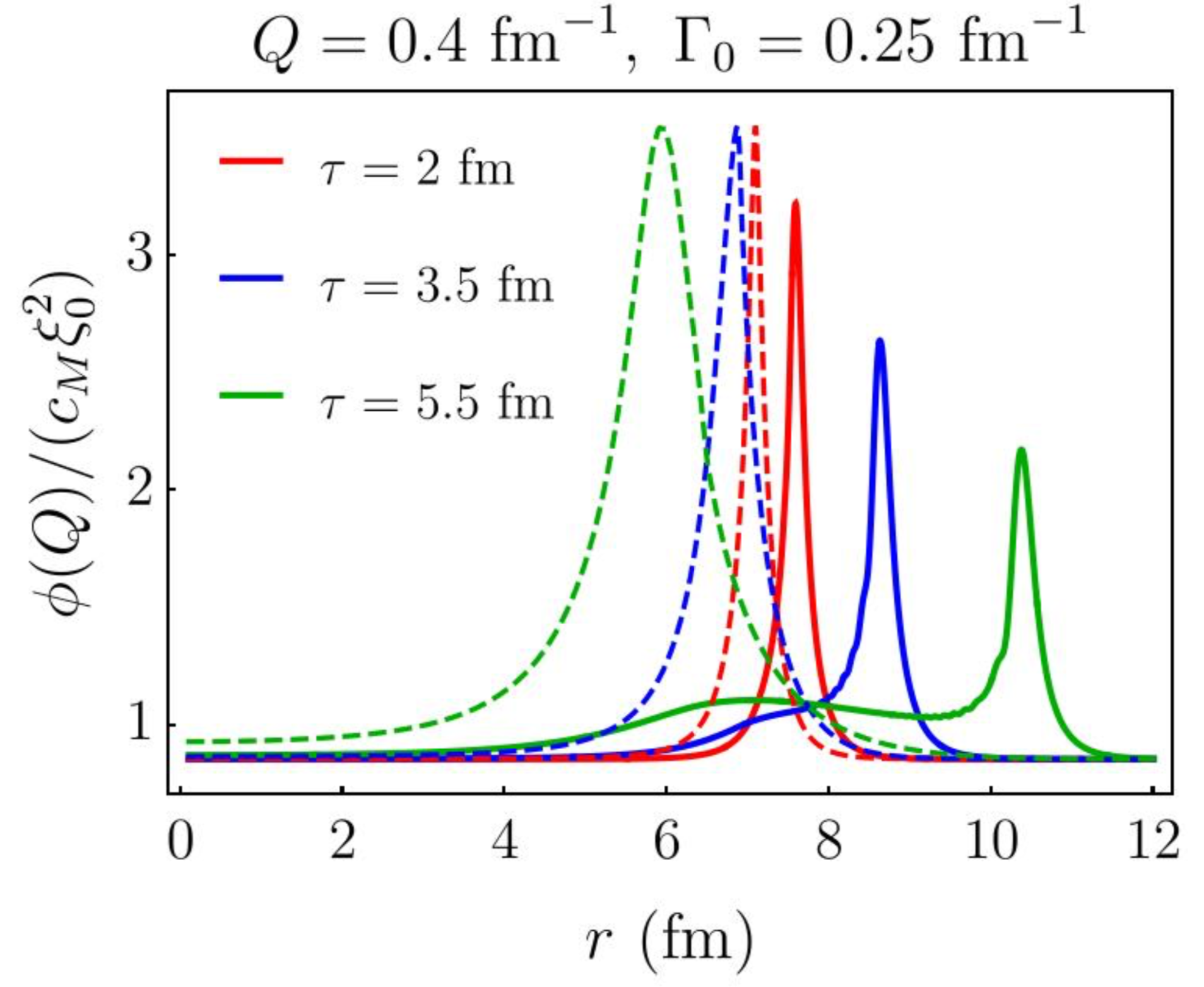}
\caption{
\label{fig:phi_evolution}
The magnitude of the critical fluctuations, $\phi(Q)$, 
plotted as a function of radius $r$ at two values of the wave vector, 
$Q=1.5~\fm^{-1}$ (left plot) and the longer wavelength $Q=0.4~\fm^{-1}$ (right two plots), obtained from Ref.~\cite{Rajagopal:2019xwg}. 
This simulation assumed azimuthal symmetry perpendicular to the collision ($\hat{z}$) axis and boost invariance along the collision axis, allowing all quantities to be plotted as a function of $r$ and $\tau$.
In all plots, solid and dashed curves show $\phi(Q)$ and $\phieq(Q)$ respectively and 
the red, blue and green curves show results at $\tau=2$, $3.5$, and $5.5$~fm, respectively.
The fluid dynamical simulation started at $\tau=1$fm with the initial condition
that $\phi(Q)$ started in equilibrium,
$\phi(Q, \tau=1\text{fm})=\phieq(Q)$. 
These results were obtained by solving Eqns.~\ref{hydro-eq-we-use} with two different values of $\G_0$, an unknown parameter determined by microscopic physics that controls the rate at which $\phi(Q)$ relaxes to its equilibrium value, which was set either to $\G_{0}=1~\fm^{-1}$ (left two plots) or a slower relaxation rate $\G_{0}=0.25~\fm^{-1}$ (right plot).
These plots demonstrate key features of the Hydro+ equations. 
$\phi(Q)$ lags behind its equilibrium value $\phieq(Q)$ because its relaxation rate $\G_{\vQ}$ is finite.
Due to critical slowing down, with all else fixed, larger wavelength modes relax slower than smaller wavelength modes do.
Additionally, due to the bulk radial outflow of the fluid, $\phi(Q)$ is advectively carried radially outward.
For sufficiently small relaxation rates, this advection leads to memory effects, demonstrated in the right two plots by the radially outflowing peak. 
The peak originated in the initial condition for $\phi(Q)$ that was chosen in this simulation. Figures taken from~\cite{Rajagopal:2019xwg}.
}
\end{figure}

One fortunate practical aspect of Eqns.~\ref{hydro-eq-we-use} is that the addition of Eqn.~\eqref{eqn:phi_relaxation} 
and the $(+)$ substitutions do not add much more computational complexity to the fluid dynamical simulation. 
Naively, Eqn.~\eqref{eqn:phi_relaxation} constitutes an addition of infinitely more variables to keep track of, one for each $\vQ$. 
To solve these equations on a computer, one must discretize momentum space and keep track of only a finite number of modes, say $N$ modes. 
The continuous $\vQ$ variable is then replaced by a finite list of momenta, $Q_i$. 
Since the time derivative of $\phi_{Q_i}$ only depends on $\phi_{Q_j}$ if $i = j$, 
each of these $N$ modes can be evolved forward in time independently of one another at each time step. 
If the derivative of $\phi_{Q_i}$ depended on $\phi_{Q_j}$ for $i \neq j$, then one would have found that $A_{ij} \partial_\tau \phi_{Q_i} = \phi_{Q_j}$ for some matrix $A_{ij}$, meaning that each time step of an Euler method would require the inversion of an $N \times N$ matrix, following the method described in \cite{Baier:2006gy}. 
The fact that $A_{ij}$ is diagonal seems to be a result of the fact that Hydro+ is currently only formulated up to two-point functions~\cite{Stephanov:2017ghc}.
We then ask: to what extent will this simplification remain true when higher-point functions are incorporated into Hydro+, 
and is there an argument why the off-diagonal terms in $A_{ij}$ are negligible?

Additionally, when simulating Eqns.~\ref{hydro-eq-we-use} in a heavy-ion inspired, though very simplified and phenomenologically inapplicable model, 
the authors of Ref.~\cite{Rajagopal:2019xwg} found that the deviations caused by the feedback of the out-of-equilibrium 
$\phi_{\vQ}$ modes on $\varepsilon$ and $u^\mu$, which are due to the $(+)$ subscripts in Eqns.~\ref{hydro-eq-we-use}, were at the percent level or below.  
Those authors argued that while the critical fluctuations from a single order parameter degree of freedom are enhanced near a critical point, 
the thermodynamics of the bulk of the QGP comes from a strongly coupled liquid built from 16 bosonic degrees of freedom and 36 fermionic degrees of freedom. 
Therefore, unless the QGP passes exactly through the critical point, the thermodynamics are dominated by the more numerous non-critical degrees of freedom, 
and the effects of the out-of-equilibrium $\phi_{\vQ}$ modes on the bulk evolution of the fluid are small. 
If it remains true that the effects of backreaction are small for more realistic heavy-ion simulations, 
then the implementation of Hydro+ in these simulations will be greatly simplified.
One could first perform a standard fluid dynamical simulation, and then, with its outputs, 
solve Eqn.~\eqref{eqn:phi_relaxation} independently to determine the evolution of the $\phi_{\vQ}$ modes.
Our next question is therefore: are the effects of backreaction negligible for phenomenologically relevant heavy-ion fluid dynamical simulations?

Other open questions in the Hydro+ formalism involve higher-point functions, initial conditions, and freeze-out. 
How can we generalize Hydro+ to incorporate 3-point and higher-point functions? 
Were we to naively generalize Eq.~\eqref{phi-def} we would introduce another insertion of $\delta M$, and with it another momentum and spacetime dimension, 
leading to a proliferation of $\phi$ modes that need to be followed during the course of a simulation. 
How many modes must be tracked in order to accurately describe a heavy-ion simulation?  Also, what are the initial conditions of these modes? 
Finally, what is the proper way to implement freeze-out for these modes?
}

\subsection{Relevant scales for transits of the critical point
\label{sec:transits} }

 The deterministic method described in Section~\ref{sec:hydrokin} can 
be used to obtain estimates of the length and time scales involved
in transits of the critical region in a heavy-ion collision. The
basic issue is that in a collision of heavy nuclei the trajectory 
of the system in the phase diagram is likely to miss the critical
point by some amount, and to only spend a finite amount of time 
in the critical region. Combined with the expansion of the system,
and the effects of critical slowing down this implies that the 
correlation length cannot become very large. The effects of critical
slowing manifest themselves differently depending on the spatial and 
momentum scales at which correlations are being studied. In 
this section we will present simple estimates of these effects,
following the work of \cite{Akamatsu:2018vjr}. 

 We consider the two-point function of the entropy per particle 
$\hat{s}=s/n$, which serves as an order parameter near the critical 
endpoint. Following Section~\ref{sec:hydrokin} we can derive 
a relaxation equation for the two-point function $W_{\hat{s}\hat{s}}
(t,x,k)$. For simplicity we will focus on a fluid undergoing locally
homogeneous isotropic expansion so that $W_{\hat{s}\hat{s}}(t,k)$ 
does not depend on $x$. The evolution equation for $W_{\hat{s}\hat{s}}$
has the form
\begin{equation}
\partial_t W_{\hat{s}\hat{s}}(t,k) = 
  -2\Gamma_{\hat{s}}(t,k) \left[ W_{\hat{s}\hat{s}}(t,k)
     -W^0_{\hat{s}\hat{s}}(t,k) \right], 
\end{equation}
where $\Gamma_{\hat{s}}$ is a relaxation rate, and $W^0_{\hat{s}\hat{s}}
(t,k)$ is the equilibrium correlation function. In a non-critical 
fluid the correlation length is small and $W^0_{\hat{s}\hat{s}}
(t,k)$ is approximately independent of $k$. Indeed, thermodynamic
identities predict that $W^0_{\hat{s}\hat{s}}(t,k)=C_p(t)$, where 
$C_p$ is the specific heat at constant pressure. 

  The relaxation rate is related to the diffusion constant, 
$\Gamma_{\hat{s}}=Dk^2$. The diffusion constant can be written as
$D=l_{\text micro}^2/\tau_0$, where $l_{\text micro}$ is the 
microscopic length scale introduced above, and $\tau_0$ is the 
non-critical relaxation time. The maximum wavelength of a fluctuation
that can be equilibrated in a fluid that is expanding at a rate
$1/\tau_Q$ is 
\begin{equation}
 l_{\text max} = l_{\text micro}\sqrt{\frac{\tau_Q}{\tau_0}}
  \equiv \frac{l_{\text micro}}{\sqrt{\epsilon}} \, , 
\end{equation}
where we have introduced a small parameter $\epsilon\equiv \tau_0/\tau_Q$, i.e. the product of the microscopic relaxation time $\tau_0$ and the macroscopic expansion rate $1/\tau_Q$. 

 In the vicinity of the critical point the correlation length 
and the specific heat diverge. We can take the effect of the 
correlation length into account by taking the equilibrium 
correlation function to be of the form
\begin{equation}
W^0_{\hat{s}\hat{s}}(t,k)=   \frac{C_p(t)}{(1+(k\xi)^{2-\eta})},  
\end{equation}
where $\eta$ is the correlation length exponent in the $3$-dimensional
Ising model. We can also incorporate the effect of critical slowing down
by modifying the relaxation rate as
\begin{equation}
\label{Gamma:crit}
    \Gamma_{\hat{s}}(t,k) = \frac{\lambda_T}{C_p\xi^2}
     (k\xi)^2 (1+(k\xi)^{2-\eta}),
\end{equation}
where $\lambda_T$ is the thermal conductivity. Eq.~(\ref{Gamma:crit})
is a simple model that corresponds to the model B dynamics discussed
in Section~\ref{sec:stodiff}. 

 Consider now the time evolution in the vicinity of a critical 
point. We will define $t=0$ to be the time at which the system 
reaches the critical value of the baryon density. Near $t=0$ the 
equilibrium correlator evolves rapidly, $(\partial_t C_p)/C_p\sim 1/t$.
However, because of critical slowing down, the equilibration rate 
of long wavelength fluctuations cannot keep up with this rapid 
evolution, and these modes necessarily fall out of equilibrium. 
Equating the rate of change of $C_p$ and the relaxation rate
\begin{equation}
\frac{\partial_t C_p(t)}{C_p(t)}\sim \frac{1}{t} 
   \sim \Gamma_{\hat{s}}(t,k)     
\end{equation}
determines a characteristic time, known as the Kibble-Zurek time 
$t_{\text KZ}$. The correlation length $\xi$ at this time is 
the Kibble-Zurek length, $l_\text{ KZ}=\xi(t_\text{ KZ})$. We 
can estimate the Kibble-Zurek length using the scaling form of 
the relaxation rate, and the critical scaling of the specific heat.
We find $l_\text{ KZ}\sim l_\text{ micro}\epsilon^{1/(a\nu z+1)}
\sim l_\text{ micro}\epsilon^{-0.19}$, where we have used the 
model B value for the dynamical exponent $z$, and Ising critical
exponents for $a=1/(1-\alpha)$ and $\nu$. This establishes a 
hierarchy
\begin{equation}
l_\text{ micro} 
 \ll \left(l_\text{ KZ}\sim l_\text{ micro}\epsilon^{-0.19}\right)
 \ll \left(l_\text{ max}\sim l_\text{ micro}\epsilon^{-0.5}\right)\,.
\end{equation}
Reference \cite{Akamatsu:2018vjr} provides numerical estimates
for $l_\text{ micro}$ and $\epsilon$ under conditions relevant to 
a possible critical endpoint, $T\simeq 155$ MeV and $n/s\simeq 1/25$.
The authors find $l_\text{ micro}\simeq 1.2$ fm and $\epsilon\simeq
0.2$. This corresponds to a hierarchy of scales
\begin{equation}
\label{lkz:est}
 1.2\, \text{ fm} \ll 1.6\, \text{ fm} \ll 2.7\, \text{ fm}.    
\end{equation}
These results indicate that the correlation length does not 
become very large, and that the enhancement in the two-particle 
correlation function in the critical regime remains modest, on 
the order of a factor of 2. 

 The methods discussed in Section~\ref{sec:hydro-kinetics} can also be 
used to study the rapidity structure of fluctuations in a QGP 
undergoing longitudinal expansion. For simplicity we consider 
Bjorken expansion. From the Green function of the diffusion equation in 
a Bjorken background we find that the width of a momentum fluctuation
localized in rapidity at time $\tau_0$ will increase to 
\cite{Gavin:2006xd,Kapusta:2011gt,Akamatsu:2016llw}
\begin{equation}
\label{eta:width}
  \sigma_\eta   \simeq \sqrt{\frac{6\eta}{sT(\tau_0)\tau_0}}\, ,
\end{equation}
where we have assumed that the shear viscosity to entropy density 
ratio is approximately constant. A similar formula can be derived 
for baryon number diffusion. Eq.~(\ref{eta:width}) shows that in 
the regime in which fluid dynamics is a good approximation the rapidity 
width of an initial state fluctuation is small, $\sigma_\eta\lesssim 1$.
We can also obtain a very rough estimate of the rapidity width of 
a critical fluctuation. Using the expansion rate to convert 
longitudinal distance to space time rapidity Eq.~(\ref{lkz:est})
gives
\begin{equation}
    \left(\sigma_{\eta}(\text{KZ})\sim \epsilon^{0.81} \right)
    \ll \left(\sigma_{\eta}(\text{max})\sim \epsilon^{0.5} \right)\,  .
\end{equation}
The estimates discussed in this section indicate that in heavy-ion 
collisions the correlation length remains modest, even if the system 
passes close to a critical point in the QCD phase diagram, and that 
critical fluctuations are localized in specific regions in momentum
space. Quantifying these statements requires the results of fluid dynamical
simulations to be converted to particle spectra in momentum space, 
which will be addressed in the following section. 

\subsection{Implementation of fluid to particle conversion
\label{sec:conversion}}

After performing either stochastic fluid dynamics or hydro-kinetics the question arises
how to compare the fluid dynamical and order parameter fields and their fluctuations to experimentally observed quantities, which are constructed from measured
particle spectra in a given, experiment specific, kinematics, and not 
from the fluid dynamical fields directly. Therefore, direct model to data
comparisons require conversion of correlations in fluid fields to finite
statistics particle
correlations. For non-relativistic fluids this problem has been addressed in several ways~\cite{Donev:2009}. One of them is to exactly match the fluxes at the interface, which in the relativistic case corresponds to local event-by-event conservation laws, or in other words, micro-canonical sampling. The Cooper-Frye (CF) particlization used in relativistic models (see e.g.~\cite{Huovinen:2012is}), on the other hand, is based on a grand-canonical local phase-space distribution. It combines the Cooper-Frye  formula for the momentum distribution in a hypersurface cell~\cite{Cooper:1974mv} with Poissonian sampling of the multiplicity distributions. As discussed in \cite{Steinheimer:2017dpb} this procedure adds additional fluctuations to those obtained from stochastic fluid dynamics. (This method and thus the subsequent discussion are relevant only for stochastic fluid dynamics. At the moment it remains unclear how to freeze-out after hydro-kinetics.)

In order to see this let us consider for simplicity the correlations of the baryon number for the case where we can ignore anti-baryons, i.e. for collisions at low energies. 
Stochastic fluid dynamics provides an ensemble of hydro events or configurations which reflect the fluctuations of the system. In addition particlization of a given event typically provides an ensemble of particle configurations. Therefore, for a given cell $i$ and a given fluid dynamical (FD) event we have the following
\begin{align*}
B_{H}(i) & =\mathrm{baryon\ number\ from\ FD\ in\ cell\ i}\\
B_{S}(i) & =\mathrm{baryon\ number\ after\ CF\ sampling\ in\ cell\ i}\\
\delta B(i) & =\mathrm{fluctuation\ of\ B\ in\ cell\ i\ due\ to\ CF\ sampling}
\end{align*}
The final baryon number (in terms of particles) in cell $i$ is then obtained by averaging over all fluid dynamical events as well as by averaging over the particle configurations for a given fluid dynamical event. Let us denote these averages as follows: 
\begin{align*}
\ave{\ldots} & =\mathrm{average\ over\ many\ CF\ particle\ configs}\\
\overline{\ldots} & =\mathrm{average\ over\ FD\ configs}\\
\ave{\ave{\ldots}} & =\mathrm{average\ over\ CF \ sampling\ AND\ over\ FD\ configs}
\end{align*}
Thus, if for a given fluid dynamical event we average over the particle samples we get 
\begin{eqnarray}
\ave{B_{S}(i)} & = & \ave{B_{H}(i)+\delta B(i)}=B_{H}(i)\,.
\end{eqnarray}
Further averaging over the fluid dynamical ensemble results in 
\begin{align}
\ave{\ave{B_{S}(i)}} & = \overline{B_{H}(i)}.
\end{align} 
Since the Cooper-Frye sampling preserves the mean everything works out. However this is not the case if we look at correlations. For a given fluid dynamical event upon averaging over the particle configurations we get 
\begin{align}
\ave{B_{S}(i)\,B_{S}(j)} &= B_{H}(i)B_{H}(j)+\ave{\delta B(i)\delta B(j)}\nonumber \\
 & =  B_{H}(i)B_{H}(j) + \delta_{i,j}\ave{\delta B(i)^{2}} \nonumber \\
 & = B_{H}(i)B_{H}(j)+ \delta_{i,j}B_{H}(i)\,,
 \end{align}
where in the last line we used the fact that for Poisson sampling we have $\ave{\delta B(i)^{2}}= \ave{B_i}=B_H(i)$. Thus we get  
for the correlation function
\begin{align}
C_{S}(i,j) & =\ave{\ave{B_{S}(i)B_{S}(j)}}-\ave{\ave{B_{S}(i)}}\ave{\ave{B_{S}(j)}}\nonumber \\
 & =\overline{B_{H}(i)B_{H}(j)}-\overline{B_{H}(i)}\,\overline{B_{H}(j)}+\delta_{i,j}\overline{B_{H}(i)}\nonumber \\
 & =C_{H}(i,j)+\delta_{i,j}\overline{B_{H}(i)}\,.
 \label{sec2:eq:c2}
\end{align}
Therefore, for all non-identical cells the correlations are reproduced
correctly, but we get spurious contributions from identical cells. If correlations could be measured in configuration space one could simply ignore the problem for identical cells, which is due to correlations of particles with themselves. However, in experiment, we look at correlations in momentum space and it is not clear how to remove this spurious contribution in  this case. The problem gets even more apparent if one looks at cumulants. Given the above expression for the correlation function the second order cumulant, $K_2$, is given by
\begin{align}
K_{2,S} & =\sum_{i,j}C_{S}(i,j)=K_{2,H}+\sum_{i}B_{H}(i)\,,
 \label{sec2:eq:kappa2}
\end{align}
where we sum over a certain subset of cells of the freeze-out hypersurface. In addition to the true second order cumulant which reflects the fluctuation of the stochastic fluid dynamical simulation we have an extra, spurious term $\sim \sum_{i}B_{H}(i)$ which arises from the Poisson sampling of the standard Cooper-Frye particlization. For example, in the case where we use stochastic fluid dynamics to simulate an ideal gas, where the fluctuations follow a Poisson distribution, we would simply double count the fluctuations so that in this case the resulting second order cumulant would be $K_2 = 2 \ave{B}$. 

From this simple example it should be clear that particlization of stochastic fluid dynamics has to ensure that the conserved quantum numbers are conserved locally and event by event. This can be achieved by sampling the particles from a micro-canonical ensemble instead of a grand-canonical ensemble as it is done in the  standard Cooper-Frye procedure. Such an algorithm has been developed and implemented in~\cite{Oliinychenko:2019zfk}. As discussed in some detail in this paper, in case of the systems created in heavy-ion collisions the micro-canonical sampling requires some extra considerations, because contrary to typical non-relativistic fluids, one deals with a rather small number of particles of the order of $10^4$ or so. At the same time, the computational grid is made of rather small cells in order to minimize numerical viscosity. As a consequence, the typical number of particles in a cell of the computational grid is much smaller than one. Micro-canonical sampling, however, requires integer quantum numbers and, preferably, that the number of particles is large compared to one. To address this issues the authors of~\cite{Oliinychenko:2019zfk} introduced so-called ``patches". These patches are larger than the computational cells and their size should be such that each patch has a sufficient number of particles for the micro-canonical sampling to be sensible. Thus the patch size introduces another scale, $\lpt$. 
Since the conserved quantum numbers are not resolved within a patch, one can determine the  correlation of conserved charges only for distances $d>\lpt$. 

The obvious question is how the new scale $\lpt$ compares with the other scales in the problem such as $l_{\mathrm{noise}}$ or $l_{\mathrm{filter}}$ and $l_{\mathrm{hydro}}$. The condition for the patch size is that one has a sufficiently large number of particles in the
patch, $N_{\mathrm{patch}} = \lpt^3 \rho \gg 1$, where $\rho$ is the particle density. Since after particlization one typically evolves the system with Boltzmann transport, the mean free path 
$\mfp$ 
should be larger than the inter-particle distance, i.e. 
$\mfp > 1/\rho^{1/3}$. 
Since fluid dynamics should be still valid at the point of particlization, the patch size should also be larger than the mean free path, $\lpt \gg \mfp$. This will automatically ensure that we have sufficiently many particles in the patch since  $N_{\mathrm{patch}} = \lpt^3 \rho \gg \mfp^3 \rho \gg 1$. Finally, of course the patch size needs to be smaller than the fluid dynamical scale, $\lpt\ll l_\mathrm{hydro}$ and larger than the cutoff or filter scale required to regularize stochastic fluid dynamics. Thus we have
\begin{equation}
  l_\text{grid}, l_\text{mfp}< l_\text{filter} \ll \lpt  \ll l_\text{hydro} 
   .
\end{equation}
Note that $l_\text{grid}$ is the size of the discretized fluid cell, which is not really a physical scale. In order to resolve the correlations we should have $\lpt \ll l_\text{KZ}$, where $\lpt$ is limited by the inter-particle spacing and therefore this condition is only marginally fulfilled in model estimates, see Eq.~(\ref{lkz:est}).

Contrary to stochastic fluid dynamics, which provides an ensemble of fluid dynamical events encoding the fluctuations and correlation of conserved charges, in deterministic hydro-kinetics one calculates the time evolution of the means and $n$-particle correlation functions. Therefore, in this case particlization will involve sampling particles such that these correlation functions are faithfully reproduced in terms of particles. At present there is no algorithm available to address this problem.


\section{Experimental challenges}
\label{sec:Exp}

The experimental measurement of fluctuation observables is currently of high interest to the heavy-ion physics community. Naturally, one of the main topics of the discussions during the RRTF meeting was the search for a QCD critical point in the SPS energy range as well as in the RHIC beam energy scan (BES) program. However, an even larger fraction of the discussions was focused on the comparison of ALICE data with lattice QCD calculations. While the QCD critical point is not accessible via measurements at LHC energies, pseudo-critical fluctuations at higher orders are measurable. In addition, also the lower order fluctuation observables are of interest as they provide the unique opportunity to test lattice QCD results against experimental data.

Nevertheless, one should be cautious when making direct comparisons to results from lattice QCD calculations. On the one hand, actual systems in high-energy heavy-ion experiments are dynamical, finite, come in different sizes, and the plasma formed is very noisy, fluctuates considerably and is indirectly measured within a given acceptance. On the other hand, lattice QCD simulations are probing equilibrium in the thermodynamic limit, and are still very constrained by the sign problem. Still, several statistical mechanics techniques successfully applied in lattice simulations can be useful in analyzing the experimental data \cite{Fraga:2003mu,Palhares:2009tf,Fraga:2011hi}. 
Near the critical region, one can systematically incorporate spurious contributions (resonances, acceptance limitations, finite size, finite lifetime and critical slowing down) expected to affect the fluctuations in the BES in a way that can be systematically improved or adapted \cite{Hippert:2015rwa,Hippert:2017xoj}.

\subsection{Matching between experimental observables and theoretical quantities}

The physics program of fluctuation studies in heavy-ion collisions is characterised by a plethora of existing observables with different sensitivities to the underlying physics phenomena. From the theory side, only some quantities are directly accessible by ab-initio lattice QCD calculations.
In any case, all observables must be properly matched between theory and experiment in order to allow for an apples-to-apples comparison.

The thermodynamic susceptibilities $\chi^{BSQ}_{lmn}$ of order $l+m+n$ for baryon number $B$, strangeness $S$, and electric charge $Q$, are given by the derivatives of the pressure $P$ with respect to the corresponding chemical potentials $\mu$ \cite{Koch:2008ia}:
\begin{equation}
\chi^{BSQ}_{lmn} = \frac{\partial^{l+m+n}(P/T^{4})}{  \partial(\mu_{B}/T)^{l} \, \partial(\mu_{S}/T)^{m} \, \partial(\mu_{Q}/T)^{n} } \;.
\end{equation}
They can be calculated in lattice QCD from first principles using imaginary time (for details see e.g. \cite{Philipsen:2010gj}). The pressure in a system of volume $V$ is connected with the partition function via \cite{Karsch:2010ck}
\begin{equation}
\frac{P}{T^{4}} = \frac{\ln\mathcal{Z}}{ VT^{3}} \; .
\end{equation}
From the experimental side, the susceptibilities of the conserved quantities are accessible via the measurement of event-by-event fluctuations in the particle production. For a quantitative comparison, the cumulants $K_{i}$ of order $i$ of the measured particle multiplicity distributions are analysed. They can be calculated from the central moments $\mu_{i} = \langle (\delta N)^{i} \rangle$ with $\delta N = N - \langle N \rangle$. In a traditional nomenclature, statistical distributions are often described with the mean $M$, the variance $\sigma$, the skewness $S$, and the kurtosis $\kappa$ which corresponds to an equivalent parameter set of the first four central moments or cumulants, respectively. With respect to the thermodynamic susceptibilities $\chi$, the following relations are found \cite{Karsch:2012wm}:
\begin{center}
\begin{math}
 \begin{array}{lcccccccc}
M          &=& K_{1} &=& \mu &=& \langle N \rangle &=& VT^{3}\cdot \chi_{1}\,, \\
\sigma^{2} &=& K_{2} &=& \mu_{2} &=& \langle (\delta N)^{2} \rangle &=& VT^{3}\cdot \chi_{2}\,, \\
S          &=& \frac{K_{3}}{ \sigma^{3}} &=& \mu_{3} / \sigma^{3} &=& \langle (\delta N)^{3} \rangle / \sigma^{3} &=& \frac{VT^{3}\cdot \chi_{3} }{ (VT^{3}\cdot \chi_{2})^{3/2}}\,, \\
\kappa     &=& \frac{K_{4} }{ \sigma^{4}} &=& (\mu_{4} - 3\mu_{2}^{2}) / \mu_{2}^{2} &=& \langle (\delta N)^{4} \rangle / \sigma^{4} - 3 &=& \frac{(VT^{3}\cdot \chi_{4}) }{ (VT^{3}\cdot \chi_{2})^{2}} \,.
 \end{array}
\end{math}
\end{center}

There are several significant effects, however, which must be carefully considered in the comparison of the theoretically calculated susceptibilities and the experimentally measured cumulants of identified particle multiplicity distributions, which will be discussed in more detail in the following sections: 
\begin{enumerate}
\item The susceptibilities of the conserved quantities in QCD are calculable on the lattice while experimentally only net-charge, net-pion, net-kaon, net-proton, and net-$\Lambda$ distributions are accessible.  The correspondence between, for example, the cumulants of the net-proton distribution and the susceptibilities $\chi^B_n$ is discussed in Sec.~\ref{sec:proton_baryonnumber}.  
\item While the susceptibilities are calculated on the lattice in a fixed volume at a fixed temperature, which enter into the equations above as $VT^3$ terms, in heavy-ion collisions these quantities are unmeasurable.  Therefore a common approach is to form combinations 
of the cumulants in order to cancel these unknown factors and compare them to ratios of the susceptibilities, such as $S\sigma = \chi_3/\chi_2$ and $\kappa\sigma^2 = \chi_4/\chi_2$.  However, as detailed in Sec.~\ref{sec:volumefluct}, the volume and temperature in heavy-ion collisions are related to the number and positions of the participating nucleons and therefore are not only unknown but fluctuate event-by-event.  These additional fluctuations mean that the $VT^3$ terms do not cancel precisely.  
\item While the lattice QCD calculations are performed for a fixed volume in the infinite limit, and the correspondence to multiplicity fluctuations of conserved charges is done within the grand-canonical ensemble picture, heavy-ion collisions occur within a finite volume over which local and global conservation laws must hold.  The effects of conservation laws can be experimentally probed by investigating the dependence of the multiplicity cumulants on the kinematic acceptance of the measurement, as described in Sec.~\ref{sec:deltaEta}.  
\item Another major topic of discussion at the RRTF was the influence of resonance decays on fluctuation observables.  For example, the feeddown of $\Lambda$ and $\overline{\Lambda}$ baryons into the proton and anti-proton multiplicities (e.g. $\Lambda \rightarrow p\pi^-$, $\overline{\Lambda} \rightarrow \overline{p}\pi^+$) and the influence of rho meson decays in the net-pion measurement ($\rho \rightarrow \pi^+\pi^-$) are of particular concern (see Sec.~\ref{sec:feeddown}).  
\end{enumerate}

Several authors argue in addition that the study of factorial cumulants provides a cleaner way to access possible non-trivial dynamics in heavy-ion collisions~\cite{Bzdak:2016sxg}. 

\subsubsection{Additional experimental observables}

\begin{enumerate}

\item $\nu_{dyn}$

In addition to cumulants of net-charge distributions, many experiments also measure fluctuations via the observable $\nu_{dyn}$.  The fluctuations between two particle types $A$ and $B$, which may represent particles and anti-particles or different particle species, can be quantified by

\begin{align}
\label{eq:nu}\nu &= \left\langle \left( \frac{N_A}{\langle N_A\rangle} -  \frac{N_B}{\langle N_B\rangle} \right)^2 \right\rangle\\
&= \frac{\langle N_A^2 \rangle}{\langle N_A\rangle^2} + \frac{\langle N_B^2 \rangle}{\langle N_B\rangle^2} - 2\frac{\langle N_A N_B\rangle}{\langle N_A\rangle \langle N_B\rangle}.\label{eq:nu2}
\end{align} 

The independent statistical fluctuations of $N_A$ and $N_B$ are then subtracted to obtain a measure of the dynamical fluctuations,
\begin{align}
\label{eq:nu3}\nu_{dyn} &= \nu - \left(\frac{1}{\langle N_A\rangle} + \frac{1}{\langle N_B\rangle}\right)\\
&= \frac{\langle N_A\left( N_A -1\right) \rangle}{\langle N_A\rangle^2} + \frac{\langle N_B \left( N_B - 1\right) \rangle}{\langle N_B\rangle^2} - 2\frac{\langle N_A N_B\rangle}{\langle N_A\rangle \langle N_B\rangle}.
\label{eq:nudyn}
\end{align}
If $N_A$ and $N_B$ have Poisson distributions and are uncorrelated, then $\nu_{dyn} = 0$. An important feature of $\nu_{dyn}$ is that it is robust against particle detection efficiency losses in the case that the detector response can be described by a binomial distribution. 

At LHC energies, where particles and anti-particles are produced in equal amounts, $\nu_{dyn}[A,\overline{A}]$ is related to the second order moments via the relation

\begin{equation}
 \nu_{dyn}[A,\overline{A}] = \frac{K_{2}(N_{A} - N_{\overline{A}})}{ \langle N \rangle^{2} } - \frac{2}{ \langle N \rangle}\,,
\end{equation}

\noindent where $\langle N_{A} \rangle \approx \langle N_{\overline{A}} \rangle \approx \langle N \rangle$.

One should note that the $\nu_{dyn}$ measure, by definition, has an intrinsic multiplicity dependence which has to be taken into account. Several scaling prescriptions are investigated in the literature such as  
charged-particle multiplicity density at mid-rapidity, number of participants \cite{Abelev:2009ai,Acharya:2017cpf} and mean multiplicities of accepted particles \cite{Koch:2009dg}. 

\item {\em Balance functions}

One of the additional elements which surfaced in the discussions was the possibility to experimentally measure the balance function $B(\Delta\eta,\Delta\varphi)$, defined by the difference between the two-particle correlations of like- and unlike-sign pairs of particles.  The correlation function itself can be written as the ratio of the particle pair density to the single particle densities, 
 \begin{equation}
C^{\alpha\beta}(\eta_1,\eta_2,\varphi_1,\varphi_2) 
=\frac{\rho^{\alpha\beta}(\eta_1, \eta_2, \varphi_1,\varphi_2)}{\rho^{\alpha}(\eta_1,\varphi_1)},
\end{equation}
where $\rho^{\alpha\beta}$ is the distribution of pairs of particles of types $\alpha$ and $\beta$ at angles $(\eta_1,\varphi_1)$ and $(\eta_2,\varphi_2)$, respectively, and $\rho^{\alpha}$ is the single particle distribution for particles of type $\alpha$ at the angle $(\eta_1,\varphi_1)$.  The correlation function can be further condensed by considering only the relative angle between the two particles in the pair
\begin{equation}
C^{\alpha\beta}(\Delta\eta,\Delta\varphi) =\frac{\rho^{\alpha\beta}(\Delta\eta,\Delta\varphi)}{N_{\alpha}} \; .
\end{equation}
When the correlation function is constructed for like-sign ($C^{++}+C^{--}$) and unlike-sign ($C^{+-}+C^{-+}$) particle pairs, then the balance function is defined as $B(\Delta\eta,\Delta\varphi) = (C^{+-}+C^{-+}-C^{++}-C^{--})/2$.  The integral of the balance function is directly related to $\nu_{dyn}$ and thus to the second cumulant $K_2$ of the net-particle distribution~\cite{Pruneau:2019baa}. 

Measuring the balance function has several advantages in that it clearly encodes the rapidity dependence of the measurement, which gives access to the correlation length $\xi$ and also allows one to see the influence of other physical effects such as resonance decays, flow, and non-thermal particle production due to jets, etc. 
However, a precise measurement of balance functions naturally requires additional statistics, and while it is more straightforward to correct correlation functions for experimental efficiency it is unclear how to account for the effects of volume fluctuations, which would need to be understood before they could be interpreted as a measurement of second moments.

\item {\em Intensive and strongly intensive quantities}

Additional fluctuation observables were also discussed, such as the strongly intensive quantities $\Sigma$ and $\Omega$~\cite{Gazdzicki:1992ri,Gorenstein:2011vq,Gazdzicki:2013ana}, which are insensitive to both the volume and volume fluctuations within models of independent particle sources
(e.g. the Wounded Nucleon Model~\cite{Bialas:1976ed} and the grand-canonical ensemble of an ideal Boltzmann gas). The scaled variance, an intensive quantity, can be written as
\begin{equation}
    \omega[N] = \frac{\langle \left(N-\langle N\rangle\right)^2\rangle}{\langle N\rangle}.
\end{equation}
Within the Wounded Nucleon Model, the scaled variance can be calculated as $\omega[N] = \omega[N]_W + \langle N \rangle / \langle W \rangle \cdot \omega[W]$, where $\omega[N]_W$ stands for the scaled variance at any fixed number of wounded nucleons, and $W = W_P + W_T$ is the sum of the number of projectile and target nucleons.  Here the first term is considered to be the physically relevant quantity, whereas the second one is unwanted.  To isolate the first term of $\omega[N]$, one can furthermore construct the strongly intensive scaled variance,
\begin{equation}
\Omega[N] = \omega[N] - ( \langle N \cdot E_P \rangle - \langle N \rangle \cdot \langle E_P \rangle ) / 
\langle E_P \rangle,
\end{equation}
where $E_P = E_{beam} - E_F$, the difference between the beam energy ($E_{beam}$) and the energy carried forward by spectators from the projectile ($E_F$).  

In the search for critical behavior, it is most interesting to construct observables which are insensitive to both the volume and volume fluctuations, called strongly intensive quantities~\cite{Gorenstein:2011vq,Gazdzicki:2013ana}.  Some examples include
\begin{equation}
\Delta[P_{T},N] =
\frac {1}{\langle N \rangle \omega[p_{T}]}[\langle N \rangle \omega[P_{T}] - 
\langle P_{T} \rangle \omega[N]]
\label{eq:delta1}
\end{equation}
and
\begin{equation}
\Sigma[P_{T},N] =
\frac{1}{\langle N \rangle \omega[p_{T}]}[\langle N \rangle \omega[P_{T}] +
\langle P_{T} \rangle \omega[N] - 2(\langle P_{T}N \rangle - 
\langle P_{T} \rangle \langle N \rangle )]\ ,
\label{eq:sigma1}
\end{equation}
where $N$ is the number of particles of a given type and $P_{T}$ is the sum of the absolute values of their transverse momenta $p_{T}$.
Another example involves measuring fluctuations and correlations for numbers from two non-overlapping sets of particles:
\begin{align}
\Delta[N_1,N_2] & =
\frac {1}{\langle N_2 \rangle - \langle N_1 \rangle}[\langle N_1 \rangle \omega[N_2] - 
\langle N_2 \rangle \omega[N_1]],
\label{eq:delta2}
\\
\Sigma[N_1,N_2] & =
\frac{1}{\langle N_1 \rangle + \langle N_2 \rangle}[\langle N_1 \rangle \omega[N_2] +
\langle N_2 \rangle \omega[N_1] 
\nonumber \\
& \quad
- 2(\langle N_1 \, N_2 \rangle - 
\langle N_1 \rangle \langle N_2 \rangle )]\ .
\label{eq:sigma2}
\end{align}

One interesting observation is that $\Sigma[N_1,N_2]$ in Eq.~\eqref{eq:sigma2} reduces, in the special case $\langle N_1 \rangle = \langle N_2 \rangle$, to the ratio of the variance $K_2(N_1-N_2)$ to the Skellam baseline $\langle N_1 \rangle + \langle N_2 \rangle$, which represents the limiting case of independent Poissonian particle and anti-particle multiplicity distributions.
The condition $\langle N_1 \rangle = \langle N_2 \rangle$ is realized to a high precision for measurements of particle and antiparticle distributions at the LHC.
The variance-over-Skellam baseline ratio for the difference of particle and antiparticle numbers measured by ALICE~(see Sec.~\ref{sec:ALICEflucs}) for various particle types belongs therefore to the class the strongly intensive quantities $\Sigma$.
While the strongly intensive quantities have been or are being measured by several experimental collaborations, the question of how to relate them to lattice QCD and other theoretical predictions remains open.  

\item {\em Scaled factorial moments and intermittency}

Finally, observables related to intermittency were discussed.  In the grand-canonical ensemble the correlation length $\xi$  diverges at the critical point (or second order phase transition line) and the system becomes scale invariant. 
This leads to large multiplicity fluctuations with special properties. They can be conveniently exposed using scaled factorial moments $F_r(M)$~\cite{Bialas:1985jb} of rank (order) $r$: 

\begin{equation}
F_r(M)=\frac{ \langle \displaystyle{\frac{1}{M}\sum_{i=1}^{M}} 
	N_i(N_i-1)...(N_i-r+1) \rangle }
{\langle \displaystyle{\frac{1}{M}\sum_{i=1}^{M}} N_i \rangle^r } ~,
\label{eq:facmom}
\end{equation}
where
$M = \Delta / \delta$ is the number of the subdivision intervals of size
$\delta$ of the momentum phase space region $\Delta$.

At the second order phase transition the matter properties 
strongly deviate from the ideal gas.  The system is a simple
fractal and the $F_r(M)$ possess a power law dependence on $M$:
\begin{equation}
F_r(M) = F_r(1) \cdot M^{-\phi_r} ~.
\label{eq:p1}
\end{equation}
Moreover the exponent (intermittency index) $\phi_r$ satisfies the relation:
\begin{equation}
\phi_r = ( r - 1 ) \cdot d_r ~,
\label{eq:p2}
\end{equation}
with the anomalous fractal dimension $d_r$ being independent 
of $r$~\cite{Bialas:1990xd}.

It should be noted that $F_r(M)$ is sensitive to both volume fluctuations and conservation laws. A formulation of a new method to study intermittency using strongly intensive quantities is needed.

The question how well these assumptions are fulfilled in realistic heavy-ion collisions was also discussed. The finite size of the created system limits naturally the growth of the correlation length $\xi$. Therefore, finite-size corrections modify the predictions made for an infinite system. In addition, it was argued that due to dynamical effects the correlation length $\xi$ is not expected to exceed $2-3$~fm~\cite{Berdnikov:1999ph}, which is small compared to the size of the system. In this case, the analysis, which assumes scale invariance and requires $\xi$ to be as large as the system itself, would not be directly applicable. Dynamical modeling of heavy-ion collisions is necessary to quantify the magnitude of these effects.

\item {\em Light nuclei production}

Light nuclei production is also discussed as an observable related to spatial density fluctuations. The latter are expected to be enhanced in the vicinity of the critical point, but they are not measurable directly. It was suggested recently that they can be inferred from the light nuclei production~\cite{Sun:2017xrx,Sun:2018jhg}. In a simple coalescence model the yields of deuterons and tritons can be expressed as 
\begin{eqnarray}
 N_d \approx & \frac{3}{2^{1/2}} \left( \frac{2\pi}{mT}\right)^{3/2} \int d^3x \, \rho_p(x) \rho_n(x) \sim  \left\langle \rho_n \right\rangle N_p (1 + C_{np}) \,, \\
 N_t \approx & \frac{3^{1/2}}{4} \left( \frac{2\pi}{mT}\right)^3 \int d^3x \, \rho_p(x) \rho_n^2(x) \sim  \left\langle \rho_n \right\rangle^2 N_p (1 + 2 C_{np} +   
 \Delta \rho_n) \,,
\end{eqnarray}
where proton and neutron densities are allowed to fluctuate in space:
\begin{eqnarray}
 \rho_n(x) = & \left\langle \rho_n \right\rangle + \delta \rho_n(x) \,,\\
 \rho_p(x) = & \left\langle \rho_p \right\rangle + \delta \rho_p(x) \,,
\end{eqnarray}
and proton-neutron density correlations and neutron density fluctuations are denoted as
\begin{eqnarray}
 C_{np} \equiv & \left\langle \delta\rho_n(x) \delta\rho_p(x) \right\rangle / (\left\langle \rho_n \right\rangle \left\langle \rho_p \right\rangle) \,,\\
 \Delta\rho_n \equiv & \left\langle \delta\rho_n(x)^2\right\rangle / \left\langle \rho_n^2 \right\rangle \,.
\end{eqnarray}
Constructing the following ratio
\begin{equation}
 \frac{N_t N_p }{ N_d^2} = \frac{1}{2\sqrt{3}} \frac{1+2C_{np} + \Delta \rho_n}{(1+C_{np})^2} 
\end{equation}
one can see that the coalescence model predicts it to be independent of collision system, energy or centrality, but sensitive to spatial density fluctuations.
In the vicinity of the critical point this ratio should exhibit a peak. Combining the data from NA49 \cite{Anticic:2010mp,Blume:2007kw,Anticic:2016ckv}, STAR \cite{Adam:2019wnb,Zhang:2019wun}, and ALICE \cite{Adam:2015vda} one indeed can see two clearly pronounced peaks in the dependence of $(N_t N_p)/N_d^2$ on collision energy in central collisions. However, the interpretation of these peaks is currently not possible for two reasons: (1) available models disagree on the $(N_t N_p)/N_d^2$ ratio without critical point, (2) there is currently no model that would include a critical point, spatial density fluctuations emerging from it and light nuclei production. It was argued that the structures in $(N_t N_p)/N_d^2$ may be related to decays of excited $^4\mathrm{He}$ states \cite{Shuryak:2019ikv}, and 4-particle correlations represented by the enhanced $K_4/K_2$ ratio are originating from the same source. The observable (and similar related ratios, such as $(N_{{}^3\mathrm{He}} N_t)/(N_d N_{\alpha})$) seems to be very promising, but requires further investigation, especially from the theory side.
\end{enumerate}

\subsubsection{A new observable: $\chi^{B}_2/\chi^{Q}_2$?}

\begin{figure}[t]
 \centering
 \includegraphics[width=0.52\textwidth]{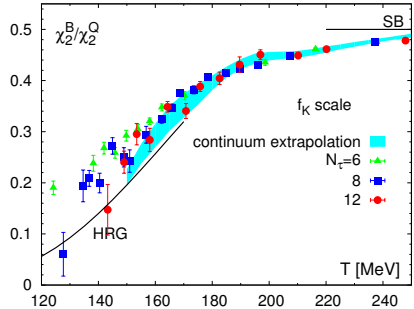}
 \caption{Susceptibility ratio $\chi^{B}_2/\chi^{Q}_2$ as a measure of net-baryon number to net-electric charge fluctuations obtained in lattice QCD calculations. Figure taken from~\cite{Bazavov:2012jq}. \label{fig:Chi2BQratio}}
\end{figure}

The quantity $\chi^{B}_2/\chi^{Q}_2$ was proposed as a particularly interesting observable, since lattice QCD 
calculations indicate that this ratio behaves almost linearly as a function
of $T$ near the critical temperature contrary to the behaviors of $\chi_4^B/\chi_2^B$
and other similar ratios that approach a constant value and are insensitive to $T$ below $T_c$. 
This is shown in Fig.~\ref{fig:Chi2BQratio}. 
Therefore a measurement of $K_2^B/K_2^Q$ would be more sensitive to the temperature and possible critical effects.  A further advantage is that it only requires measuring second cumulants, which are more easily within the statistical reach of experiments.  However, the net-charge cumulants are the most difficult to measure for several reasons, including the large mean multiplicities which increase the statistical uncertainties and the very significant effects of resonance decays which must be carefully controlled.

\subsection{Isospin and strangeness randomisation across collision energies\label{sec:proton_baryonnumber}}

In the hadronic phase, processes of the form $p+\pi^0 \leftrightarrow \Delta^+ \leftrightarrow n+\pi^+$ can alter the isospin-identity of nucleons. After only two of these cycles the original isospin distribution is completely randomised. This does not affect average quantities but significantly influences higher-order fluctuations. The efficiency for isospin randomisation depends strongly both on the density of pions and the regeneration plus decay time of the intermediate resonance in comparison to the duration of the hadronic stage. 
Isopsin randomisation might work efficiently at LHC energies because of the pion bath. However, the pion bath is not present at lower beam energies and the mechanism will break down at some point. 
The importance of isospin randomisation for relating the measured net-proton number cumulants to the theoretically interesting net-baryon number fluctuations has been worked out in~\cite{Kitazawa:2011wh,Kitazawa:2012at}. 
In fact, using the isospin randomisation it has been argued that the net-baryon number cumulants can be constructed from the experimentally observed proton number distribution even without the measurement of neutrons~\cite{Kitazawa:2011wh,Kitazawa:2012at}.
In~\cite{Nahrgang:2014fza}, the mechanism has been studied quantitatively and it was found that the net-proton distribution would be pushed strongly toward the Skellam limit if the distribution originally deviated from it.  

Similar questions should be investigated for the measurement of fluctuations in the net-strangeness. There are ongoing and existing measurements of net-kaon and net-$\Lambda$ fluctuations and the question is if such measurements are sufficient to reconstruct the original net-strangeness fluctuations. Analogous to isospin randomisation, future work in this direction can be based on reactions of the type $p+K \leftrightarrow \Lambda + \pi$ that might occur frequently enough in the hadronic phase.

\subsection{Volume fluctuations~\label{sec:volumefluct}}

While factors of volume appear in the relations between cumulants and susceptibilities listed above, the ``volume'' of the medium produced in a heavy-ion collision is not a very well defined quantity, although it is related to the number of nucleons that participate in the collision and their geometrical orientation and therefore to the collision centrality.  Event-by-event fluctuations of the participants, and therefore the volume, are inescapable and are an additional source of fluctuations that must be assessed in experimental measurements.  

Experimental measurements are performed in centrality classes, and the method used for estimating the centrality significantly influences the magnitude of the corresponding volume fluctuations.  Depending on the detector setup, the centrality may be estimated by energy deposited at forward pseudorapidity (for example, in the V0 detector in ALICE), the number of charged particles reconstructed at midrapidity (e.g. in the STAR TPC), or the spectators measured at zero degrees (as is done e.g. in HADES).  Each method has advantages and disadvantages and several experiments utilise different methods for cross-checks and cross-calibration. While decorrelation effects between different regions of phase space cause the participant fluctuations to be larger for a given centrality class, determining the centrality and measuring multiplicity fluctuations in the same region of phase space leads to autocorrelations.  

\begin{figure}[t]
 \centering
 \includegraphics[width=\textwidth]{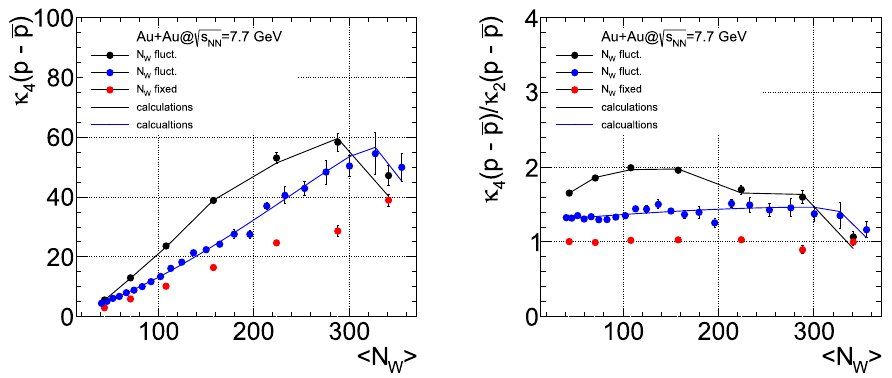}
 \caption{Left panel: Fourth cumulants of net-protons for Au+Au Collisions at $\sqrt{s_{\mathrm{NN}}} = 7.7$~GeV. Right panel: Ratio of fourth and second cumulants. Red points correspond to fixed number of wounded nucleons while, for the black points, the fluctuations of wounded nucleons are included. The centrality bin width is 2.5\% for the blue points, while for the black points variable bin widths are used. Figures taken from~\cite{Braun-Munzinger:2016yjz}.
 \label{fig:volumefluccbwc}}
\end{figure}

Within the experimental community, there are two approaches to account for volume fluctuations in measurements of higher-order multiplicity cumulants.  One approach is to attempt to correct the data by removing volume fluctuations.  One method which is currently being developed is to use a data-driven unfolding for this correction.  Another method is the centrality bin width correction (CBWC)~\cite{Luo:2013bmi}, in which the moments measured in narrow centrality bins are combined to obtain the cumulants in a wide centrality bin. It should be noted, however, that participant fluctuations will be present even in the limit of very fine centrality bins, and therefore the CBWC is only a partial correction, see Fig.~\ref{fig:volumefluccbwc}.  An alternative approach is to evaluate the impact of the volume fluctuations on the measured cumulants and then fold them into the baseline or model comparison.  Such an approach was explored in Ref.~\cite{Braun-Munzinger:2016yjz}, where the ALICE centrality estimation procedure and mean (anti-)proton multiplicities were used to evaluate the effects of the volume fluctuations on the second cumulants within the Wounded Nucleon class of models.  This approach means that no experimental information is lost in a correction procedure, but may also introduce a model-dependence into the interpretation of the results.

\subsection{The rapidity window dependence\label{sec:deltaEta}}

\begin{figure}[h]
 \centering
 \includegraphics[width=\textwidth]{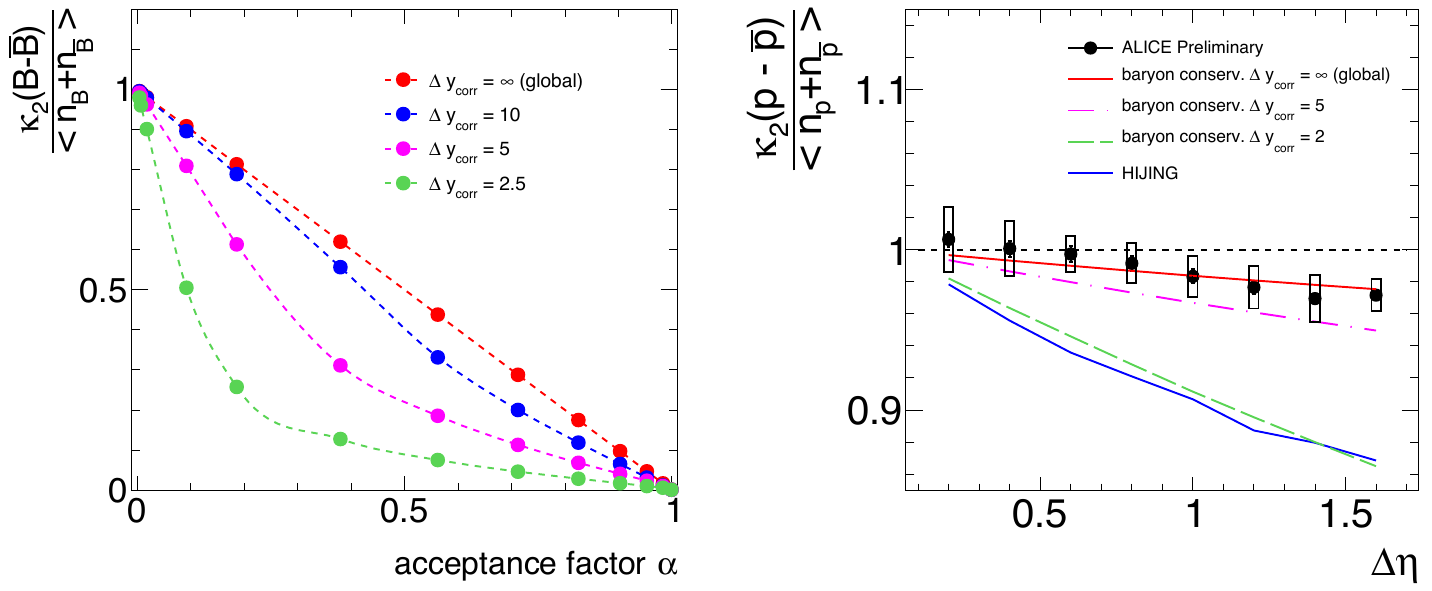}
 \caption{Left panel: the normalized values of $K_2$ for net-baryons, for different values of $\Delta y_{corr}$, as a function of  accepted  fraction  of  baryons, where $\Delta y_{corr}=2|y_{B}-y_{\Bar{B}}|$.  Right panel:  comparison of the results with the ALICE data.  Within the experimental  uncertainties,  the  data  are  best  described  by  global  baryon  number  conservation ($\Delta y_{corr}=\infty$) but are consistent with $\Delta y_{corr}\geq 5$. The blue solid line, representing the results of the HIJING generator, underestimates the experimental data and is described by the local baryon number conservation with $\Delta y_{corr}=2$. Figures taken from~\cite{Braun-Munzinger:2019yxj}.
 \label{fig:baryonnumberconservation}}
\end{figure}

The acceptance dependence can provide information about the nature
and origin of the correlations and fluctuations in heavy-ion
collisions. For instance, the effects of conservation laws on fluctuation observables can be assessed by changing the kinematic acceptance, in particular the (pseudo-)rapidity range of the measurement. When the acceptance of the measurement is small, then the fluctuations are reduced to purely statistical (Poissonian) fluctuations.  Meanwhile, when the acceptance of the measurement is large compared to the phase space of produced particles, conservation of baryon number, strangeness, and charge have an impact on the measured fluctuations.  The effect of global and local baryon number conservation laws was demonstrated in a model~\cite{Braun-Munzinger:2016yjz,Braun-Munzinger:2019yxj} to explain the rapidity window $\Delta\eta$ dependence of $K_2$ for net-protons measured by ALICE~\cite{Rustamov:2017lio}, see Fig.~\ref{fig:baryonnumberconservation}.  

Furthermore, two other major sources of 
fluctuations have characteristically distinct rapidity
window dependences: fluctuations of initial conditions and fluctuations
due to thermal noise~\cite{Kapusta:2011gt}. The 
contribution of thermal fluctuations to intensive measures (such as
  $\omega[N]$) grow with the acceptance window and saturate when the
  window width reaches the correlation rapidity range (typically, one
  unit of rapidity).  In contrast, the initial fluctuations lead to
  long-range (up to several units) rapidity correlations and their
  growth continues until the fireball boundary effects become
  important, e.g., due to conservation
  laws~\cite{Braun-Munzinger:2016yjz,Braun-Munzinger:2019yxj}.

\begin{figure}[h]
  \centering
  \includegraphics[width=.52\textwidth]{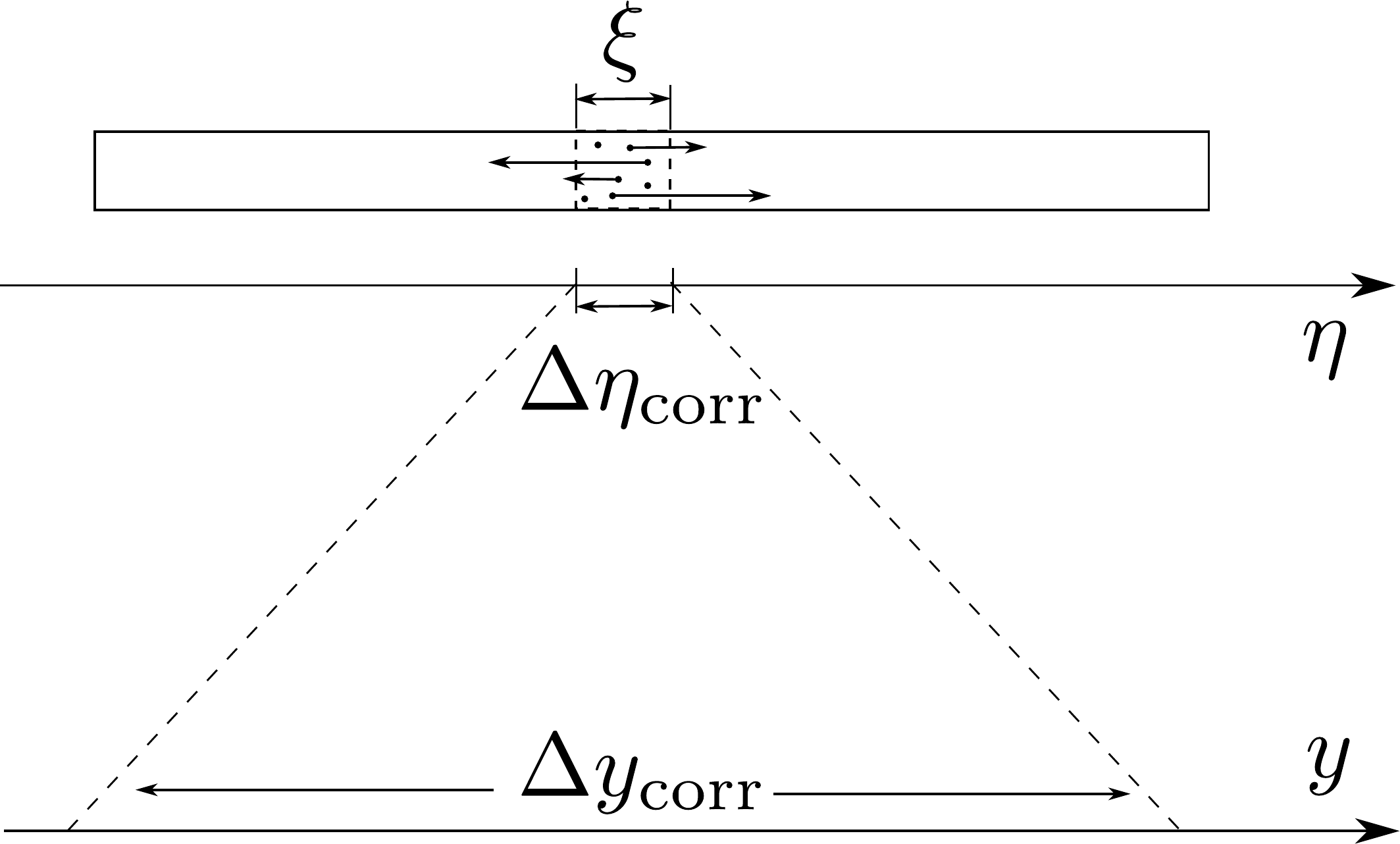}
\caption[]{\label{fig:deta-dy} Schematic illustration of the relation
  between the spatial (Bjorken) rapidity $\eta$ 
and kinematic rapidity $y$ via the effect of the thermal broadening
(freezeout smearing). The figure is from Ref.~\cite{Ling:2015yau}.}
\end{figure}

Since the fluctuations due to the QCD critical point are essentially
thermal fluctuations, their correlation range is of order one unit
of rapidty \cite{Ling:2015yau}. It should be emphasized that the spatial
correlation length, $\xi$, despite becoming anomalously large at the
critical point, has little effect on the range of the {\em kinematic} rapidity
correlations (see Fig.~\ref{fig:deta-dy}). 
The larger spatial correlation length $\xi$ translates into a larger
number of particles being correlated and thus manifests in a larger 
{\em magnitude} of the fluctuation measures \cite{Stephanov:2008qz,Stephanov:1999zu}.

\begin{figure}
\centering
  \includegraphics[width=0.325\textwidth]{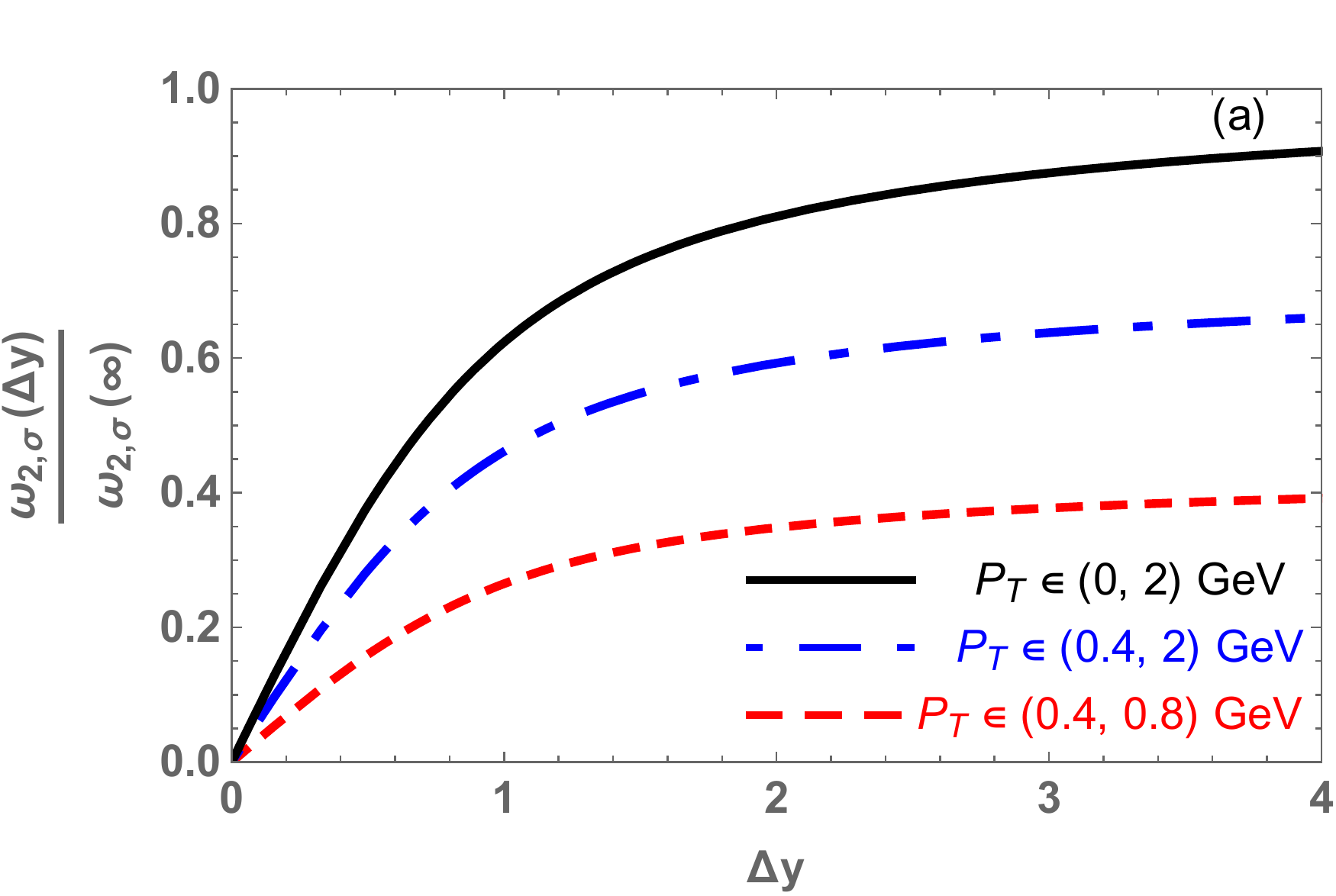}
  \includegraphics[width=0.325\textwidth]{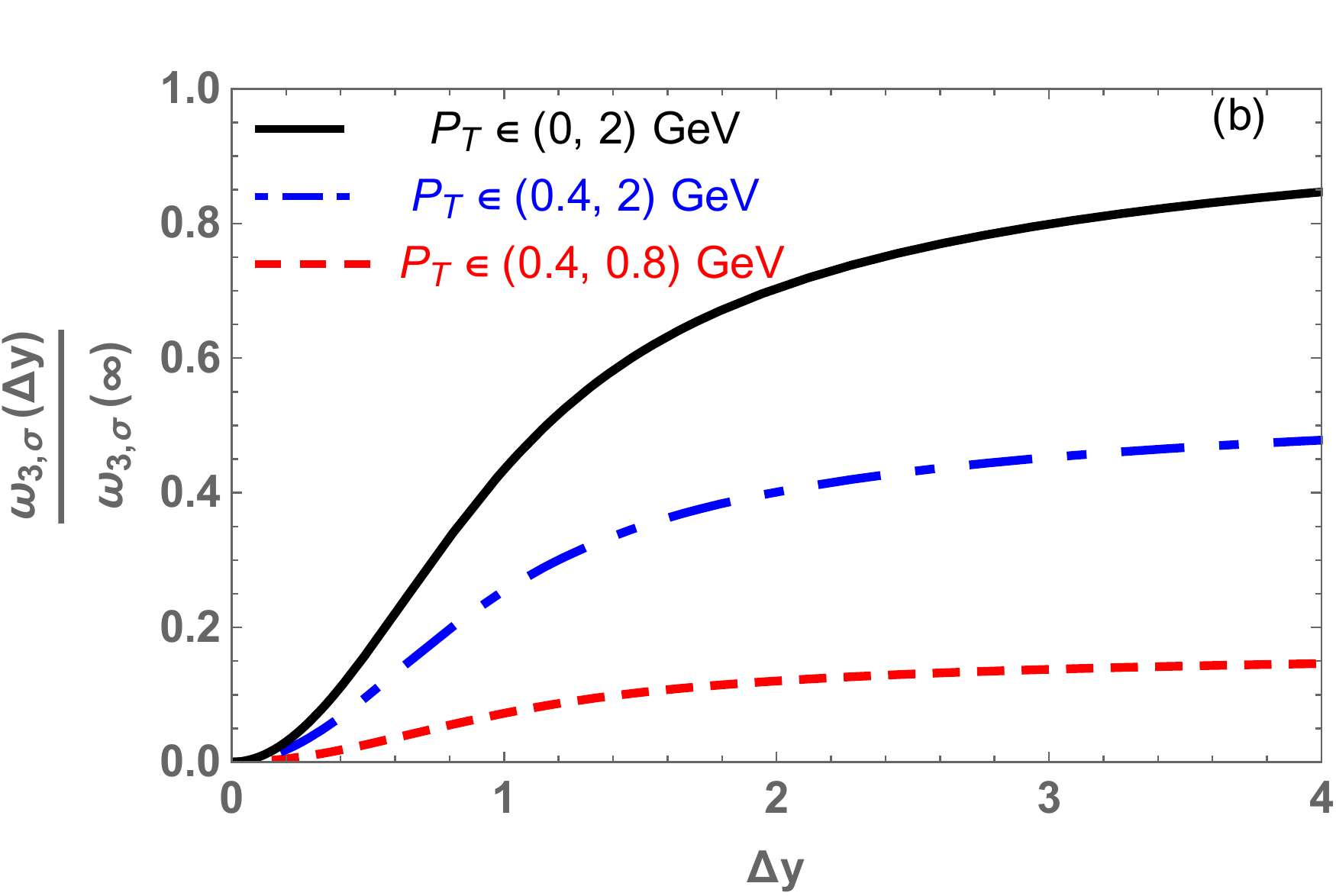}
  \includegraphics[width=0.325\textwidth]{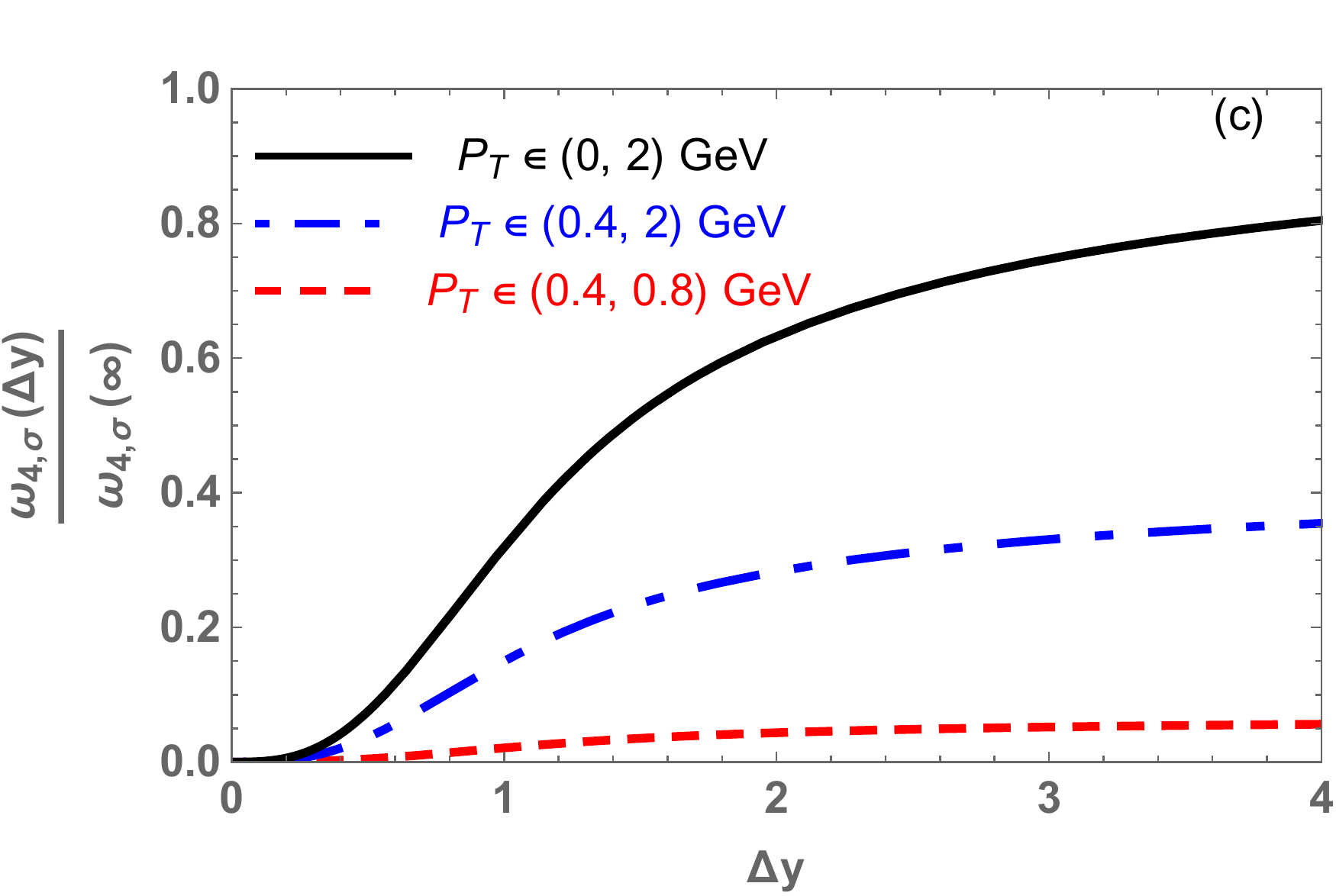}
\caption[]{Acceptance dependence of the critical contribution to the normalized cumulants of proton number. Figures taken from~\cite{Ling:2015yau}.}
\label{fig:o234}
\end{figure}

The rapidity window dependence of the cumulants (normalized by
multiplicity to make them intensive) follows the pattern expected from
thermal fluctuations~\cite{Ling:2015yau}, as shown in
Fig.~\ref{fig:o234}. For small rapidity windows
$\Delta y\ll 1$ the
normalized cumulant, $\omega_k/N$, grows as a power $\Delta
y^{k-1}$. 
Furthermore, since the critical fluctuations correlate
particles with different transverse momenta $p_T$, the magnitudes of
the fluctuation measures increase with the $p_T$ acceptance, as also
illustrated in Fig.~\ref{fig:o234}. 

Other effects which are sensitive to the width of the rapidity window were also discussed in detail, particularly resonance decays and the influence of diffusion. 
At the intermediate range of $\Delta\eta$, the dependence of higher-order cumulants on $\Delta\eta$ is determined by the correlations between observed particles, and thus can be sensitive to the diffusion process and particle production mechanisms. Various estimates on the $\Delta\eta$ dependence have been made in the literature based on models for the diffusion and particle production~\cite{Kitazawa:2013bta,Kitazawa:2015ira,Ling:2015yau,Bzdak:2017ltv}.

\subsection{Influence of resonance decays on fluctuation observables\label{sec:feeddown}}

Resonance decays play an important role in particle production. For example, thermal model estimates show that -- integrated over all transverse momenta -- about 60\% of all pions originate from the decays of heavier hadronic states~\cite{Andronic:2017pug}. 
On an event-by-event basis this can vary considerably and, thus, resonance decays may influence fluctuation observables significantly. In general, the decay of resonances follows a multinomial probability distribution~\cite{Begun:2006jf} from which the impact of the decays on the cumulants of a particle multiplicity distribution can be derived. This has been done up to the fourth order cumulant in~\cite{Fu:2013gga}. Based on those results, the influence of resonance decays on the fluctuations in the net-proton number has been estimated in~\cite{Nahrgang:2014fza}. It was found that the higher-order cumulants are stronger influenced by the decays. Moreover, a proper inclusion of the probabilistic character of the decay process turned out to be essential: for net-protons this can be an up to 20\% effect on the cumulant ratios and is expected to be even stronger for pions~\cite{Nahrgang:2014fza}. 

An important question concerns the connection between resonance decay contributions and the rapidity window dependence. 
For the thermal model interpretation of the yields (first moments), this does not play a role: if a charged pion from a $\rho$ decay leaves the acceptance window, it is quite likely that in another event a pion from a neighboring rapidity window enters the acceptance window, leaving the number of pions in the acceptance window unchanged on average. Higher-order fluctuations, however, are sensitive to resonance decays particularly if the acceptance window is smaller or of the same order as the average rapidity difference between the decay daughters.  Therefore, decays of the type $\rho \rightarrow \pi^{+}\pi^{-}$ strongly influence the measurement of net-charge fluctuations. As a matter of fact, it was previously suggested to constrain the number of produced $\rho$ and $\omega$ mesons via a measurement of net-charge fluctuations~\cite{Jeon:1999gr}. Similarly, the decay $\phi \rightarrow K^{+}K^{-}$ influences the measurement of net-strangeness measurements. Detailed studies are needed for the understanding of net-charge and net-strangeness observables. These effects are best studied using event generators which are either QCD inspired such as HIJING or based on statistical-thermal particle yields coupled to hydrodynamic expansion. 
Early estimates~\cite{Jeon:1999gr} based on a Monte Carlo study for Pb-Pb collisions at SPS energies indicated a modification of only 1\% in the resonance decay contributions to the fluctuations in the $\pi^{+}/\pi^{-}$ ratio if the acceptance in rapidity is limited to $\Delta\eta=1$. 
In any case, a proper particle decay model needs to be an integral part of the code.

Decays of the type $\Delta^{++} \rightarrow p\pi^{+}$ and $\Lambda \rightarrow p\pi^-$ are rather different because they do not change the baryon number in a correlated way.  However, the effects of feed-down in the net-proton measurements should be experimentally addressed.  In principle, including primary and secondary protons from all sources would come the closest to a measurement of the baryon multiplicity.  In practice, though, it must be noted that in every experiment the detection efficiency is different for primary particles and particles of secondary origin, and branching ratios have to be taken into account. Therefore, it would be necessary to measure the multiplicity fluctuations for each particle species individually.  Experimentally, protons from weak decays are characterized by a larger distance-of-closest-approach (DCA) to the primary vertex and this information could be used to assign probabilities for each proton if it is of primary or secondary origin (analogous to the Identity Method for particle identification). From the theoretical side, such measurements could be accompanied by the simultaneous determination of $\chi^{B}$, $\chi^{BS}$, and $\chi^{S}$ since for instance $\Lambda$, $\Xi$, and $\Omega$ decays contribute to all three of these fluctuation observables simultaneously. Setting up such a detailed experimental and theoretical framework should be one of the main goals of the future research activity in this field.

Additional complications in these studies might arise from re-generation and re-scattering processes which are likely to occur in the hadronic phase of the collision. Two primordially produced pions might pseudo-elastically re-scatter via the large cross-section process $ \pi^{+}\pi^{-} \rightarrow \rho \rightarrow \pi^{+}\pi^{-}$ and thus again leave the acceptance window. Re-generation effects are not modeled by event generators like HIJING and a correct treatment implies the usage of afterburners based on UrQMD.

\subsection{Overview of the current experimental techniques}

The experimental techniques and methodologies across experiments are currently not fully harmonised. As a matter of fact, 
one of the goals of the RRTF was to contribute to the community-wide efforts 
to establish a common approach in the measurement of fluctuation observables. 
In order to obtain an overview of the current status of fluctuation measurements in the various experiments, each experiment was asked to provide answers to the following set of questions: 
\begin{enumerate}
    \item Which fluctuation observables are measured? 
    \item Which results are already available and what are the plans for the future?
    \item What are the applied acceptance cuts?
    \item What are the available statistics, at which energies and systems, now and in the future?
    \item Are unphysical sources removed (e.g. spallation protons) and how?
    \item How are secondaries from weak decays treated (e.g. $\Lambda$ $\rightarrow$ p + $\pi^{-}$)?
    \item How is the efficiency correction performed?
    \item How are potential event-by-event fluctuations in the efficiency treated or modeled?
\end{enumerate}
The responses are summarized below.  

\subsubsection{Fluctuation measurements in ALICE\label{sec:ALICEflucs}}

ALICE has shown preliminary results on the first and second central moments of net-pion, net-kaon, net-proton, and net-$\Lambda$ distributions~\cite{Rustamov:2017lio,Ohlson:2019erm} measured using the Identity Method~\cite{Gazdzicki:2011xz,Gorenstein:2011hr,Rustamov:2012bx}, as well as the third- and fourth-order cumulants of the net-proton distribution~\cite{Behera:2018wqk} analysed with traditional cut-based particle identification.  Furthermore, the ALICE collaboration has published measurements of particle ratio fluctuations~\cite{Acharya:2017cpf} and net-charge fluctuations~\cite{Abelev:2012pv}, quantified with the observable $\nu_{dyn}$, as well as balance functions~\cite{Abelev:2013csa,Adam:2015gda}.  

\begin{enumerate}
\item {\em Second cumulants of net-proton, net-kaon, and net-pion distributions}

The analysis of the second moments of the net-pion, net-kaon, and net-proton distributions was performed with data from Pb-Pb collisions at $\sqrt{s_{NN}} = 2.76$ TeV collected in 2010 by the ALICE detector.  The kinematic acceptance of the measurement is $|\eta| < 0.8$ and $0.6 < p_T < 1.5~\text{GeV}/c$.  In this momentum range, the tracking efficiency for protons (anti-protons) in the Time Projection Chamber (TPC) is roughly constant at approximately 78\% (70\%), which confers a technical advantage of allowing the analysis to be performed in a single inclusive momentum bin. In this analysis, the efficiency correction was performed at the level of the first and second moments using simulated Monte Carlo events passed through a GEANT model of the ALICE detector. Two Monte Carlo generators, AMPT and HIJING, were used for the efficiency correction; the small difference between the generators was used to estimate the corresponding systematic uncertainty. The procedure was also cross-checked by assuming binomial track loss \cite{Nonaka:2017kko}. The accuracy of the correction procedure was estimated to be on the percent level and was included in the systematic uncertainties.
Secondary protons, mainly from the decay of $\Lambda$ baryons, were not explicitly removed from the measurement, but their influence on the final results was evaluated by varying the selection on the DCA between the tracks and the primary vertex, and the observed small deviations were included in the systematic uncertainties.  

\item {\em Second cumulants of net-$\Lambda$ distributions}

The analysis of the second cumulants of the net-$\Lambda$ distribution was performed for $\sqrt{s_{NN}} = 5.02$ TeV Pb-Pb collisions, using the data set collected in 2015.  $\Lambda$ and $\overline{\Lambda}$ baryons were reconstructed via their decay to (anti-)protons and charged pions.  To account for the background of combinatoric proton-pion pairs, the Identity Method was applied along the invariant mass ($m_{p\pi}$) axis by evaluating the probability at each value of $m_{p\pi}$ that a proton-pair corresponds to a true $\Lambda$ baryon decay or a combinatoric pair.  Since the $\Lambda$ reconstruction efficiency depends strongly on $p_T$ throughout the kinematic range, from a minimum of 10\% at $p_T = 1$ GeV/$c$ to a maximum of roughly 30\% at $p_T = 4$ GeV/$c$, the $p_T$-dependent efficiency correction was performed assuming binomial efficiency loss according to the prescription in Ref.~\cite{Pruneau:2017fim}.  The secondary contamination of $\Lambda$ and $\overline{\Lambda}$ baryons originating from the decay of $\Xi$ baryons is also incorporated into the efficiency correction procedure.  

\item {\em Third and fourth cumulants of net-proton distributions}

The analysis of the third and fourth moments of the net-proton multiplicity distributions was performed in Pb-Pb collisions at $\sqrt{s_{NN}} = 2.76$ and 5.02 TeV; the results at both energies are consistent within statistical and systematic uncertainties.  Protons in the kinematic range $|\eta| < 0.8$ and $0.4 < p_{T} < 1~\text{GeV}/c$ are identified according to tight selection cuts on specific energy loss in the TPC.  Due to the stricter particle identification cuts used in this analysis, the reconstruction efficiency for protons (anti-protons) is approximately 65\% (60\%).  The moments are corrected for efficiency according to Ref.~\cite{Nonaka:2017kko}; the centrality bin width correction is also applied. 
\end{enumerate}

\noindent Each of the cumulant measurements described above is compared to the Skellam baseline.  Small deviations from the Skellam baseline are observed for the net-proton and net-$\Lambda$ second cumulants; within the precision of the measurement these deviations can be fully described by a model which includes the effects of baryon number conservation~\cite{Braun-Munzinger:2016yjz,Braun-Munzinger:2019yxj}. The effects of baryon number conservation were further tested by performing these measurements differentially with respect to the pseudorapidity acceptance ($\Delta\eta$).  It was observed that Poissonian/Skellam behavior is recovered for small $\Delta\eta$, and the measured cumulants decrease with respect to the Skellam limit as $\Delta\eta$ increases, consistent with expectations.  In the most central Pb-Pb collisions the higher-order cumulants are consistent with a Skellam distribution; the significant statistical and systematic uncertainties do not yet make it possible to observe any deviations.  

An ongoing analysis of the third- and fourth-order cumulants of net-proton distributions using the Identity Method will allow the kinematic range and precision of the measurement to be extended.  A phenomenological evaluation of the effects of volume fluctuations and baryon number conservation on the higher moments is also underway, which will make a precise and quantitative test of lattice QCD possible.  Furthermore, the data collected by the ALICE experiment in Runs 3 and 4 at the LHC will make it possible to measure the fourth moments of identified particles with unprecedented precision, and the sixth moments are also foreseen to come within experimental reach.  

\subsubsection{Fluctuation measurements in STAR}

The STAR experiment at RHIC has measured a range of fluctuation observables, including the higher moments of net-charge~\cite{Adamczyk:2014fia,Abelev:2008jg}, net-proton~\cite{Adamczyk:2013dal,Adam:2020unf,Aggarwal:2010wy}, and net-kaon~\cite{Adamczyk:2017wsl} multiplicity distributions, as well as event-by-event fluctuations of identified particle ratios~\cite{Abdelwahab:2014yha}, mean $\pt$ fluctuations~\cite{Adam:2019rsf}, and balance functions~\cite{Adamczyk:2015yga}.  As part of the RHIC beam energy scan (BES) program, these measurements have been performed in \AuAu~collisions across a wide range of collision energies, from $\sqrt{s_{\mathrm{NN}}} = 7.7$ GeV to 200 GeV.  The available statistics from the BES Phase I and the top RHIC energies are listed in Table~\ref{tab:bes}, as well as the projected statistics which will be collected in BES Phase II, to take place between 2018 and 2021. 

\begin{table}
\begin{tabular}{|c|c|c|} \hline
$\sqrt{s_{\mathrm{NN}}}$ (GeV) & Available statistics & Expected statistics in BES-II \\
 & (millions of events) & (millions of events) \\ \hline
7.7 & 4 & 100 \\
9.1 & -- & 160 \\
11.5 & 12 & 230 \\
14.5 & 20 & 300 \\
19.6 & 36 & 400 \\
27 & 70 & 500 \\ 
39 & 130 & -- \\ 
54.4 & 1200 & -- \\ 
62.4 & 67 & -- \\ 
200 & $>850$ & -- \\ \hline
\end{tabular}
\caption{\label{tab:bes} The collected statistics in the 0-80\% centrality range from BES Phase I and the expected statistics from BES Phase II, listed in millions of events.}  
\end{table}

Of particular interest are the higher moments of the net-charge, net-kaon, and net-proton multiplicity distributions measured across the full range of BES energies.  The measurements are performed in the (pseudo)rapidity windows $|\eta| < 0.5$ and $|y| < 0.5$ for unidentified and identified particles, respectively, and in the following transverse momentum ranges: $0.2 < \pt < 2$~GeV/$c$ (net-charge), $0.2 < \pt < 1.6$~GeV/$c$ (net-kaon), $0.4 < \pt < 0.8$~GeV/$c$ (net-proton); since publication the net-proton kinematic range has been extended to $0.4 < \pt < 2$~GeV/$c$~\cite{Luo:2015ewa}.  

Particle identification is performed using the specific energy loss in the TPC and time of flight from the TOF detector.  The finite tracking efficiency is corrected under the assumption of binomial track loss, which has been extensively tested in Monte Carlo simulations~\cite{Adamczyk:2017wsl}.  Efforts to apply an unfolding procedure in the efficiency correction are underway.  The CBWC is also applied.  Decay products from weak decays and spallation protons are rejected with experimental cuts on the transverse momentum and distance of closest approach to the primary vertex, although there is no explicit correction for the residual contamination.  

The centrality of each event is determined from the charged-particle multiplicity at mid-rapidity, not including the particle under study (i.e. in the net-proton measurement, protons are excluded from the centrality determination).  While this avoids maximally correlating the observable with the centrality, remaining autocorrelations may still be present.  

Future work will include the analysis of fluctuation observables in BES Phase II, where a massive increase in available statistics is anticipated (see Table~\ref{tab:bes}).  Furthermore, the beam energy scan program will be extended by inserting a gold target into the STAR detector, such that fixed-target collision events can be recorded.  For example, when the collider energy is $\sqrt{s_{\mathrm{NN}}} = 62.4$ GeV, the fixed-target energy is $\sqrt{s_{\mathrm{NN}}} = 7.7$ GeV.  Similarly, for a collider energy of $\sqrt{s_{\mathrm{NN}}} = 7.7$ GeV, the corresponding fixed-target center-of-mass energy is $\sqrt{s_{\mathrm{NN}}} = 3$ GeV.  By utilizing the flexibility and performance of the STAR detector and RHIC beams, fluctuation studies will be able to be performed across a wide range of the phase diagram in $T-\mu_B$ space. 

\subsubsection{Fluctuation measurements in HADES}
With the HADES detector, located at the SIS18 at GSI, the proton number fluctuations have been investigated in \AuAu{} collisions at $\sqrt{s_{\mathrm{NN}}} = 2.41$~GeV.  HADES operates at low energies where there is no anti-proton production, hence net-proton fluctuations are measured as proton number fluctuations.  However, approximately $^1/_3$ of the protons are bound in light nuclear clusters produced either thermally or via coalescence. While so far only the fluctuations in the number of free protons have been analyzed, future work will also include the protons bound in deuterons, tritons, and He nuclei into the fluctuation signals.  

Protons are selected within the HADES geometrical acceptance with $0.4 < \pt < 1.6~\text{GeV}/c$ and $y = y_{0} \pm 0.5$, where $y_0 = 0.74$ is the center-of-mass rapidity.  For the analysis, the \AuAu{} data is subdivided into four centrality classes: 0-10\%, 10-20\%, 20-30\%, and 30-40\%; the resulting volume fluctuations also produce a fluctuation signal and must therefore be investigated carefully \cite{Braun-Munzinger:2016yjz,Skokov:2013abc}. The effects of detector (in)efficiency have been extensively investigated in simulations.  Corrections are applied on a bin-by-bin and event-by-event basis in 240 phase-space bins to account for the dependence of the reconstruction efficiency on rapidity, transverse momentum, and detector load, assuming binomial efficiency loss~\cite{Bzdak:2012ab,Nonaka:2017kko,Bzdak:2013pha}.  Event-by-event fluctuations of the detector efficiency are of order 10-15\% in HADES and must be taken into account.  A linear model, adjusted phase-space bin by bin to simulated events, is used to recalculate the efficiency in each event and for each bin.  These event-by-event efficiencies are incorporated into Kitazawa's efficient scheme~\cite{Nonaka:2017kko}.  Unfolding techniques based on a simulated 2d detector response matrix using regularization or singular value decomposition (SVD) schemes have also been investigated.  Furthermore, the newly proposed moment-expansion scheme~\cite{Nonaka:2018mgw} has been implemented and tested.  Both, in simulations and in data, the three methods agree well for the first (mean), second (variance), and third order (skewness) moments, and still reasonably well in the fourth order (kurtosis).  

The analyzed data set consists of $2 \cdot 10^8$ \AuAu{} events at $\sqrt{s_{\mathrm{NN}}} = 2.41$~GeV. Protons from weak decays are not fully subtracted, but partially suppressed by track vertex cuts.  However, since strangeness production is subthreshold at SIS energies, these protons contribute on the $<10^{-3}$ level.  Contributions from spallation protons are being investigated in GEANT3 simulations, but are expected to be weak due to the fixed target setup and the low beam energies used.  Event pile-up is at a $<2 \cdot 10^{-4}$ level.  

In addition to proton number fluctuations, HADES can measure net-charge fluctuations by considering the free and bound protons as well as the charged pions.  (Strangeness production is subthreshold and does not contribute much to the charge.)  In the future, fluctuation observables will be analyzed in the high-statistic data sets recently recorded from $5 \cdot 10^8$ \AgAg{} collisions at $\sqrt{s_{\mathrm{NN}}} = 2.41$~GeV and $6.5 \cdot 10^9$ events at $\sqrt{s_{\mathrm{NN}}} = 2.55$~GeV.  Within the FAIR phase-0 stage HADES may also request a more complete energy scan at SIS18 with Au beams of 0.2 - 1.0 GeV/u to probe the liquid-gas phase transition region.  Finally, beyond 2027, HADES is expected to extend the excitation function of fluctuation signals at SIS100 using various heavy-ion beams of 3.5~GeV/u.

\subsubsection{Fluctuation measurements in NA61/SHINE}

The NA61/SHINE experimental program encompasses a diverse set of colliding beams (\pp{}, \pPb{}, \BeBe{}, \ArSc{}, \XeLa{}, \PbPb{}) over a wide range of beam momenta from 13$A$ to 150$A$~$\text{GeV}/c$.  The detector offers a unique opportunity to study a wide range of fluctuation observables due to its high tracking efficiency ($>90\%$ down to $\pt = 0~\text{GeV}/c$) and particle identification capabilities.  

The impact of volume fluctuations on fluctuation observables is reduced by analyzing only the most central collisions, where few spectators are detected in the Projectile Spectator Detector, thus maximizing the number of wounded nucleons participating in the collisions and limiting fluctuations.  Furthermore, the use of strongly-intensive quantities~\cite{Gazdzicki:1992ri,Gorenstein:2011vq,Gazdzicki:2013ana} is preferred which are independent of the volume and volume fluctuations within models of independent particle sources, for example the Wounded Nucleon Model~\cite{Bialas:1976ed} and the grand-canonical ensemble of an ideal gas of Boltzmann particles.  

The intensive quantity $\omega$ as well as the strongly intensive $\Omega$, $\Delta$, and $\Sigma$ observables have been measured across a wide range of collision systems and energies.  Additionally, a systematic study of intermittency has also been performed.  For a summary of recent results, see~\cite{Grebieszkow:2019yjd}. 

\begin{enumerate}

\item{\em Intensive observables}

The scaled variance, $\omega$, has been measured in \pp{}, \BeBe{} and \ArSc{} across a range of energies~\cite{Czopowicz:2015mfa,Aduszkiewicz:2015jna,Seryakov:2017sss}.  In \pp{} interactions, and also in \BeBe{} collisions, multiplicity fluctuations are larger than predicted by statistical models.  However, they are close to statistical model predictions for large volume systems in central \ArSc{} and \PbPb{} collisions~\cite{Begun:2006uu}.  The observed rapid change of hadron production properties that start when moving from \BeBe{} to \ArSc{} collisions may be interpreted as the beginning of the creation of large clusters of interacting matter, or the onset of the fireball~\cite{Aduszkiewicz:2287091}. 

\item {\em Strongly intensive observables}

The observable $\Sigma[P_T,N]$ has been measured for both positively and negatively charged hadrons in \pp{}, \BeBe{}, and \ArSc{} collisions at beam momenta of 20$A$, 31$A$, 40$A$, 80$A$, and 158$A$ $\text{GeV}/c$~\cite{Andronov:2016ddd}.  In this scan of the phase diagram, no "fluctuation hill", or increase of fluctuations which would indicate the presence of a critical point, has been observed.  

\item {\em Intermittency}

The intermittency $F_2(M)$ has been measured by NA61/SHINE~\cite{Davis:2017puj,MMP:QM2019} and NA49~\cite{Anticic:2012xb} in \BeBe{} (NA61/SHINE), \CC{} (NA49), \SiSi{} (NA49), \ArSc{} (NA61/SHINE), and \PbPb{} (NA49) collisions with beam momenta in the range 150$A$-158$A$~$\text{GeV}/c$.  
NA49 reported an indication for critical fluctuations in \SiSi{} collisions (a power-law enhancement with respect to the mixed events baseline). The NA61/SHINE results on \ArSc{} show no convincing indication of critical fluctuations so far. 
The results for \BeBe{}, \CC{} and \PbPb{} collisions are consistent with the  mixed events baseline.
Analysis of data for all reactions recorded by NA61/SHINE is ongoing.

\end{enumerate}

\subsubsection{Future common standards to be followed by experiments}

As the experimental measurements of higher-order moments are extremely challenging and sensitive, and high precision is needed for informative comparisons to theoretical calculations, it is important to establish a set of basic quality checks that can be performed in order to give confidence in the results.  As a starting point, some suggestions are listed here: 
\begin{enumerate}
    \item Since the correction for detector (in)efficiency is a critical part of a higher-moments analysis, it is important to verify the underlying assumptions implicit in the correction procedure.  In particular, if the correction procedure relies on the assumption that particle loss occurs according to a binomial distribution (as in Refs.~\cite{Bzdak:2012ab,Bzdak:2013pha}), then it should be demonstrated that the detector response is, in fact, binomial.  If it is not, the effect of deviations from a binomial track loss should be investigated in, for example, a Monte Carlo closure test (see below).  
    \item At the LHC and at top RHIC energies, where stopped baryons lie at far forward rapidities and not within the experimental acceptance, the odd net-particle moments ($K_1$, $K_3$, $K_5$, ...) should be zero.  Experimental verification of this is critical to determine if the detector efficiency correction procedure is under control.  
    \item A ``Monte Carlo closure test'' is an extremely useful tool for validating the accuracy of an analysis procedure.  To check for closure, the observable of interest is calculated in a Monte Carlo model using generator-level particles.  Then the generated events are passed through a simulated model of the detector and subsequent data reconstruction chain.  The analysis can then be run on the reconstruction-level particles (in the form of tracks, etc), just as it is in real data.  If the full analysis chain and all correction procedures are working well, then the corrected reconstruction-level results should match that obtained from the generator-level particles.  Note that, to first order, it does not matter whether the Monte Carlo reproduces the physics seen in the real data -- the Monte Carlo closure test is only a test of the analysis method, not a physics result.  As such, however, it is a \textit{necessary but not sufficient} proof of the robustness of an analysis.  
\end{enumerate}

\section{Fluctuations in atomic gases and other related systems}
\label{sec:Other}


Fluctuations and correlations play an important role
in many other systems, ranging from the microscale as 
in ultra-cold atomic gases and condensed matter systems, to
the largest structures in the universe, galaxies and 
clusters of galaxies. Indeed, the idea that fluctuations
can serve as a signal of critical behavior goes back
to the explanation of critical opalescence near the 
endpoint of the liquid-gas phase transition in water
in terms of large fluctuations by Smoluchowski and Einstein. 
Since then, fluid mixtures and condensed matter systems 
have served as important testing grounds for ideas about 
dynamical critical phenomena and critical transport. 
During the RRTF meeting, connections between fluctuation 
studies in heavy-ion collisions and in other physical 
systems were discussed. 
In this section we will address possible avenues 
for testing the ideas presented in the previous sections
in more controlled, table-top experiment, settings. We
will also discuss how ideas about critical behavior 
in heavy-ion collisions may motivate new experiments 
in atomic or condensed matter physics. 

An important model system is given by ultra-cold atomic 
quantum gases. Atomic gases allow for a great amount of 
control, both in terms of the ability to select the initial 
state, as well as the ability to tune the strength of the
interaction between the atoms. Atomic systems can be 
exposed to a variety of external probes and monitored 
in real time. Experiments can be performed in equilibrium
and in conditions that are far away from thermal equilibrium.

Within atomic gases the BCS/BEC crossover in dilute atomic 
Fermi gases has received particular attention. Because the 
systems are dilute, details of the atomic structure are 
not important, and we can describe the gas as being composed
of non-relativistic, point-like, spin 1/2 fermions that interact
via zero-range forces. This force can be tuned using Feshbach
resonances to cover the range between weakly attractive 
(the Bardeen-Cooper-Schrieffer, BCS, regime) to very strongly 
attractive (the Bose-Einstein condensation, BEC, limit). 
In the BEC limit the system forms tightly bound pairs with
weak residual interactions. As a result the dilute Fermi gas 
is most strongly correlated at the BCS/BEC crossover. A
special case arises when a bound state first appears in 
the two-body spectrum. In that case the $s$-wave scattering
length is infinite, and the system exhibits non-relativistic
scale invariance. This is known as the unitary limit, because
the $s$-wave scattering length saturates the unitarity bound. 

\subsection{Equilibrium fluctuations and correlations}

 The unitary Fermi gas has played an important role 
in our understanding of nearly perfect fluidity because,
like relativistic heavy-ion collisions, it exhibits strong
elliptic flow when released from a spatially imhomogeneous 
initial state \cite{Schafer:2009dj,OHara:2002pqs,Cao:2010wa}.
Indeed, this phenomenon can be analyzed in much the same
way that flow is analyzed in heavy-ion collisions, using the 
known equation of state and viscous fluid dynamics. 

 More detailed information is provided by two-point 
correlation functions. The simplest correlator is the dynamic
structure factor
\begin{equation}
\begin{split}
& \frac{1}{2}\langle \delta n(t_1, \mathbf{x}_1) 
                     \delta n(t_2, \mathbf{x}_2) 
 + \delta n(t_2, \mathbf{x}_2) \delta n(t_1, \mathbf{x}_1) \rangle \\
& = \int \frac{d\omega d^3 k}{(2\pi)^4} 
   e^{-i\omega(t_1-t_2)+i\mathbf{k} (\mathbf{x}_1-\mathbf{x}_2)} 
   \Delta_{nn}^\text{S}(\omega, \mathbf{k}).
\end{split}\label{eq:symmetricDensityResponse}
\end{equation}
Here, $\delta n(t,\mathbf{x})=n(t,\mathbf{x})-\bar n$ with $\bar n = 
\langle n(t,\mathbf{x})\rangle$ is a fluctation in the particle density. 
The zero frequency limit of $\Delta_{nn}^S(\omega,\mathbf{k})$ is 
known as the static structure factor. A closely related quantity is
the retarded response function $\Delta_{nn}^\text{R}(\omega, \mathbf{k})$,
\begin{equation}
\begin{split}
& i \theta(t_1-t_2) 
\langle \delta n(t_1, \mathbf{x}_1) \delta n(t_2, \mathbf{x}_2) 
      - \delta n(t_2, \mathbf{x}_2) \delta n(t_1, \mathbf{x}_1) 
      \rangle \\
& = \int \frac{d\omega d^3 k}{(2\pi)^4} 
e^{-i\omega(t_1-t_2)+i\mathbf{k} (\mathbf{x}_1-\mathbf{x}_2)} 
\Delta_{nn}^\text{R}(\omega, \mathbf{k}).
\end{split}
\end{equation}
In thermal equilibrium the symmetric correlation function and the 
retarded response function are related through a fluctuation-dissipation 
relation,
\begin{equation}
    \Delta_{nn}^\text{S}(\omega, \mathbf{k}) = \left[\frac{1}{2} + \frac{1}{e^{\omega/T}-1}  \right] 2 \, \text{Im} \, \Delta_{nn}^\text{R}(\omega, \mathbf{k}).
\end{equation}
The response function of ultra-cold gases has been measured using 
Bragg scattering \cite{Gaebler_2010}. In these experiments one uses two
crossed laser beams, where the differences in frequency and wave number
between the two beams determine the $(\omega,\mathbf{k})$ at which the 
response is measured. The two laser beams drive two-photon transitions
where one photon is absorbed from the first beam, and the second
photon is emitted into the second beam. The rate is proportional
to the response function. Early measurements focused on the tail
of the structure factor, which is a measure of the short range 
structure of the many-body wave function. In the unitary Fermi gas, 
the short range correlations can be characterized in terms of a 
quantity known as the contact density \cite{Tan_2008}.

 First generation measurements suffered from the fact that the average
density of the cloud was not constant, so that Bragg spectroscopy 
measures an average of the structure factor over the density profile 
of the atomic gas. More recently experimentalists have succeeded in
generating confining box potentials in which the equilibrium density
is approximately constant and the response of a homogeneous gas can 
be studied. Recent experiments have also investigated the dynamic
structure factor  in the hydrodynamic regime, where we expect the
response to be dominated by the Rayleigh (diffusive) and Brillouin
(sound) peaks. For this purpose a homogeneous gas is perturbed by a 
time and space-dependent external potential, and the density response 
is measured by taking images of the cloud. Experiments by groups at 
North Carolina State University and MIT \cite{baird:2019,patel:2019} 
have demonstrated that this method can be used to extract the sound
attenuation constant of the gas from the width of the Brillouin 
peak in the response function.

 Static fluctuations of atomic gases can be measured more directly, 
by studying intensity fluctuations in absorption images of the cloud. 
This method was explored by a group at MIT \cite{Sanner:2010}, which
demonstrated that Poissonian fluctuations in the density are 
suppressed in the quantum degenerate regime. 

\subsection{Fluctuations and transport phenomena in critical
systems}

 The phase diagram of the dilute Fermi gas has a second order 
superfluid phase transition in the whole BCS/BEC regime. In the 
BCS limit this phase transition occurs at an exponentially small 
temperature, and is difficult to observe. In the strong coupling 
regime the critical temperature is of the same order as the 
degeneracy temperature $T_F$, and the phase transition has been 
studied in some detail. In particular, experiments have observed 
the predicted critical behavior in the specific heat \cite{Ku_2012}.

 It would be interesting to extend the fluctuation measurements
discussed in the previous section to the critical regime. In a 
harmonically trapped gas the temperature is constant but the density
varies as a function of position, so that only a small part of the 
cloud is critical. This means that we do not expect to see a strong
enhancement of fluctuation probes. However, as discussed above, 
recent experiments have employed box potentials. In connection 
with the heavy-ion program it would be particularly interesting
to see if non-Gaussian density fluctuations are observable.

 Note that the order parameter for the superfluid transition is not the 
density but the phase of the condensate. Gradients of the phase 
correspond to the superfluid velocity, which is difficult to 
measure. However, the density is a conserved quantity and the 
coupling to the critical equation of state is determined by 
thermodynamic relations. In liquid helium density fluctuations 
are small compared to temperature fluctuations \cite{onuki_2002}.
This issue has not been carefully studied in cold gases, but the
measured compressibility does show an enhancement near the 
critical temperature \cite{Ku_2012}. 

 In cold gases it is also possible to measure the momentum
distribution, which is defined as the spatial Fourier transform of the 
density matrix 
\begin{equation}
n_k(\mathbf{x},t) = \int d^3\mathbf{y}\,
e^{i\mathbf{k}\cdot\mathbf{y}}
\langle\psi^\dagger(\mathbf{x}+\mathbf{y}/2,t)
    \psi(\mathbf{x}-\mathbf{y}/2,t)\rangle\,  .
\end{equation}
This quantity can be determined using radio-frequency (RF) spectroscopy 
or time-of-flight analysis \cite{Regal_2005}. In RF spectroscopy an
external RF source drives a transition from one of the trapped spin
states to a non-interacting state (a state that can be thought of
as a third spin component). The total absorption rate is proportional 
to the off-diagonal density matrix of the interacting state. In 
time-of-flight analysis the gas is rapidly swept to a non-interacting 
gas, in which the momentum distribution can be measured by simple 
expansion experiments. These methods have been used to study the large
momentum tail of the momentum distribution, but they have not been used 
to study critical fluctuations  with $|\mathbf{k}|\sim\xi^{-1}$, where 
$\xi$ is the correlation length.

 The dynamical theory of critical behavior predicts that transport
coefficients exhibit critical scaling in the vicinity of a 
phase transition \cite{Hohenberg:1977ym}. There are some differences
between the quark gluon plasma and ultra-cold gases. The critical 
endpoint in the QCD phase diagram is expected to be governed by 
model H in the classification of Hohenberg and Halperin 
\cite{Son:2004iv}, whereas the superfluid transition in ultra-cold
gases is expected to be in the same universality class as the 
$\lambda$-transition in liquid helium (model F). However, both 
models predict a very weak singularity in the shear viscosity, and
stronger effects in the thermal conductivity. QCD may also exhibit
a strong divergence in the bulk viscosity \cite{Martinez:2019bsn},
whereas the unitary Fermi gas is scale invariant, and the bulk 
viscosity vanishes. 

  Critical behavior in the sound attenuation constant has been 
observed in liquid helium \cite{onuki_2002}, but recent measurements in 
the unitary Fermi gas performed by the MIT group do not show any 
non-analytic behavior \cite{patel:2019}. It would be interesting 
to understand why this is the case. The experiment was performed 
in a box potential, with $N\sim O(10^5)$ atoms. If this number is too
small to observe critical transport, then this observation would 
clearly hold important lessons for heavy-ion collisions. 

 The dynamical theory of critical phenomena also predicts
enhanced long-time tails and critical slowing down. These 
phenomena should be visible in the long-time response of the 
density to an applied external potential, similar to what was
done in the experiments of the North Carolina State University group 
\cite{baird:2019}. However, so far no dedicated experiment 
of this type has been performed. This is an interesting problem 
beyond its relevance to the heavy-ion program, because aside
from checks of the scaling behavior of the attenuation
rate \cite{Hohenberg:1977ym} there are no direct measurements
of long-time tails in the literature. 

\subsection{Dynamical evolution in critical systems}

  Ultimately, one may envision using cold atomic gases as a 
testbed for dynamical theories of the time evolution of a near-critical
system. As mentioned above, experiments have studied the time evolution
of elliptic flow in a unitary gas released from a deformed harmonic
trap \cite{OHara:2002pqs}. These experiments can be analyzed using 
Navier-Stokes hydrodynamics, although care has to be taken in order 
to take into account effects of the dilute corona, which does 
not behave fluid dynamically \cite{Bluhm:2017rnf}. 

  Existing experiments cover the critical regime, but if the gas
is released from a harmonic trap then the fraction of the gas that
is critical at any point in time is always small. One might try 
to address this issue by releasing the cloud from a trap with a 
flat bottom, or by seeding fluctuations in the initial state. 
This would also correspond to an initial state that more 
closely resembles the Glauber initial conditions in a heavy-ion 
collision. 

 The superfluid transition is a second order phase transition
in the entire BCS/BEC crossover regime. First order transitions
appear in spin imbalanced gases or Bose/Fermi mixtures. A typical
example is a spin imbalanced cloud in the unitary limit. At 
sufficiently low temperature the center of the cloud is a fully 
paired (spin balanced) superfluid state, separated by a first 
order transition from a polarized corona in the normal fluid state. 
There are some studies of collective oscillations in a spin
imbalanced cloud \cite{Nascimbene_2009}, but the expansion after
release from a harmonic trap has not been studied. The expectation
is that the cloud would remain fluid dynamical, and the first order 
discontinuity expands with the gas. In order to study spinodal 
decomposition one might consider quenching the gas, for example 
by sweeping the scattering length across the first order transition. 

\subsection{Other physical systems}
 
  A classic nuclear system in which fluctuations have been investigated
is the nuclear liquid-gas phase transition. Multi-fragmentation experiments
indicate that the endpoint of the liquid-gas phase transition occurs at
a temperature $T_c=17.9\pm 0.4\,{\rm MeV}$ and a baryon density $n_c=0.06
\pm 0.01\, {\rm fm}^{-3}$ \cite{Elliott:2013}. These numbers come from
model fits to the fragment distribution for different system sizes and
energies that explore the spinodal region. Model analyses suggest 
significant equilibrium fluctuations of the nucleon number close to 
the endpoint~\cite{Vovchenko:2015pya}, which may  have an influence on 
the higher-order cumulants of the net-proton number distribution in relativistic 
heavy-ion collisions~\cite{Vovchenko:2017ayq}. It remains somewhat unclear 
what the correct dynamical theory of this transition is, and whether 
dynamical fluctuations of the type discussed in this report play an 
important role. It would appear that the criteria for the applicability 
of stochastic fluid dynamics or related theories are not met, because the 
mean free path is too long and the system size and life time too small.
However, researchers have investigated kinetic theories that include mean
field potentials and stochastic forces \cite{Danielewicz:2019mvp}.

 Neutron star mergers explore even higher temperatures in the baryon 
rich regime. Temperatures as high as $T\sim 100\,\text{MeV}$ and chemical
potentials of $\mu_Q \sim 400\,\text{MeV}$ might be
reached~\cite{Hanauske:2019qgs,Perego:2019adq} and could
potentially provide an astrophysical test for the existence of a 
first-order phase transition in the QCD phase diagram
\cite{Most:2018eaw,Bauswein:2018bma} and, thus, an indirect proof of a critical 
point. The first order region may also be accessible in future heavy-ion
collision experiments such as HADES and CBM at FAIR or the STAR at RHIC BES fixed target plans.
For a discussion of the chemical freeze-out conditions in 
the low-energy region see e.g.~\cite{Floerchinger:2012xd}. 

\section{Summary and outlook}

In this report we summarize the presentations and discussions at the EMMI Rapid
Reaction Task Force "Dynamics of critical fluctuations: Theory -- phenomenology --
heavy-ion collisions" held at GSI, Darmstadt, Germany in April 2019. Both 
theoretically and experimentally this field is actively developing and many 
discussions still have exploratory character. This is reflected in the 
diversity of the approaches and models that are presented in this report.

Efforts to study the QCD phase diagram using fluctuation observables have  
to go hand in hand with developing a fully dynamical treatment of the fluctuations, both critical and non-critical. Non-critical fluctuations provide a crucial baseline, and without understanding this part of the dynamics we cannot reliably address critical behavior. Only after a reliable framework for treating the dynamics of fluctuations has been developed can we hope to constrain critical behavior in the thermodynamics of QCD based on experimental observables.

On the theory side, two main avenues emerged in the discussions, both relying on 
the success of fluid dynamical simulations of heavy-ion collisions: Stochastic fluid dynamics and hydro-kinetics. The first propagates fluid dynamical fluctuations
explicitly in an event-by-event setup, while the second propagates correlation
functions, which are already averages over thermal fluctuations, coupled to fluid
dynamics. 

Understanding the dynamics of thermal fluctuations is important even if one
is not specifically interested in the vicinity of a critical point. Indeed, 
fluctuation-dissipation relations imply that a consistent treatment of 
dissipative fluid dynamics always has to include fluctuations. The critical 
point can then be included in the framework via an appropriate equation of state.

During the discussions we identified the following three aspects which deserve 
major theoretical attention:
\begin{itemize}
\item Are the relevant scale relations, which separate the thermal noise,
non-equilibrium fluctuations and the fluid dynamical evolution, fulfilled in a
heavy-ion collision? This question is especially important when it comes to technical
length scales, like the numerical regulator $l_{\rm filter}$ or the patch size for
particlization $l_{\rm patch}$.

\item How can stochastic fluid dynamics be properly regulated such that the averages
and the fluctuation observables are independent of the length scale $l_{\rm filter}$
but the essential critical physics is preserved?
    
\item How do we interface stochastic fluid dynamics with hadronic afterburners, 
in particular, how can we particlize fluctuations of conserved charges? Are 
there schemes that can be applied to both stochastic fluid dynamics as well as
hydro-kinetics? What is the shortest length-scale that can be resolved in such a
procedure?

\end{itemize}

The key experimental challenges which were discussed are:
\begin{itemize}
\item What are the underlying physical phenomena that affect the dependence on 
the rapidity window? Experiments only resolve the rapidity structure of 
fluctuations at freezeout. How do correlations in rapidity evolve over 
the course of the collision?

\item What is the influence of hadronic resonance decays on the final observable? 
Experiments cannot directly measure fluctuations of certain observables, such 
as net-baryon number, and standard proxies, like net-proton number are 
affected by resonance decays and hadronic rescatterings. 

\item What are the observables that are the most sensitive to criticality? What are
their advantages and disadvantages in terms of experimental feasibility and of
theoretical accessibility from first-priniciple calculations and dynamical models?

\end{itemize}

Intersections with other physical systems, notably ultra-cold atomic gases, provide
opportunities for fluctuation studies. Here, ideas could be tested in more controlled
settings or motivate new experiments in atomic and condensed matter physics, which
could help validate dynamical theories describing the dynamics of fluctuations in
heavy-ion collisions.

\section*{Acknowledgments}

We gratefully acknowledge the support and organisation provided by the ExtreMe
Matter Institute EMMI at GSI, Darmstadt, which made the Rapid Reaction Task 
Force "Dynamics of critical fluctuations: Theory -- phenomenology -- heavy-ion collisions" possible. 

We thank K.~Rajagopal for his useful comments on the report.

This work has been supported by the Region Pays de la Loire, France, under an
"Etoiles montantes" grant (MB, MN, NT), 
by the Deutsche Forschungsgemeinschaft (DFG)  Collaborative  Research  Centre
”SFB1225 (ISOQUANT)" (SF, AM, JMP, NW), 
by CAPES (Finance Code 001), CNPq, FAPERJ, and INCT-FNA, 
Process No. 464898/2014-5 (ESF),
by the Director, Office of Science, Office of High Energy and Nuclear Physics,
Division of Nuclear Physics, and by the Office of Basic Energy Sciences, 
Division of Nuclear Sciences, of the U.S. Department of Energy under Contract 
No.~DE-AC03-76SF00098 (VK), under Contract No.~DE-FG02-03ER41260 (TS), and 
under Contract No.~DE-FG02-01ER41195 (MS), 
by the CLASH project under the grant number KAW 2017-0036 (CP), 
by the National Science Centre Poland under grant number 2018/30/A/ST2/00226 (MG),
by a U.S. National Science Foundation Graduate Research Fellowship (GWR), 
by the SUT-CHE-NRU project of Thailand (CH), and 
by the project Centre of Advanced Applied Sciences with number 
CZ.02.1.01/0.0/0.0/16-019/0000778 (IK) which 
is co-financed by the European Union. 
\appendix
\section{Approaches to the theoretical description of the dynamics of fluctuations}
\label{sec:app}

\subsection{Effective kinetic theory of hydrodynamic fluctations\label{sec:hydro-kinetics}}

The Quark-Gluon Plasma (QGP) created in heavy-ion collisions can be well modeled by relativistic viscous hydrodynamics\footnote{Based on the proceeding for  the talk given  at Critical
Point and Onset of Deconfinement 2017 \cite{Mazeliauskas:2017wyz}.}. As required by the fluctuation-dissipation theorem such models should consistently include
thermal fluctuations. In equilibrium, the resulting two-point correlation functions of hydrodynamic fields, e.g. momentum  or energy density, obtain the well known equilibrium values.
In evolving systems, the expansion can drive these two-point correlation functions away from the equilibrium values and the out-of-equilibrium evolution of noise correlators must be calculated. Although the present model does not include the effects of the QCD critical point directly, the relaxation dynamics of near-equilibrium fluctuations tells us how fast the fluctuations would respond to the presence of criticality.

\begin{figure}
\centering
\includegraphics[width=0.7\linewidth]{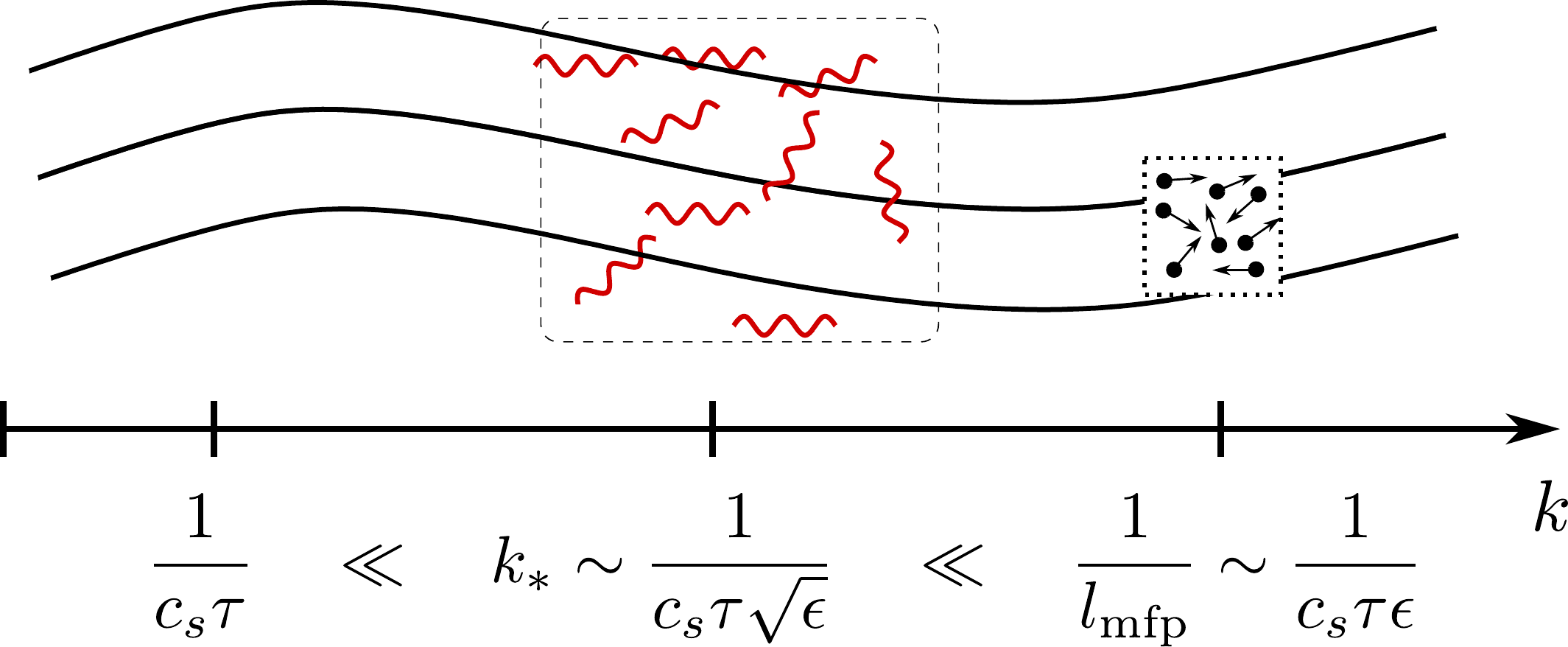}
\caption{The separation of long-wavelength hydrodynamic modes $1/(c_s\tau)$, the dominant out-of-equilibrium hydrodynamic modes $k_{*}\sim 1/(c_s \tau\sqrt{\epsilon })$ and the microscopic modes $1/l_\text{mfp}\sim1/(c_s \tau \epsilon)$, where $l_\text{mfp}$ is the mean free path in the system, $1/\tau$ the Bjorken expansion rate and $\epsilon = l_\text{mfp}/(c_s \tau)\ll 1$ is the hydrodynamic gradient expansion parameter. Figure taken from~\cite{Mazeliauskas:2017wyz}.}
\label{fig:scales2}
\end{figure}

In this model we consider the effective kinetic description of hydrodynamic fluctuations~\cite{Akamatsu:2016llw,Akamatsu:2017rdu,Mazeliauskas:2017wyz}, which is based on the separation of scales (see \Fig{fig:scales2}) of small wave-number hydrodynamic modes (which are never in thermal equilibrium and are determined by initial conditions), and large wave-numbers $k> k_*={1}/{c_s\tau \sqrt{\epsilon}}$, which are damped and excited at the comparable rate to the background expansion $\partial_\mu u^\mu = 1/\tau$.
Here $\epsilon$ stands for the hydrodynamic expansion parameter $\epsilon=l_\text{mfp}/(c\tau)\ll1$. 
The equation of motion in relativistic hydrodynamics with noise is given by the conservation of the energy-momentum tensor
and stochastic constitutive equations.
For linear perturbations in energy and momentum
$
	\phi_a(\tau,\vec k)\equiv (c_s\delta e, \vec g)
$  around a homogenenous Bjorken expanding background one can derive the Langevin type equation~\cite{Akamatsu:2016llw,Akamatsu:2017rdu}
\begin{equation}
-\dot\phi_a(\tau,\vec k)
	=\ubr{i\mathcal{L}_{ab}\phi_b}{ideal}
	+\ubr{\mathcal{D}_{ab}\phi_b}{viscous} +\ubr{ \xi_a}{noise}
	+\ubr{\mathcal{P}_{ab}(\tau)\phi_b}{expansion}\label{single} \,.
\end{equation}
Fluctuations can be decomposed into two propagating sound modes and two transverse diffusive modes.
The two-point correlation functions $N_{AB}$ for these eigenmodes, defined as
$
\langle \phi_A(t,\vec k) \phi_B(t,-\vec k')\rangle \equiv N_{AB}(t,\vec k)(2\pi)^3\delta(\vec k - \vec k')
$,
then satisfy the relaxation type kinetic equations, e.g.\
\begin{align}
\partial_\tau N_{++}
&= \ubr{-\frac{4}{3}\gamma_{\eta}K^2 \left[N_{AA} - N_\text{eq}\right]}{equilibration}
-\ubr{\frac{1}{\tau}\left[2+c_{s}^2+\cos^2\theta_K \right]N_{++}}{expansion} \,,
\end{align}
where $\gamma_\eta=\eta/(e_0+p_0)$ and  
$\vec K \equiv (k_x,k_y,k_{\eta}/\tau)$ .
The two-point correlations are relaxing to the instantaneous thermal equilibrium value with the rate $\sim \gamma_\eta K^2$ and driven away from the equilibrium by the expansion term $\sim \frac{1}{\tau}$. 

In the presence of hydrodynamic noise, the effective long wavelength energy momentum tensor is modified by the contributions coming from the two-point correlators of out-of-equilibrium noise at scale $k\sim k_*$. Specifically, the energy and longitudinal pressure are increased by the non-linear contributions of  momentum fluctuations  $\vec{g} = (T^{\tau x}, T^{\tau y}, \tau T^{\tau \eta})$
\begin{align}
    \label{eave}
    \langle T^{\tau\tau} \rangle = e  +  \frac{\langle \vec{g}^2 \rangle}{e +p} , \quad
     \langle T^{zz} \rangle = c_s^2 e - \frac{4\eta}{3\tau}  + 
     \frac{\langle (
g^{ z})^2 \rangle}{e +p}.
\end{align}
The non-linear noise expectation can be written as an integral over the phase space of hydrodynamic modes, which is divergent 
due to the equilibrium expectation value of $N_{AB}$ and the leading large $K^2$ expansion term~\cite{Akamatsu:2016llw}. Regulating the integral by a UV cutoff $\Lambda$, the universal divergent terms can be computed explicitly and agree with previous computations using diagrammatic approaches~\cite{Kovtun:2011np}. The divergent contributions reflect the fact that the initial bare pressure and viscosity are also cut-off dependent, but their sum is independent of $\Lambda$.
After absorbing divergent terms in the physical pressure and viscosity,
the remaining finite term for longitudinal pressure is
\begin{equation}
\frac{\langle T^{zz}(\tau) \rangle}{e+p}
= \frac{p}{e + p} - \frac{4 \gamma_\eta}{3\tau} 
  + \frac{1.08318}{s\, (4\pi \gamma_{\eta}\tau)^{3/2} }.
\end{equation}
The finite correction (also known as long time tail) comes with the characteristic fractional power, which can be understood from the simple estimate of the phase space of modes around the critical scale $k_*$ and the
 equipartition of energy:
$
\langle T^{zz} \rangle_\text{fluct.} \sim T k_*^3 \sim T\left(\frac{1}{\gamma_{\eta}\tau}\right)^{3/2}
$

In the presence of noise the evolution of the average energy density of the system obeys
\begin{equation}
\frac{d \langle\langle T^{\tau\tau}\rangle \rangle}{d\tau} = -\frac{ \langle\langle T^{\tau\tau}\rangle\rangle  +  \langle\langle \tau^2T^{\eta\eta} \rangle\rangle}{\tau},\ \ \
\end{equation}
where the double brackets notate an average over (long range in rapidity)
initial conditions and thermal noise. To close the system of equations, the relationship between average energy density $\langle\langle T^{\tau\tau} \rangle\rangle$ and the average rest frame energy  density $e(\tau)$ must be specified~\cite{Akamatsu:2016llw}. This relation receives contributions from the two-point correlation functions of noise and therefore the simultaneous solution of the background hydrodynamic and the hydro-kinetic equations is required for the effective description of hydrodynamics with noise.

The presented framework of hydro-kinetic equations is a general and extendable way of calculating the physics of out-of-equilibrium noise in expanding systems. We successfully reproduce the universal renormalizations of bare energy, pressure and shear viscosity $\eta$ in agreement with previous diagrammatic calculations for conformal systems. We also  calculate corrections to bulk viscosity $\zeta$ in non-conformal systems due to hydrodynamic fluctuations. 
The hydro-kinetic equations is an alternative way of studying hydrodynamics with noise and can be profitably applied to a versatile range of hydrodynamic systems (see recent work of \cite{An:2019osr,Martinez:2018wia,Akamatsu:2018vjr}). 

\subsection{Stochastic diffusion of critical net-baryon density fluctuations\label{app:stochasticdiff}}

The free energy functional near the QCD critical point studied in the numerics~\cite{Nahrgang:2018afz}, which discusses the dynamics of critical fluctuations in $n_B$ (see Section~\ref{sec:stodiff}), has the following polynomial form 
\begin{multline}
 {\cal F}[n_B] = T\int{\rm d}^3 x \Bigg(\frac{m^2}{2n_c^2}(\Delta n_B)^2 + \frac{K}{2n_c^2} 
 (\nabla n_B)^2 + \\
 \frac{\lambda_3}{3n_c^3}(\Delta n_B)^3 + \frac{\lambda_4}{4n_c^4}(\Delta n_B)^4 + \frac{\lambda_6}{6n_c^6}(\Delta n_B)^6 \Bigg) \,,
\label{eq:GLpotential}
\end{multline}
where $\Delta n_B=n_B-n_c$ denotes the difference of $n_B$ from the critical density $n_c$. In general, this Ginzburg-Landau form of ${\cal F}$ still needs to be supplemented by further regular contributions in line with the equation of state. 

The coefficients in Eq.~\eqref{eq:GLpotential}, i.e.~their scaling with the correlation length $\xi$, may be obtained by mapping the $3$-dimensional Ising model to a universal effective potential~\cite{Tsypin:1994nh,Tsypin:1997zz}. In this way criticality is included in the approach in line with the assumed underlying static universality class, which gives the following dependencies 
\begin{align}
 \label{eq:couplingsA}
 m^2 &= \frac{{\tilde m^2}}{\xi_0^3},\quad \tilde{m}=\frac{1}{\xi/\xi_0}\, ,\\
 K &= \tilde K/\xi_0\, ,\\
 \lambda_3 &= n_c\, \tilde\lambda_3\, (\xi/\xi_0)^{-3/2}\, ,\\
 \lambda_4 &= n_c\, \tilde\lambda_4\,(\xi/\xi_0)^{-1}\,, \\
 \lambda_6 &= n_c\, \tilde\lambda_6\,,
 \label{eq:couplings}
\end{align}
with $\xi_0$ the correlation length far away from the transition temperature $T_c$. 

The inclusion of a term proportional to $\lambda_6$ in Eq.~\eqref{eq:GLpotential} turns out to be important for understanding 
Monte Carlo simulation results of the probability distribution in the Ising model~\cite{Tsypin:1994nh,Tsypin:1997zz}. Moreover, the term proportional to $K$ represents a surface tension contribution to ${\cal F}$. Special limits of the free energy functional represent the Gauss model form~\cite{Nahrgang:2017hkh} with $K=\lambda_i=0$ and the Gauss$+$surface model form~\cite{Bluhm:2019yfb} with $\lambda_i=0$. 

In general, the dimensionless couplings $\tilde\lambda_i$ are universal. Their values in the QCD phase diagram can be determined by translating the Ising model variables to $T$ and $\mu_B$ in QCD and comparing the cumulants of the critical mode as explained in~\cite{Bluhm:2016trm}. This translation of variables is, however, non-universal and introduces uncertainties into the description. In the numerics~\cite{Nahrgang:2018afz}, constant values for the non-linear couplings $\tilde\lambda_i$ were used for simplicity, see section~\ref{sec:stodiff}. The free energy functional ${\cal F}$ still depends implicitly on the thermal variables via the correlation length $\xi$. This dependence is obtained from comparing the variance of the critical mode from the effective potential to the one obtained from the parametric representation~\cite{Guida:1996ep} of the scaling equation of state of the Ising model as explained in~\cite{Bluhm:2016byc}. In the numerics~\cite{Nahrgang:2018afz}, $\xi$ as a function of $T$ for a constant $\mu_B$ close to the critical point on the crossover side is considered.

\subsection{Nonequilibrium chiral fluid dynamics (N$\chi$FD)}
\label{app:nchifd}

\subsubsection{Quark-meson model}
\label{sec:qm}

The Lagrangian of the quark-meson model is the foundation of the dynamical nonequilibrium fluid dynamics model. It reads
\begin{align}
\label{eq:Lagrangian}
 {\cal L}&=\overline{q}\left(i \gamma^\mu \partial_\mu-g_{\rm q} \sigma\right)q + \frac{1}{2}\left(\partial_\mu\sigma\right)^2- U(\sigma)~, \\
 U(\sigma)&=\frac{\lambda^2}{4}\left(\sigma^2-f_{\pi}^2\right)^2-f_{\pi}m_{\pi}^2\sigma~,
\end{align}
with the light quark doublet $q=(u,d)$ and the chiral condensate $\sigma$ which dynamically generates the mass of the constituent quarks. 
We can fix the quark-meson coupling constant $g$ from the vacuum nucleon masses $m=\SI{940}{\MeV}$ to $g=3.37$. The additional parameters used here are the pion decay constant of $f_\pi=93$~MeV and the pion mass $m_\pi=138$~MeV. The term proportional to $\sigma$ accounts for the small explicit symmetry breaking due to the finite current quark masses. The self-coupling constant $\lambda$ is related to the sigma mass $m_\sigma=\SI{600}{\MeV}$ through $\lambda^2=\frac{m_\pi^2-m_\sigma^2}{2f_\pi^2}$. 

This model is well studied and we can immediately write down the mean-field effective thermodynamic potential as
\begin{align} 
\label{eq:thermpot}
\Omega(\sigma)&=\frac{T}{V}\Gamma[\sigma]=U(\sigma)-\Omega_{\rm q\bar q}(T,\mu; \sigma)~,\\
\Omega_{\rm q\bar q}(T,\mu; \sigma)&=
d_q T\int\frac{\mathrm d^3 p}{(2\pi)^3} \left\{\ln\left[1+\mathrm e^{-\frac{E_{\rm q}-\mu}{T}}\right]
+\ln\left[1+\mathrm e^{-\frac{E_{\rm q}+\mu}{T}}\right]\right\}~,
\end{align}
with the degeneracy factor $d_q=12$ and the quasiparticle energy $E_{\rm q}=\sqrt{p^2+m_{\rm q}^2}$. In this notation, the quark chemical potential $\mu=\mu_{\rm B}/3$ is used. 

\subsubsection{Nonequilibrium chiral fluid dynamics}
\label{sec:nxfd}

From the quark-meson Lagrangian, Eq.\ \eqref{eq:Lagrangian}, together with the thermodynamic potential, Eq.\ \eqref{eq:thermpot}, we are able to obtain the full nonequilibrium dynamics, where we explicitly propagate the chiral order parameter with a Langevin equation of motion, derived from the 2PI effective action,
\begin{equation}
\label{eq:eomsigma}
 \partial_\mu\partial^\mu\sigma+\eta\partial_t \sigma+\frac{\delta \Omega}{\delta\sigma}=\xi~.
\end{equation}
The damping coefficient $\eta$ arises from the $\sigma\leftrightarrow q\bar q$ reaction and has been evaluated as
\begin{equation}
\label{eq:dampingcoeff}
  \eta=\frac{12 g^2}{\pi}\left[1-2n_{\rm F}\left(\frac{m_\sigma}{2}\right)\right]\frac{1}{m_\sigma^2}\left(\frac{m_\sigma^2}{4}-m_{\rm q}^2\right)^{3/2}~.
\end{equation}
The stochastic noise term $\xi$ is assumed to be Gaussian and white, and its width is determined by the dissipation-fluctuation relation
\begin{equation}
\label{eq:dissfluctsigma}
 \langle\xi(t,\vec x)\xi(t',\vec x')\rangle=\delta(\vec x-\vec x')\delta(t-t')m_\sigma\eta\coth\left(\frac{m_\sigma}{2T}\right)~.
\end{equation}
To avoid unphysical dependences on the lattice spacing, we model a spatial correlation of the noise field over a correlation length of $1/m_\sigma$. Hereby, the mass of sigma can be determined as a function of temperature and chemical potential equal to the curvature of the thermodynamic potential in equilibrium, 
\begin{equation}
 \label{eq:corrl}
 m_\sigma^2=\frac{\partial^2 \Omega}{\partial\sigma^2}\bigg|_{\sigma=\langle\sigma\rangle}~.
\end{equation}

The locally equilibrated quark plasma acts as a heat bath in which the field $\sigma$ evolves. The local pressure of this heat bath is given by
\begin{equation}
\label{eq:pressure}
 p(T,\mu; \sigma) = -\Omega_{\rm q\bar q}(T,\mu; \sigma)~,
\end{equation}
allowing us to calculate the local net-baryon and energy densities in the standard fashion as
\begin{equation}
 n=\frac{\partial p}{\partial\mu}~,~~e=T\frac{\partial p}{\partial T}-p+\mu n~.
\end{equation}

As the total energy and momentum of the coupled system of fluid and field are conserved, we obtain the following expressions for the divergences of the ideal energy-momentum tensor of the fluid $T^{\mu\nu}=(e+p)u^{\mu}u^{\nu}-pg^{\mu\nu}$ and the net-baryon current $N^{\mu}=n u^{\mu}$ with the local four-velocity $u^{\mu}$, 
\begin{align}
\label{eq:fluidT}
\partial_\mu T^{\mu\nu}&=-\partial_\mu T_\sigma^{\mu\nu}~,\\
\label{eq:fluidN}
\partial_\mu N^{\mu}&=0~.
\end{align}
It is worth pointing out that the stochastic nature of the source term on the right hand side of Eq.\ \eqref{eq:fluidT} constitutes a stochastic evolution for the fluid dynamical medium. 

\subsubsection{Expanding medium}

Modeling an expanding medium can be achieved by defining an initial profile $T(\vec x)$ and $\mu(\vec x)$ which determines the field in equilibrium and the hydrodynamic quantities, i.e. $\sigma$, $e$, $n$, $p$ at $t=0$ at each point in space. Hereby, spherical or ellipsoidal shapes with a smoothed edge are possible but also more realistic initial conditions which can be obtained e.g. from the UrQMD transport model. The expansion and cooling is then described selfconsistently by solving the coupled equations \eqref{eq:eomsigma}, \eqref{eq:fluidT}, \eqref{eq:fluidN}.

If one is not interested in spatial fluctuations, then a more simple approach can be used, reducing the dynamics to a Bjorken-type expansion along the beam direction. Contracting Eq.~\eqref{eq:fluidT} with the four-velocity $u^\nu$ gives the equation for the evolution of the energy density, 
\begin{equation}
\label{eq:eom_eden}
 \dot e=-\frac{e+p}{\tau}+\left[\frac{\delta\Omega_{q \bar q}}{\delta\sigma}+\left(\frac{D}{\tau}+\eta\right)\dot\sigma\right]\dot\sigma~,
\end{equation}
with the constant $D=1$ in the hubble term. The net-baryon density then follows the equation
\begin{equation}
 \label{eq:eom_nden}
 \dot n = -\frac{n}{\tau}~.
\end{equation}

As a benchmark test, the equilibration in a box to specific values of $T$ and $\mu$ has been tested and shown to reproduce the proper behavior of the cumulants in $\sigma$ in comparison to the corresponding susceptibilities that are obtained from functional derivatives. 

Further observables that have been studied in the past are: 
\begin{itemize}
 \item Trajectories in the phase diagram
 \item Density fluctuations in single events, azimuthal distributions
 \item Net-proton fluctuations during a crossover at small $\mu$ (after a Cooper-Frye particlization)
 \item Cumulants of the sigma field during a crossover at small $\mu$ 
 \item Production of entropy as a function of the initial condition
\end{itemize}

\subsection{QCD assisted transport}\label{app:QCDassisted}

In order to understand the connection between experimental results obtained in heavy-ion collisions and the underlying phase structure of QCD, we require an approach that connects both at a fundamental level. This is a necessary prerequisite to establish the existence of a critical endpoint (CEP) in the phase diagram of QCD. Approaches working towards this direction have initially been put forward in \cite{Nahrgang:2018afz,Nahrgang:2011mg,Nahrgang:2011mv,Herold:2016uvv,Bluhm:2018qkf}.

Within the approach outlined in this section, as put forward in \cite{Bluhm:2018qkf}, we require an accurate description of the equilibrium phase structure of QCD. Since correlation functions over the entire phase diagram are not obtainable from first principles yet, we resort to their calculation in low energy effective theories. In particular the 2+1 flavour Quark-Meson (QM) model provides a quantitatively reliable description at small chemical potentials. Additionally, it features a CEP that is believed to be in the same universality class as the potential CEP of QCD. The treatment of this effective theory within the framework of the Functional Renormalization Group (FRG) allows for a systematic embedding within QCD, cf.~the discussion in \cite{Alkofer:2018guy}. Moreover, we are able to obtain correlation functions not only in Euclidean space-time, but also in Minkowski space-time.

Based on the equilibrium linear response functions of the low energy effective description, we use the associated transport equation to calculate the cumulants of the critical mode. 
Therefore this section is split into two parts, the first part describing the calculation of the required equilibrium correlation functions, and the second one focusing on the time evolution of the critical mode around a given set of equilibrium correlation functions.

The FRG is utilized to calculate all required equilibrium correlation functions for the subsequent transport evolution. Being a versatile, first-principle tool, the FRG has been applied successfully to QCD, see e.g.~\cite{Cyrol:2017ewj}, and low-energy effective versions thereof, see e.g.~\cite{Herbst:2010rf,Rennecke:2016tkm}. Its advantage in the present context is that it allows for the computation of the phase structure, i.e. the effective potential, and momentum-dependent correlation functions in a unified framework. The equilibrium part of our work, i.e.~the equation of state and the equilibrium correlation functions, are based on a 2+1 flavour study of a low-energy effective description of QCD, where the dynamics of constituent quarks as well as the lowest scalar and pseudoscalar meson nonets, including their wave function renormalizations, are taken into account~\cite{Rennecke:2016tkm}. It
captures, by design, the relevant physical effects at small chemical potential $\mu$ and temperatures \mbox{$T \lesssim T_c$}. Additionally, it features a critical endpoint which is in the same static universality class as the one potentially present in QCD. Therefore this model provides a well-suited base for studying how dynamical non-equilibrium effects manifest themselves in observables.

\begin{figure}[t]
	\centering
	\includegraphics[width=0.5\textwidth]{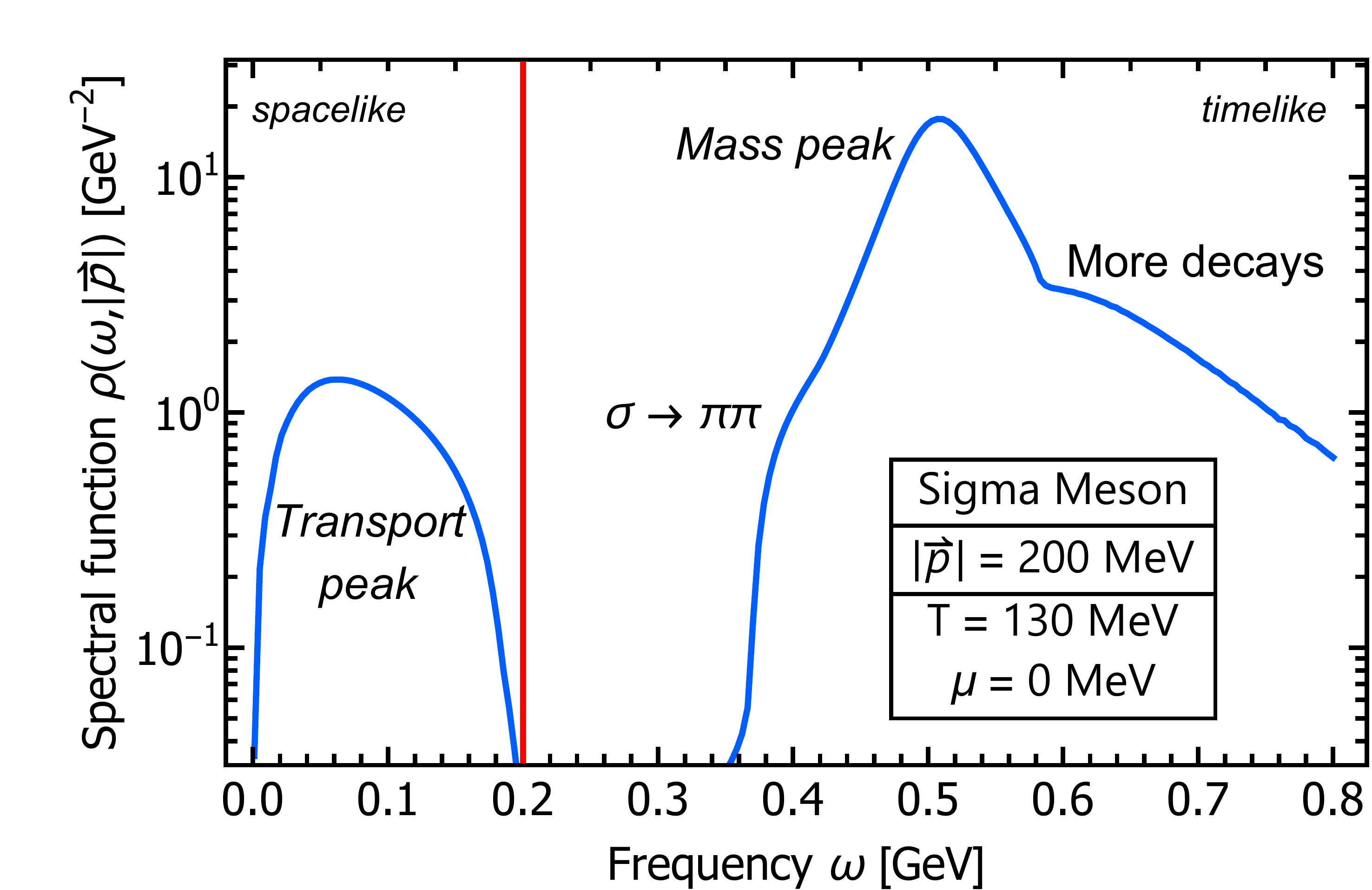}
	\caption{ Spectral function of the sigma meson at $T=130$~MeV,
		$\mu=0$~MeV in the phase diagram. The transport peak and the
		mass peak are associated with the diffusion in the transport
		equation. A detailed discussion of the seen structures can
		be found in e.g.~\cite{Pawlowski:2017gxj}. 
		Figure taken from~\cite{Bluhm:2018qkf}.}
	\label{fig:sigma_spectral}
\end{figure}

In general, spectral functions can be obtained either via analytically continuing numerical data, see e.g.~\cite{Cyrol:2018xeq}, or via a direct computation from analytically continued flow equations, see e.g.~\cite{Floerchinger:2011sc, Kamikado:2013sia}. If possible, the latter is preferred and also the option utilized in this work. The spectral functions of the sigma meson are calculated similarly to \cite{Tripolt:2013jra, Pawlowski:2017gxj} with suitable modifications in order to take the non-trivial wave-function renormalizations into account. As a result we have access to the two-point correlator $\Gamma_{\sigma\sigma}^{(2)}(\omega,|\vec{p}|)$, depending on an external frequency $\omega$ and an external momentum $\vec{p}$, as well as momentum-independent vertices $\Gamma_{\sigma^n}^{(n)}$ which are extracted from the full effective potential computed in~\cite{Rennecke:2016tkm}. An exemplary spectral function is shown in~\autoref{fig:sigma_spectral}. The two main features that influence the behaviour of the dynamical evolution are the transport peak and the mass peak. The transport peak, if present at small frequencies $\omega<|\vec{p}|$, dominates the long range behaviour of the sigma field. The mass peak, instead, becomes the driving force for the evolution dynamics when the transport peak is absent, e.g.\ in the vacuum.

We are now in the position to study the time-evolution of the critical mode and its event-by-event fluctuations. For this purpose, we solve the Langevin-type transport equation
\begin{align}
\label{equ:Langevin}
	\frac{\mathrm{d}\Gamma}{\mathrm{d}\sigma} = \xi
\,.
\end{align}
Above, the equation of motion contains a kinetic term related to the real part of $\Gamma^{\, (2)}_{\sigma\sigma}$, a diffusion term sensitive to the imaginary part of $\Gamma^{\, (2)}_{\sigma\sigma}$, and the effective potential mentioned above, while $\xi$ represents the noise field chosen such that the fluctuation-dissipation balance is guaranteed.

For the numerical results presented in \autoref{fig:equilibration} in Section \ref{sec:chiral} we consider the critical mode to be spatially isotropic, i.e.~$\sigma(\vec{x},t)=\sigma(r,t)$, where we split $\sigma=\sigma_0+\delta\sigma$. We study the time-evolution of the critical fluctuations for a system subject to a sudden quench from
high temperatures to a specific point in the QCD phase diagram. Accordingly, the system is initialized such that $\sigma(r,t=0)=0$ and $\partial_t\sigma(r,t=0)=0$ which implies that the initial fluctuations $\delta\sigma(t=0)$ are of the magnitude of the equilibrium value $\sigma_0$ after the quench. Moreover, we consider spatially constant Gaussian white noise, with zero mean and a variance given as~\cite{Nahrgang:2011mg}
\begin{align}
\label{equ:noisevariance}
	\langle\xi(t)\xi(t')\rangle=\frac1V\delta(t-t')m_\sigma\eta\coth\left(\frac{m_\sigma}{2T}\right)
\, ,
\end{align}
where the diffusion coefficient $\eta$ is extracted from the imaginary part of $\Gamma^{\, (2)}_{\sigma\sigma}$.

\subsection{Critical dynamics from small, noisy, fluctuating systems}

\subsubsection{Including spurious effects near criticality}

We describe long-range fluctuations of the order parameter $\sigma$ by a probability distribution
\begin{equation}
 \mathcal{P}[\sigma] \sim e^{-\Omega[\sigma]/T} \approx \displaystyle  e^{ \int d^3 x \; \left[ \frac{1}{2}(\nabla \sigma)^2 + \frac{1}{2}m^2_\sigma \sigma^2 + \frac{1}{3}\lambda_3 \sigma^3 + \frac{1}{4}\lambda_4 \sigma^4 + \,\dots \right]/T}
\,,
\label{eqPsigma2}
\end{equation}
assuming fluctuations of small amplitude, so that we can use a Gaussian approximation by considering only the mass term, where $m_\sigma \sim \xi^{-1}$. 
We also assume fluctuations to be homogeneous and use $\sigma_0 = \int d^3x \, \sigma(x)/V$, and couple them to observable particles via mass corrections, i.e.
\begin{equation}
 \mathcal{L}_{int} = -G\,\sigma_0 \, \vec \pi \cdot \vec \pi - g \,\sigma_0\, \bar\psi_p\, \psi_p\,,
\end{equation}
where we illustrate the couplings to pions and protons \cite{Stephanov:2008qz,Stephanov:1999zu}. The pion-sigma coupling can be roughly estimated to be 
around $G\sim 300$ MeV \cite{Stephanov:1999zu}.

Fluctuations of the order parameter are then coupled to observable particles and will have an impact, for instance, in fluctuations of 
particle multiplicities. The effects of these fluctuations can be calculated by looking at the modification of the single-particle energy levels, due 
to fluctuations of the order parameter, i.e.
\begin{eqnarray}
  \omega  &=&  \sqrt{p^2 + m_0^2 + \delta m^2 } 
       \approx   \omega_0\left[1+\dfrac{1}{2} \dfrac{\delta m^2}{\omega_0^2} - \dfrac{1}{8} \dfrac{(\delta m^2)^2}{\omega_0^4} + \cdots \right] 
\,,
\label{dwds}
\end{eqnarray}
where we have used a Taylor expansion over the mass corrections $\delta m$ from fluctuations of the order parameter. 
Expanding quantities in powers of the shift in the single-particle energies $\delta \omega_{\vec p}$ and taking averages over the fluctuations of $\sigma_0$, denoted by $\overline{(\cdots)}$, it is possible to calculate critical contributions to averages and correlations. For instance,
\begin{equation}
 \overline{\langle Q \rangle} = \langle Q \rangle_0 \displaystyle +  \sum_{\vec p} \dfrac{\partial\;}{\partial \omega_{\vec p}}\langle \Delta Q \rangle_0\;\overline{\delta \omega_{\vec p}}  + 
\dfrac{1}{2} \,  \sum_{\vec p,\vec p^\prime} \dfrac{\partial\;}{\partial \omega_{\vec p}}  \dfrac{\partial\;}{\partial \omega_{\vec p^\prime}} \langle Q_1\rangle_0 \;\overline{\delta \omega_{\vec p} \, \delta \omega_{\vec p^\prime}}\,,
\label{eq-crit-Q}
\end{equation}
where $\langle \cdots \rangle_0$ denotes the usual equilibrium averages in a grand-canonical ensemble and $Q$ is a generic quantity \cite{Hippert:2017xoj}. 

Near criticality, the equilibration timescale of the system also diverges with some power of $\xi$ due to critical slowing-down, which limits the growth of $\xi$ and, hence, of possible signatures which scale with $\xi$ to some power. It is implemented in the ansatz equation \cite{Hippert:2015rwa,Berdnikov:1999ph}
\begin{equation}
 \dfrac{{d} \xi}{{  d} t} = A\; \left(\dfrac{\xi}{\xi_0} \right)^{2-z}\,\left(\dfrac{\xi_0}{\xi} - \dfrac{\xi_0}{\xi_{eq}(t)}\right)\, ,
\label{modBerdnikov1}
\end{equation}
where $\xi_{eq}(t) = \xi_0 \;|{t}/{\tau} |^{-\nu/\beta \delta}$, $\xi_0 \sim 1.6$ fm fixes the initial correlation length at proper time $t=-\tau$ and $\tau$ is the 
typical cooling time before reaching the neighborhood of the critical point. The critical exponents are given by $\alpha = 0.11$, $\nu=0.63$, $z=2 + \alpha/\nu$, $\beta=0.326$, $\delta=4.80$, coming from universality class arguments \cite{Hohenberg:1977ym,Guida:1996ep}. 
The parameter $A$ in Eq. (\ref{modBerdnikov1}) can be constrained by imposing causality (i.e. $d\xi/dt \leq 1$), constraining $\xi/\xi_0$ to below 
$1.3$ for $\tau = 1$ fm and below $1.9$ for $\tau = 5.5$ fm and significantly restraining signatures of criticality \cite{Hippert:2017xoj}. 

The statistics to be measured in collision experiments are contaminated by spurious fluctuations, modified by acceptance and efficiency limitations and are not calculated over direct particles only. These effects can be introduced into our calculations in a simple fashion. Effects such as the dynamical expansion of the system are, for now, neglected. 

Effects from a limited acceptance window can be implemented in the calculation of multiplicity fluctuations by 
considering an acceptance probability factor $F(p)$, such that each produced particle of momentum $p$ (in modulus) 
has a probability  $F(p)$ of being detected \cite{Hippert:2017xoj}.  
For instance, if $n_p$ is the number of particles with momentum $p$, these kinematic cuts modify 
$\langle (\Delta n_p)^2 \rangle$ according to $ \langle (\Delta n_p)^2 \rangle_{acc} = F(p)^2\,\langle (\Delta n_p)^2 \rangle + F(p) \big(1-F(p)\big)\, \langle n_p \rangle$.

Resonance decays can be introduced in a similar fashion. For a decay into two particles, we consider the probabilities that one ($P_1$), both ($P_2$) or neither ($P_0$) of the particles produced in a single decay are found in the acceptance window. 
Results can be shown as a function of the momentum $p$ of the resonance and are calculated by using the phase-space volume as a measure of probability. A branching ratio of less than $100\%$ can be implemented by simply rescaling $P_1$, $P_2$ and $P_0$. 

Finally, spurious fluctuations coming from the imperfect control of the freeze-out thermodynamic variables, 
such as temperature, chemical potential and volume can also be included by shifting the one-particle energy levels $\omega_{\vec p}$. 
Considering spherically symmetric boundary conditions, for instance, momentum levels are distributed as $p_i = \alpha_i/R$, where $R$ is the system radius. 
This means that a geometric fluctuation of the radius of $\delta R$ will affect the energy levels through 
\begin{eqnarray}
 p_i= & \dfrac{\alpha_{i}}{R +\delta R} \approx p_{0\, i} \left[ 1 - \dfrac{\delta R}{R} + \left(\dfrac{\delta R}{R}\right)^2 + \cdots\right] \,.
\end{eqnarray}
Fluctuations of temperature and chemical potential can likewise be included by introducing the effective energy shift $\delta\omega_{T,\mu}$, such that 
${\omega + \delta \omega_{T,\mu}-\mu}/{T} = {\omega-(\mu + \delta\mu)}/{T+\delta T}$.
 
The results above are, then, used to calculate the average multiplicity of charged pions, $M_{\pi_{ch}}$, and its variance, $V_{\pi_{ch}}$, as a function of $\xi$. 
Then, we can compute the percentage by which the example-signature  $V_{\pi_{ch}}/M_{\pi_{ch}}$  grows with $\xi$, 
with respect to its value at $\xi = 0.4$ fm, when only critical, background and the decay of rho-meson contributions are taken into account. 
More details and results can be found in \cite{Hippert:2017xoj}, where caveats are also discussed. 
Future work will extend these results to the more interesting signatures connected to protons and higher-order moments of particle multiplicities. 

\subsubsection{Finite-size effects}

For the pseudo-critical chiral phase diagram within the linear sigma model with constituent quarks \cite{Palhares:2009tf}, it has been shown that the amplitudes of the shifts due to the finite volume are sizable for length scales probed at current experiments, so that the position of the CEP probed experimentally may differ significantly from the expected critical temperature and chemical potential in the thermodynamic limit. On the other hand, the non-monotonic behavior of correlation functions near criticality for systems of different sizes, tagged by different centralities in heavy-ion collisions, must obey finite-size scaling (FSS). In this vein, the fact that heavy-ion collisions generate data from an ensemble of systems of different sizes provides an alternative signature for the presence of a CEP. 

In the FSS regime, any correlation function $X(T,L)$ of the order parameter does not depend independently on the external parameter $T$ and on the size $L$ of the system, having the following scaling form \cite{Cardy:1996xt}:
$
X(T,L)=L^{\gamma_x/\nu}f_x(tL^{1/\nu})\, ,
$
where $t=(T-T_c)/T_c$ represents a dimensionless measure of the distance, in the external parameter domain, to the genuine CEP (in the thermodynamic limit), $\gamma_x$ is a dimension exponent and $\nu$ is the universal critical exponent defined by the divergence of the correlation length. This scaling form 
implies (and is implied by) the existence of a {\it scaling plot} in which all the curves for different system sizes collapse into a single curve. 

One can pragmatically map these quantities to experimental observables in heavy-ion collisions: the correlation functions should be related to pion multiplicity fluctuations or transverse-momentum fluctuations; the distance $t$ to the CEP is given in terms of the center-of-mass energy (which is related to a $(T,\mu)$ point in the freeze-out curve from thermal models); and the size $L$ can be obtained, e.g., via HBT analysis. To identify FSS in the data, it is then necessary to have different measurements corresponding to the same value of the scaling variable. Since the available system sizes in heavy-ion collisions are limited, the range of energies that can be compared is also restricted. Nevertheless, one can assume the presence of FSS and predict from one data set the amplitude of the fluctuations at a different energy scale, in a thorough analysis of RHIC and SPS data \cite{Fraga:2011hi}. Finite size effects can also modify considerably the dynamics in the first-order transition region \cite{Fraga:2003mu}.

\subsection{Modeling of time correlations with hydrodynamic fluctuations}

The approach to modeling time correlations and their effect on net-baryon fluctuations described here is based on the use of hydrodynamic fluctuations. The approach comprises two separate studies: the first, which uses white noise to model critical fluctuations of the baryon density \cite{Kapusta:2012zb}; and the second, which uses both white noise and colored noise to model (non-critical) electric charge fluctuations \cite{Kapusta:2017hfi}.

In Ref.~\cite{Kapusta:2012zb}, the authors consider the effects of a critical point on hydrodynamic fluctuations in heavy-ion collisions.  They apply mode-coupling theory, together with a model of the free energy (which includes 3-dimensional Ising critical exponents and amplitudes) to model the behavior of the thermal conductivity near the critical point.    Mode-coupling theory permits a rough separation of the critical and non-critical contributions to the thermal conductivity near the critical point, and the exact behavior of these contributions can be matched consistently onto an equation of state which exhibits the right critical scaling.  One special advantage of mode-coupling is that it can be readily extended outside the critical regime, and allows naturally for one to explore the effects of critical fluctuations which come to dominate non-critical fluctuations close to the critical point.

Within this formalism, the magnitude (i.e. the two-point function) of hydrodynamic fluctuations is proportional to the thermal conductivity, as a consequence of the fluctuation-dissipation theorem.  The divergence of the thermal conductivity was thus found to lead to an enhancement in the magnitude of the fluctuations close to the critical point, and to generate corresponding enhancements in charge balance functions sensitive to net-baryon fluctuations (Ref.~\cite{Kapusta:2012zb} considered both $\pi\pi$ and $pp$ balance functions).  The same approach was later applied to HBT fluctuations near the critical point \cite{Plumberg:2017tvu}, and yielded similar conclusions regarding the magnitude of effects due to critical fluctuations.

In Ref.~\cite{Kapusta:2017hfi}, the authors considered the effects of non-trivial time correlations on electric charge fluctuations at top RHIC energies.  The non-triviality was taken to be a simple, decaying exponential in proper-time separation between two correlated fluctuations in the system, containing a single free parameter $\tau_Q$ which effectively fixes the rate at which fluctuations can propagate throughout the system and become correlated with one another (the limit $\tau_Q \to 0$ corresponds to the trivial white noise case discussed above).

The authors then explored the effects of these non-trivial correlations on the (electric) charge balance functions discussed in Ref.~\cite{Kapusta:2012zb}, and found that choosing $\tau_Q \neq 0$ implied a \textit{reduction} in the speed of propagation of wave fronts in the system; in fact, setting $\tau_Q = 0$ can be shown to lead to an infinite speed of propagation which violates relativistic causality.  The requirement that $\tau_Q > 0$ then restores relativistic causality and leads to a corresponding reduction in the efficiency of conserved charge diffusion in heavy-ion collisions, and a consequent \textit{narrowing} of the charge balance functions in rapidity separation.  Similar effects should be expected near a critical point, where the phenomenon of critical slowing down results from the divergence of the system's relaxation timescale.

The comparison of white noise and colored noise in Ref.~\cite{Kapusta:2017hfi} allows one to understand the consequences of non-trivial time correlations on physical quantities such as electric charge and baryon densities.  So far, non-trivial time correlations (i.e. colored noise) have been considered only for non-critical, electric charge fluctuations.  Nevertheless, the same basic results would carry over to the case of critical fluctuations, with just a few straightforward modifications in accordance with the treatment of Ref.~\cite{Kapusta:2012zb}.  This would allow one to explore the effects of non-trivial time correlations near the critical point in heavy-ion collisions.

A weakness of the approaches described here are their inability to account for long-time tails explicitly, since they are based on a linearized version of the (fluctuating) hydrodynamic equations of motion with Bjorken expansion. In particular, these approaches take all linear fluctuating contributions to thermodynamic quantities to be vanishing on average: e.g. $\delta 
\left< T^{\mu\nu} \right> \equiv 0$.  This differs from studies such as Ref.~\cite{Akamatsu:2016llw} where $\delta \left< T^{\mu\nu} \right> \neq 0$ as a result of the non-linear constitutive relation $T^{\mu\nu} = (e+P)u^\mu u^\nu - P g^{\mu\nu}$.  Approaches such as Ref.~\cite{Akamatsu:2016llw} thus yield non-linear equations in the hydrodynamic fluctuations which generate ``long-time tails" in thermodynamic time correlation functions, a standard signature of systems governed by fluctuating hydrodynamics which the approach presented here is unable to reproduce.

This failure to produce long-time tails is compensated for somewhat by exploiting the mode-coupling approach described above, where the effects of long-time tails are essentially absorbed into the critical enhancement of the thermal conductivity, in a way which can be readily and smoothly extended away from the critical regime.  Moreover, the use of colored noise permits a natural regularization of divergences associated to the standard white noise treatment; alternative approaches typically require a renormalized treatment of thermodynamic quantities to eliminate divergences which result from white-noise correlations \cite{An:2019osr}.

Another key advantage of the approach described here is its ability to model the subtraction of so-called self-correlations from physical observables based on e.g. multi-particle correlations, where trivial correlations of a particle (or fluid cell) with itself are generally neglected.  One way to do this was considered in \cite{Ling:2013ksb} for the case of white noise, with the extension to colored noise being considered in \cite{Kapusta:2017hfi}.  It would be interesting to consider how this same subtraction could be performed in alternative approaches (or whether such a subtraction would even need to be carried out).

\subsection{Summary of approaches to critical dynamics}

Specific features of the different numerical implementations studying the dynamics of critical fluctuations are summarized in the following table: 
\begin{sidewaystable}
\begin{tabular}{|m{0.12\textwidth}|m{0.12\textwidth}|m{0.12\textwidth}|m{0.12\textwidth}|m{0.122\textwidth}|m{0.12\textwidth}|m{0.12\textwidth}|}
\hline
\hline
    Approach & Stochastic diffusion (Sec.~\ref{sec:stodiff}) & N$\chi$FD (Sec.~\ref{sec:chiral}) & QCD assisted transport (Sec.~\ref{app:QCDassisted}) & Hydro$+$\linebreak implementation (Sec.~\ref{sec:Hydro+}) & Hydrokinetics (Sec.~\ref{sec:hydrokin}) & Transits (Sec.~\ref{sec:transits}) \\[3ex]
\hline
    fluctuating or propagated quantity & net-B density & chiral\linebreak condensate & $\sigma$-meson\linebreak expectation value & order\,\,parameter two-point \linebreak function & energy\,\,and momentum density\,\,two-point functions & entropy\,\,per \linebreak baryon\,\,two-\linebreak point function \\[3ex]
\hline
    inclusion\,\,of nonlinearities (nonlinear fluctuations) & yes & yes & yes & no & no & \hspace{2cm}no \\[3ex]
\hline
    modeling\,\,of (expanding) medium & cooling, \hspace{3mm} \linebreak no expansion & ideal hydro & $T$\,-\,quench, \hspace{1mm} \linebreak no expansion & viscous hydro & viscous \linebreak Bjorken-type \linebreak expansion & \hspace{2cm}viscous hydro \\[3ex]
\hline
    coupling\,\,to\,\,the medium & N/A & yes & N/A & yes & yes & \hspace{2cm}yes \\[3ex]
\hline
    dimensionality & $1+1$d & $3+1$d & $1+1$d & $3+1$d with\hspace{2cm}\linebreak symmetries & $3+1$d with\hspace{2cm}\linebreak symmetries & \vspace{4mm}$3+1$d with\hspace{2cm}\linebreak symmetries \\[3ex]
\hline
    conservation equations & net-B\linebreak conservation & hydro\linebreak equations & N/A & hydro\linebreak equations & hydro\linebreak equations & \vspace{4mm}hydro equations \\[3ex]
\hline
    thermodynamic model & 3d Ising & $N_f=2$ QM \hspace{4mm} \linebreak (mean field) & $N_f=2+1$ QM \linebreak (beyond MF) & 3d Ising & N/A & \vspace{3mm}3d Ising \\[3ex]
\hline
    region of applicability & near CEP & small and intermediate $\mu$ & small and intermediate $\mu$ & near CEP (if at small $\mu$) & crossover domain & near CEP \\[3ex]
\hline
\hline
\end{tabular}
\end{sidewaystable}

\bibliographystyle{utphys} 
\bibliography{BiblioEmmiRRTF}  

\providecommand{\href}[2]{#2}\begingroup\raggedright\begin{thebibliography}{100}

\bibitem{Odyniec:2013aaa}
G.~Odyniec, ``{RHIC Beam Energy Scan Program: Phase I and II},''
{\em PoS} {\bfseries CPOD2013} (2013) 043.

\bibitem{Gazdzicki:995681}
{\bfseries NA61/SHINE} Collaboration, M.~Gazdzicki {\em et~al.}, ``{Study of
  Hadron Production in Hadron-Nucleus and Nucleus-Nucleus Collisions at the
  CERN SPS},'' Tech. Rep. CERN-SPSC-2006-034. SPSC-P-330, CERN, Geneva, Nov,
  2006.
\newblock \url{https://cds.cern.ch/record/995681}.

\bibitem{Luo:2015doi}
X.~Luo, ``{Exploring the QCD Phase Structure with Beam Energy Scan in Heavy-ion
  Collisions},'' \href{http://dx.doi.org/10.1016/j.nuclphysa.2016.03.025}{{\em
  Nucl. Phys.} {\bfseries A956} (2016) 75--82},
\href{http://arxiv.org/abs/1512.09215}{{\ttfamily arXiv:1512.09215 [nucl-ex]}}.

\bibitem{Bzdak:2019pkr}
A.~Bzdak, S.~Esumi, V.~Koch, J.~Liao, M.~Stephanov, and N.~Xu, ``{Mapping the
  Phases of Quantum Chromodynamics with Beam Energy Scan},''
\href{http://arxiv.org/abs/1906.00936}{{\ttfamily arXiv:1906.00936 [nucl-th]}}.

\bibitem{Stephanov:1998dy}
M.~A. Stephanov, K.~Rajagopal, and E.~V. Shuryak, ``{Signatures of the
  tricritical point in QCD},''
  \href{http://dx.doi.org/10.1103/PhysRevLett.81.4816}{{\em Phys. Rev. Lett.}
  {\bfseries 81} (1998) 4816--4819},
\href{http://arxiv.org/abs/hep-ph/9806219}{{\ttfamily arXiv:hep-ph/9806219
  [hep-ph]}}.

\bibitem{Heinz:2013th}
U.~Heinz and R.~Snellings, ``{Collective flow and viscosity in relativistic
  heavy-ion collisions},''
  \href{http://dx.doi.org/10.1146/annurev-nucl-102212-170540}{{\em Ann. Rev.
  Nucl. Part. Sci.} {\bfseries 63} (2013) 123--151},
\href{http://arxiv.org/abs/1301.2826}{{\ttfamily arXiv:1301.2826 [nucl-th]}}.

\bibitem{Teaney:2009qa}
D.~A. Teaney, \href{http://dx.doi.org/10.1142/9789814293297_0004}{``{Viscous
  Hydrodynamics and the Quark Gluon Plasma},''} in {\em Quark-gluon plasma 4},
  R.~C. Hwa and X.-N. Wang, eds., pp.~207--266.
\newblock 2010.
\newblock
\href{http://arxiv.org/abs/0905.2433}{{\ttfamily arXiv:0905.2433 [nucl-th]}}.
\newblock

\bibitem{deSouza:2015ena}
R.~Derradi~de Souza, T.~Koide, and T.~Kodama, ``{Hydrodynamic Approaches in
  Relativistic Heavy Ion Reactions},''
  \href{http://dx.doi.org/10.1016/j.ppnp.2015.09.002}{{\em Prog. Part. Nucl.
  Phys.} {\bfseries 86} (2016) 35--85},
\href{http://arxiv.org/abs/1506.03863}{{\ttfamily arXiv:1506.03863 [nucl-th]}}.

\bibitem{Schenke:2010nt}
B.~Schenke, S.~Jeon, and C.~Gale, ``{(3+1)D hydrodynamic simulation of
  relativistic heavy-ion collisions},''
  \href{http://dx.doi.org/10.1103/PhysRevC.82.014903}{{\em Phys. Rev.}
  {\bfseries C82} (2010) 014903},
\href{http://arxiv.org/abs/1004.1408}{{\ttfamily arXiv:1004.1408 [hep-ph]}}.

\bibitem{Nagle:2018nvi}
J.~L. Nagle and W.~A. Zajc, ``{Small System Collectivity in Relativistic
  Hadronic and Nuclear Collisions},''
  \href{http://dx.doi.org/10.1146/annurev-nucl-101916-123209}{{\em Ann. Rev.
  Nucl. Part. Sci.} {\bfseries 68} (2018) 211--235},
\href{http://arxiv.org/abs/1801.03477}{{\ttfamily arXiv:1801.03477 [nucl-ex]}}.

\bibitem{Schlichting:2019abc}
S.~Schlichting and D.~Teaney, ``The first fm/c of heavy-ion collisions,''
  \href{http://dx.doi.org/10.1146/annurev-nucl-101918-023825}{{\em Annual
  Review of Nuclear and Particle Science} {\bfseries 69} no.~1, (2019)
  447--476},
  \href{http://arxiv.org/abs/https://doi.org/10.1146/annurev-nucl-101918-023825}{{\ttfamily
  https://doi.org/10.1146/annurev-nucl-101918-023825}}.
  \url{https://doi.org/10.1146/annurev-nucl-101918-023825}.

\bibitem{Florkowski:2017olj}
W.~Florkowski, M.~P. Heller, and M.~Spalinski, ``{New theories of relativistic
  hydrodynamics in the LHC era},''
  \href{http://dx.doi.org/10.1088/1361-6633/aaa091}{{\em Rept. Prog. Phys.}
  {\bfseries 81} no.~4, (2018) 046001},
\href{http://arxiv.org/abs/1707.02282}{{\ttfamily arXiv:1707.02282 [hep-ph]}}.

\bibitem{Romatschke:2017ejr}
P.~Romatschke and U.~Romatschke,
  \href{http://dx.doi.org/10.1017/9781108651998}{{\em {Relativistic Fluid
  Dynamics In and Out of Equilibrium}}}.
\newblock Cambridge Monographs on Mathematical Physics. Cambridge University
  Press, 2019.
\newblock
\href{http://arxiv.org/abs/1712.05815}{{\ttfamily arXiv:1712.05815 [nucl-th]}}.
\newblock

\bibitem{Hirano:2005wx}
T.~Hirano and M.~Gyulassy, ``{Perfect fluidity of the quark gluon plasma core
  as seen through its dissipative hadronic corona},''
  \href{http://dx.doi.org/10.1016/j.nuclphysa.2006.02.005}{{\em Nucl.Phys.}
  {\bfseries A769} (2006) 71--94},
  \href{http://arxiv.org/abs/nucl-th/0506049}{{\ttfamily arXiv:nucl-th/0506049
  [nucl-th]}}.

\bibitem{Schafer:2009dj}
T.~Schäfer and D.~Teaney, ``{Nearly Perfect Fluidity: From Cold Atomic Gases
  to Hot Quark Gluon Plasmas},''
  \href{http://dx.doi.org/10.1088/0034-4885/72/12/126001}{{\em Rept. Prog.
  Phys.} {\bfseries 72} (2009) 126001},
\href{http://arxiv.org/abs/0904.3107}{{\ttfamily arXiv:0904.3107 [hep-ph]}}.

\bibitem{Heller:2013oxa}
M.~P. Heller, D.~Mateos, W.~van~der Schee, and M.~Triana, ``{Holographic
  isotropization linearized},''
  \href{http://dx.doi.org/10.1007/JHEP09(2013)026}{{\em JHEP} {\bfseries 09}
  (2013) 026},
\href{http://arxiv.org/abs/1304.5172}{{\ttfamily arXiv:1304.5172 [hep-th]}}.

\bibitem{Heller:2015dha}
M.~P. Heller and M.~Spalinski, ``{Hydrodynamics Beyond the Gradient Expansion:
  Resurgence and Resummation},''
  \href{http://dx.doi.org/10.1103/PhysRevLett.115.072501}{{\em Phys. Rev.
  Lett.} {\bfseries 115} no.~7, (2015) 072501},
\href{http://arxiv.org/abs/1503.07514}{{\ttfamily arXiv:1503.07514 [hep-th]}}.

\bibitem{Stock:2010hoa}
R.~Stock, ed., \href{http://dx.doi.org/10.1007/978-3-642-01539-7}{{\em
  {Relativistic Heavy Ion Physics}}}, vol.~23 of {\em Landolt-Boernstein -
  Group I Elementary Particles, Nuclei and Atoms}.
\newblock Springer,
2010.
\newblock

\bibitem{Braun-Munzinger:2015hba}
P.~Braun-Munzinger, V.~Koch, T.~Sch{\"a}fer, and J.~Stachel, ``{Properties of
  hot and dense matter from relativistic heavy ion collisions},''
  \href{http://dx.doi.org/10.1016/j.physrep.2015.12.003}{{\em Phys. Rept.}
  {\bfseries 621} (2016) 76--126},
\href{http://arxiv.org/abs/1510.00442}{{\ttfamily arXiv:1510.00442 [nucl-th]}}.

\bibitem{Jeon:2000wg}
S.~Jeon and V.~Koch, ``{Charged particle ratio fluctuation as a signal for
  QGP},'' \href{http://dx.doi.org/10.1103/PhysRevLett.85.2076}{{\em Phys. Rev.
  Lett.} {\bfseries 85} (2000) 2076--2079},
\href{http://arxiv.org/abs/hep-ph/0003168}{{\ttfamily arXiv:hep-ph/0003168
  [hep-ph]}}.

\bibitem{Asakawa:2000wh}
M.~Asakawa, U.~W. Heinz, and B.~Muller, ``{Fluctuation probes of quark
  deconfinement},'' \href{http://dx.doi.org/10.1103/PhysRevLett.85.2072}{{\em
  Phys. Rev. Lett.} {\bfseries 85} (2000) 2072--2075},
\href{http://arxiv.org/abs/hep-ph/0003169}{{\ttfamily arXiv:hep-ph/0003169
  [hep-ph]}}.

\bibitem{Alver:2010gr}
B.~Alver and G.~Roland, ``{Collision geometry fluctuations and triangular flow
  in heavy-ion collisions},''
  \href{http://dx.doi.org/10.1103/PhysRevC.82.039903,
  10.1103/PhysRevC.81.054905}{{\em Phys. Rev.} {\bfseries C81} (2010) 054905},
  \href{http://arxiv.org/abs/1003.0194}{{\ttfamily arXiv:1003.0194 [nucl-th]}}.
[Erratum: Phys. Rev.C82,039903(2010)].

\bibitem{Adamczyk:2013dal}
{\bfseries STAR} Collaboration, L.~Adamczyk {\em et~al.}, ``{Energy Dependence
  of Moments of Net-proton Multiplicity Distributions at RHIC},''
  \href{http://dx.doi.org/10.1103/PhysRevLett.112.032302}{{\em Phys. Rev.
  Lett.} {\bfseries 112} (2014) 032302},
\href{http://arxiv.org/abs/1309.5681}{{\ttfamily arXiv:1309.5681 [nucl-ex]}}.

\bibitem{Adamczyk:2014fia}
{\bfseries STAR} Collaboration, L.~Adamczyk {\em et~al.}, ``{Beam energy
  dependence of moments of the net-charge multiplicity distributions in Au+Au
  collisions at RHIC},''
  \href{http://dx.doi.org/10.1103/PhysRevLett.113.092301}{{\em Phys. Rev.
  Lett.} {\bfseries 113} (2014) 092301},
\href{http://arxiv.org/abs/1402.1558}{{\ttfamily arXiv:1402.1558 [nucl-ex]}}.

\bibitem{Rustamov:2017lio}
{\bfseries ALICE} Collaboration, A.~Rustamov, ``{Net-baryon fluctuations
  measured with ALICE at the CERN LHC},''
  \href{http://dx.doi.org/10.1016/j.nuclphysa.2017.05.111}{{\em Nucl. Phys.}
  {\bfseries A967} (2017) 453--456},
\href{http://arxiv.org/abs/1704.05329}{{\ttfamily arXiv:1704.05329 [nucl-ex]}}.

\bibitem{Adam:2020unf}
{\bfseries STAR} Collaboration, J.~Adam {\em et~al.}, ``{Net-proton number
  fluctuations and the Quantum Chromodynamics critical point},''
\href{http://arxiv.org/abs/2001.02852}{{\ttfamily arXiv:2001.02852 [nucl-ex]}}.

\bibitem{Bazavov:2012vg}
A.~Bazavov, H.~Ding, P.~Hegde, O.~Kaczmarek, F.~Karsch, {\em et~al.},
  ``{Freeze-out Conditions in Heavy Ion Collisions from QCD Thermodynamics},''
  \href{http://dx.doi.org/10.1103/PhysRevLett.109.192302}{{\em Phys.Rev.Lett.}
  {\bfseries 109} (2012) 192302},
\href{http://arxiv.org/abs/1208.1220}{{\ttfamily arXiv:1208.1220 [hep-lat]}}.

\bibitem{Borsanyi:2014ewa}
S.~Borsanyi, Z.~Fodor, S.~D. Katz, S.~Krieg, C.~Ratti, and K.~K. Szabo,
  ``{Freeze-out parameters from electric charge and baryon number fluctuations:
  is there consistency?},''
  \href{http://dx.doi.org/10.1103/PhysRevLett.113.052301}{{\em Phys. Rev.
  Lett.} {\bfseries 113} (2014) 052301},
\href{http://arxiv.org/abs/1403.4576}{{\ttfamily arXiv:1403.4576 [hep-lat]}}.

\bibitem{Jeon:2003gk}
S.~Jeon and V.~Koch, ``{Event by event fluctuations},'' in {\em Quark Gluon
  Plasma 3}, X.~W. R.~Hwa, ed.
\newblock World Scientific, 2004.
\newblock
\href{http://arxiv.org/abs/hep-ph/0304012}{{\ttfamily arXiv:hep-ph/0304012
  [hep-ph]}}.
\newblock

\bibitem{Schuster:2009jv}
M.~Nahrgang, T.~Schuster, M.~Mitrovski, R.~Stock, and M.~Bleicher,
  ``{Net-baryon-, net-proton-, and net-charge kurtosis in heavy-ion collisions
  within a relativistic transport approach},''
  \href{http://dx.doi.org/10.1140/epjc/s10052-012-2143-6}{{\em Eur. Phys. J.}
  {\bfseries C72} (2012) 2143},
\href{http://arxiv.org/abs/0903.2911}{{\ttfamily arXiv:0903.2911 [hep-ph]}}.

\bibitem{Bzdak:2012an}
A.~Bzdak, V.~Koch, and V.~Skokov, ``{Baryon number conservation and the
  cumulants of the net proton distribution},''
  \href{http://dx.doi.org/10.1103/PhysRevC.87.014901}{{\em Phys. Rev.}
  {\bfseries C87} no.~1, (2013) 014901},
\href{http://arxiv.org/abs/1203.4529}{{\ttfamily arXiv:1203.4529 [hep-ph]}}.

\bibitem{Kitazawa:2011wh}
M.~Kitazawa and M.~Asakawa, ``{Revealing baryon number fluctuations from proton
  number fluctuations in relativistic heavy ion collisions},''
  \href{http://dx.doi.org/10.1103/PhysRevC.85.021901}{{\em Phys. Rev.}
  {\bfseries C85} (2012) 021901},
\href{http://arxiv.org/abs/1107.2755}{{\ttfamily arXiv:1107.2755 [nucl-th]}}.

\bibitem{Kitazawa:2012at}
M.~Kitazawa and M.~Asakawa, ``{Relation between baryon number fluctuations and
  experimentally observed proton number fluctuations in relativistic heavy ion
  collisions},'' \href{http://dx.doi.org/10.1103/PhysRevC.86.024904,
  10.1103/PhysRevC.86.069902}{{\em Phys. Rev.} {\bfseries C86} (2012) 024904},
  \href{http://arxiv.org/abs/1205.3292}{{\ttfamily arXiv:1205.3292 [nucl-th]}}.
[Erratum: Phys. Rev.C86,069902(2012)].

\bibitem{Kitazawa:2013bta}
M.~Kitazawa, M.~Asakawa, and H.~Ono, ``{Non-equilibrium time evolution of
  higher order cumulants of conserved charges and event-by-event analysis},''
  \href{http://dx.doi.org/10.1016/j.physletb.2013.12.008}{{\em Phys.Lett.}
  {\bfseries B728} (2014) 386--392},
\href{http://arxiv.org/abs/1307.2978}{{\ttfamily arXiv:1307.2978 [nucl-th]}}.

\bibitem{Mukherjee:2015swa}
S.~Mukherjee, R.~Venugopalan, and Y.~Yin, ``{Real time evolution of
  non-Gaussian cumulants in the QCD critical regime},''
  \href{http://dx.doi.org/10.1103/PhysRevC.92.034912}{{\em Phys. Rev.}
  {\bfseries C92} no.~3, (2015) 034912},
\href{http://arxiv.org/abs/1506.00645}{{\ttfamily arXiv:1506.00645 [hep-ph]}}.

\bibitem{Cooper:1974mv}
F.~Cooper and G.~Frye, ``{Comment on the Single Particle Distribution in the
  Hydrodynamic and Statistical Thermodynamic Models of Multiparticle
  Production},''
\href{http://dx.doi.org/10.1103/PhysRevD.10.186}{{\em Phys. Rev.} {\bfseries
  D10} (1974) 186}.

\bibitem{Oliinychenko:2019zfk}
D.~Oliinychenko and V.~Koch, ``{Particlization with local event-by-event
  conservation laws},''
\href{http://arxiv.org/abs/1902.09775}{{\ttfamily arXiv:1902.09775 [hep-ph]}}.

\bibitem{Bzdak:2012ab}
A.~Bzdak and V.~Koch, ``{Acceptance corrections to net baryon and net charge
  cumulants},'' \href{http://dx.doi.org/10.1103/PhysRevC.86.044904}{{\em Phys.
  Rev.} {\bfseries C86} (2012) 044904},
\href{http://arxiv.org/abs/1206.4286}{{\ttfamily arXiv:1206.4286 [nucl-th]}}.

\bibitem{Bzdak:2016qdc}
A.~Bzdak, R.~Holzmann, and V.~Koch, ``{Multiplicity dependent and non-binomial
  efficiency corrections for particle number cumulantsMultiplicity-dependent
  and nonbinomial efficiency corrections for particle number cumulants},''
  \href{http://dx.doi.org/10.1103/PhysRevC.94.064907}{{\em Phys. Rev.}
  {\bfseries C94} no.~6, (2016) 064907},
\href{http://arxiv.org/abs/1603.09057}{{\ttfamily arXiv:1603.09057 [nucl-th]}}.

\bibitem{Nonaka:2018mgw}
T.~Nonaka, M.~Kitazawa, and S.~Esumi, ``{A general procedure for
  detector--response correction of higher order cumulants},''
  \href{http://dx.doi.org/10.1016/j.nima.2018.08.013}{{\em Nucl. Instrum.
  Meth.} {\bfseries A906} (2018) 10--17},
\href{http://arxiv.org/abs/1805.00279}{{\ttfamily arXiv:1805.00279
  [physics.data-an]}}.

\bibitem{Kovtun:2003vj}
P.~Kovtun and L.~G. Yaffe, ``{Hydrodynamic fluctuations, long time tails, and
  supersymmetry},'' \href{http://dx.doi.org/10.1103/PhysRevD.68.025007}{{\em
  Phys. Rev.} {\bfseries D68} (2003) 025007},
\href{http://arxiv.org/abs/hep-th/0303010}{{\ttfamily arXiv:hep-th/0303010
  [hep-th]}}.

\bibitem{Gavin:2006xd}
S.~Gavin and M.~Abdel-Aziz, ``{Measuring Shear Viscosity Using Transverse
  Momentum Correlations in Relativistic Nuclear Collisions},''
  \href{http://dx.doi.org/10.1103/PhysRevLett.97.162302}{{\em Phys. Rev. Lett.}
  {\bfseries 97} (2006) 162302},
\href{http://arxiv.org/abs/nucl-th/0606061}{{\ttfamily arXiv:nucl-th/0606061
  [nucl-th]}}.

\bibitem{Kovtun:2014hpa}
P.~Kovtun, G.~D. Moore, and P.~Romatschke, ``{Towards an effective action for
  relativistic dissipative hydrodynamics},''
  \href{http://dx.doi.org/10.1007/JHEP07(2014)123}{{\em JHEP} {\bfseries 07}
  (2014) 123},
\href{http://arxiv.org/abs/1405.3967}{{\ttfamily arXiv:1405.3967 [hep-ph]}}.

\bibitem{Crossley:2015evo}
M.~Crossley, P.~Glorioso, and H.~Liu, ``{Effective field theory of dissipative
  fluids},'' \href{http://dx.doi.org/10.1007/JHEP09(2017)095}{{\em JHEP}
  {\bfseries 09} (2017) 095},
\href{http://arxiv.org/abs/1511.03646}{{\ttfamily arXiv:1511.03646 [hep-th]}}.

\bibitem{Haehl:2018lcu}
F.~M. Haehl, R.~Loganayagam, and M.~Rangamani, ``{Effective Action for
  Relativistic Hydrodynamics: Fluctuations, Dissipation, and Entropy Inflow},''
  \href{http://dx.doi.org/10.1007/JHEP10(2018)194}{{\em JHEP} {\bfseries 10}
  (2018) 194},
\href{http://arxiv.org/abs/1803.11155}{{\ttfamily arXiv:1803.11155 [hep-th]}}.

\bibitem{LandauStatPart1}
L.~Landau and E.~Lifshitz, {\em Statistical Physics}, vol.~5 of {\em Course of
  theoretical physics}.
\newblock Pergamon Press, 1980.

\bibitem{LandauStatPart2}
E.~Lifshitz and L.~Pitaevskii, {\em Statistical Physics}, vol.~9 of {\em Course
  of theoretical physics}.
\newblock Pergamon Press, 1980.

\bibitem{landau2013fluid}
L.~Landau and E.~Lifshitz, {\em Fluid Mechanics: Landau and Lifshitz: Course of
  Theoretical Physics}.
\newblock No.~v. 6. Elsevier Science, 2013.
\newblock \url{https://books.google.ch/books?id=eOBbAwAAQBAJ}.

\bibitem{Karpenko:2013wva}
I.~Karpenko, P.~Huovinen, and M.~Bleicher, ``{A 3+1 dimensional viscous
  hydrodynamic code for relativistic heavy ion collisions},''
  \href{http://dx.doi.org/10.1016/j.cpc.2014.07.010}{{\em Comput. Phys.
  Commun.} {\bfseries 185} (2014) 3016--3027},
\href{http://arxiv.org/abs/1312.4160}{{\ttfamily arXiv:1312.4160 [nucl-th]}}.

\bibitem{DelZanna:2013eua}
L.~Del~Zanna, V.~Chandra, G.~Inghirami, V.~Rolando, A.~Beraudo, A.~De~Pace,
  G.~Pagliara, A.~Drago, and F.~Becattini, ``{Relativistic viscous
  hydrodynamics for heavy-ion collisions with ECHO-QGP},''
  \href{http://dx.doi.org/10.1140/epjc/s10052-013-2524-5}{{\em Eur. Phys. J.}
  {\bfseries C73} (2013) 2524},
\href{http://arxiv.org/abs/1305.7052}{{\ttfamily arXiv:1305.7052 [nucl-th]}}.

\bibitem{bell2007numerical}
J.~B. Bell, A.~L. Garcia, and S.~A. Williams, ``Numerical methods for the
  stochastic {Landau-Lifshitz Navier-Stokes} equations,''
  \href{http://dx.doi.org/10.1103/PhysRevE.76.016708}{{\em Phys. Rev. E}
  {\bfseries 76} (Jul, 2007) 016708}.
  \url{http://link.aps.org/doi/10.1103/PhysRevE.76.016708}.

\bibitem{donev2011diffusive}
A.~Donev, J.~B. Bell, A.~de~la Fuente, and A.~L. Garcia, ``Diffusive transport
  by thermal velocity fluctuations,''
  \href{http://dx.doi.org/10.1103/PhysRevLett.106.204501}{{\em Phys. Rev.
  Lett.} {\bfseries 106} (May, 2011) 204501}.
  \url{http://link.aps.org/doi/10.1103/PhysRevLett.106.204501}.

\bibitem{balboa2012staggered}
F.~B. Usabiaga, J.~B. Bell, R.~Delgado-Buscalioni, A.~Donev, T.~G. Fai, B.~E.
  Griffith, and C.~S. Peskin, ``Staggered schemes for fluctuating
  hydrodynamics,'' \href{http://dx.doi.org/10.1137/120864520}{{\em Multiscale
  Modeling \& Simulation} {\bfseries 10} no.~4, (2012) 1369--1408}.
  \url{http://dx.doi.org/10.1137/120864520}.

\bibitem{Kovtun:2012rj}
P.~Kovtun, ``{Lectures on hydrodynamic fluctuations in relativistic
  theories},'' \href{http://dx.doi.org/10.1088/1751-8113/45/47/473001}{{\em J.
  Phys.} {\bfseries A45} (2012) 473001},
\href{http://arxiv.org/abs/1205.5040}{{\ttfamily arXiv:1205.5040 [hep-th]}}.

\bibitem{Kapusta:2011gt}
J.~I. Kapusta, B.~Muller, and M.~Stephanov, ``{Relativistic Theory of
  Hydrodynamic Fluctuations with Applications to Heavy Ion Collisions},''
  \href{http://dx.doi.org/10.1103/PhysRevC.85.054906}{{\em Phys. Rev.}
  {\bfseries C85} (2012) 054906},
\href{http://arxiv.org/abs/1112.6405}{{\ttfamily arXiv:1112.6405 [nucl-th]}}.

\bibitem{Xu:1999aq}
Z.~Xu and C.~Greiner, ``{Stochastic treatment of disoriented chiral condensates
  within a Langevin description},''
  \href{http://dx.doi.org/10.1103/PhysRevD.62.036012}{{\em Phys. Rev.}
  {\bfseries D62} (2000) 036012},
\href{http://arxiv.org/abs/hep-ph/9910562}{{\ttfamily arXiv:hep-ph/9910562
  [hep-ph]}}.

\bibitem{Murase:2013tma}
K.~Murase and T.~Hirano, ``{Relativistic fluctuating hydrodynamics with memory
  functions and colored noises},''
\href{http://arxiv.org/abs/1304.3243}{{\ttfamily arXiv:1304.3243 [nucl-th]}}.

\bibitem{Kapusta:2017hfi}
J.~I. Kapusta and C.~Plumberg, ``{Causal Electric Charge Diffusion and Balance
  Functions in Relativistic Heavy Ion Collisions},''
  \href{http://dx.doi.org/10.1103/PhysRevC.97.014906}{{\em Phys. Rev.}
  {\bfseries C97} no.~1, (2018) 014906},
\href{http://arxiv.org/abs/1710.03329}{{\ttfamily arXiv:1710.03329 [nucl-th]}}.

\bibitem{Murase:2016rhl}
K.~Murase and T.~Hirano, ``{Hydrodynamic fluctuations and dissipation in an
  integrated dynamical model},''
  \href{http://dx.doi.org/10.1016/j.nuclphysa.2016.01.011}{{\em Nucl. Phys.}
  {\bfseries A956} (2016) 276--279},
\href{http://arxiv.org/abs/1601.02260}{{\ttfamily arXiv:1601.02260 [nucl-th]}}.

\bibitem{Hirano:2018diu}
T.~Hirano, R.~Kurita, and K.~Murase, ``{Hydrodynamic fluctuations of entropy in
  one-dimensionally expanding system},''
  \href{http://dx.doi.org/10.1016/j.nuclphysa.2019.01.010}{{\em Nucl. Phys.}
  {\bfseries A984} (2019) 44--67},
\href{http://arxiv.org/abs/1809.04773}{{\ttfamily arXiv:1809.04773 [nucl-th]}}.

\bibitem{Nahrgang:2017oqp}
M.~Nahrgang, M.~Bluhm, T.~Schäfer, and S.~Bass, ``{Toward the description of
  fluid dynamical fluctuations in heavy-ion collisions},''
  \href{http://dx.doi.org/10.5506/APhysPolBSupp.10.687}{{\em Acta Phys. Polon.
  Supp.} {\bfseries 10} (2017) 687},
\href{http://arxiv.org/abs/1704.03553}{{\ttfamily arXiv:1704.03553 [nucl-th]}}.

\bibitem{Bluhm:2018plm}
M.~Bluhm, M.~Nahrgang, T.~Schäfer, and S.~A. Bass, ``{Fluctuating fluid
  dynamics for the QGP in the LHC and BES era},''
  \href{http://dx.doi.org/10.1051/epjconf/201817116004}{{\em EPJ Web Conf.}
  {\bfseries 171} (2018) 16004},
\href{http://arxiv.org/abs/1804.03493}{{\ttfamily arXiv:1804.03493 [nucl-th]}}.

\bibitem{Singh:2018dpk}
M.~Singh, C.~Shen, S.~McDonald, S.~Jeon, and C.~Gale, ``{Hydrodynamic
  Fluctuations in Relativistic Heavy-Ion Collisions},''
  \href{http://dx.doi.org/10.1016/j.nuclphysa.2018.10.061}{{\em Nucl. Phys.}
  {\bfseries A982} (2019) 319--322},
\href{http://arxiv.org/abs/1807.05451}{{\ttfamily arXiv:1807.05451 [nucl-th]}}.

\bibitem{Kovtun:2011np}
P.~Kovtun, G.~D. Moore, and P.~Romatschke, ``{The stickiness of sound: An
  absolute lower limit on viscosity and the breakdown of second order
  relativistic hydrodynamics},''
  \href{http://dx.doi.org/10.1103/PhysRevD.84.025006}{{\em Phys. Rev.}
  {\bfseries D84} (2011) 025006},
\href{http://arxiv.org/abs/1104.1586}{{\ttfamily arXiv:1104.1586 [hep-ph]}}.

\bibitem{Chafin:2012eq}
C.~Chafin and T.~Schäfer, ``{Hydrodynamic fluctuations and the minimum shear
  viscosity of the dilute Fermi gas at unitarity},''
  \href{http://dx.doi.org/10.1103/PhysRevA.87.023629}{{\em Phys. Rev.}
  {\bfseries A87} no.~2, (2013) 023629},
\href{http://arxiv.org/abs/1209.1006}{{\ttfamily arXiv:1209.1006
  [cond-mat.quant-gas]}}.

\bibitem{Akamatsu:2016llw}
Y.~Akamatsu, A.~Mazeliauskas, and D.~Teaney, ``{A kinetic regime of
  hydrodynamic fluctuations and long time tails for a Bjorken expansion},''
  \href{http://dx.doi.org/10.1103/PhysRevC.95.014909}{{\em Phys. Rev.}
  {\bfseries C95} no.~1, (2017) 014909},
\href{http://arxiv.org/abs/1606.07742}{{\ttfamily arXiv:1606.07742 [nucl-th]}}.

\bibitem{An:2019osr}
X.~An, G.~Basar, M.~Stephanov, and H.-U. Yee, ``{Relativistic Hydrodynamic
  Fluctuations},''
\href{http://arxiv.org/abs/1902.09517}{{\ttfamily arXiv:1902.09517 [hep-th]}}.

\bibitem{andreev1971two}
A.~Andreev, ``Two-liquid effects in a normal liquid,'' {\em SOVIET PHYSICS
  JETP} {\bfseries 32} no.~5, (1971) .

\bibitem{andreev1978corrections}
A.~Andreev, ``Corrections to the hydrodynamics of liquids,'' {\em Sov. Phys.
  JETP} {\bfseries 48} no.~3, (1978) .

\bibitem{Akamatsu:2017rdu}
Y.~Akamatsu, A.~Mazeliauskas, and D.~Teaney, ``{Bulk viscosity from
  hydrodynamic fluctuations with relativistic hydrokinetic theory},''
  \href{http://dx.doi.org/10.1103/PhysRevC.97.024902}{{\em Phys. Rev.}
  {\bfseries C97} no.~2, (2018) 024902},
\href{http://arxiv.org/abs/1708.05657}{{\ttfamily arXiv:1708.05657 [nucl-th]}}.

\bibitem{Martinez:2018wia}
M.~Martinez and T.~Schäfer, ``{Stochastic hydrodynamics and long time tails of
  an expanding conformal charged fluid},''
  \href{http://dx.doi.org/10.1103/PhysRevC.99.054902}{{\em Phys. Rev.}
  {\bfseries C99} no.~5, (2019) 054902},
\href{http://arxiv.org/abs/1812.05279}{{\ttfamily arXiv:1812.05279 [hep-th]}}.

\bibitem{Israel:1979wp}
W.~Israel and J.~M. Stewart, ``{Transient relativistic thermodynamics and
  kinetic theory},''
\href{http://dx.doi.org/10.1016/0003-4916(79)90130-1}{{\em Annals Phys.}
  {\bfseries 118} (1979) 341--372}.

\bibitem{Sakaida:2017rtj}
M.~Sakaida, M.~Asakawa, H.~Fujii, and M.~Kitazawa, ``{Dynamical evolution of
  critical fluctuations and its observation in heavy ion collisions},''
  \href{http://dx.doi.org/10.1103/PhysRevC.95.064905}{{\em Phys. Rev.}
  {\bfseries C95} no.~6, (2017) 064905},
\href{http://arxiv.org/abs/1703.08008}{{\ttfamily arXiv:1703.08008 [nucl-th]}}.

\bibitem{Nahrgang:2017hkh}
M.~Nahrgang, M.~Bluhm, T.~Schäfer, and S.~A. Bass, ``{Baryon number diffusion
  with critical fluctuations},''
  \href{http://dx.doi.org/10.1016/j.nuclphysa.2017.04.021}{{\em Nucl. Phys.}
  {\bfseries A967} (2017) 824--827},
\href{http://arxiv.org/abs/1804.02976}{{\ttfamily arXiv:1804.02976 [nucl-th]}}.

\bibitem{Bluhm:2019yfb}
M.~Bluhm and M.~Nahrgang, ``{Time-evolution of net-baryon density fluctuations
  across the QCD critical region},''
\newblock 2019.
\newblock
\href{http://arxiv.org/abs/1911.08911}{{\ttfamily arXiv:1911.08911 [nucl-th]}}.
\newblock

\bibitem{Nahrgang:2018afz}
M.~Nahrgang, M.~Bluhm, T.~Schäfer, and S.~A. Bass, ``{Diffusive dynamics of
  critical fluctuations near the QCD critical point},''
\href{http://arxiv.org/abs/1804.05728}{{\ttfamily arXiv:1804.05728 [nucl-th]}}.

\bibitem{Hohenberg:1977ym}
P.~C. Hohenberg and B.~I. Halperin, ``{Theory of Dynamic Critical Phenomena},''
\href{http://dx.doi.org/10.1103/RevModPhys.49.435}{{\em Rev. Mod. Phys.}
  {\bfseries 49} (1977) 435--479}.

\bibitem{Son:2004iv}
D.~T. Son and M.~A. Stephanov, ``{Dynamic universality class of the QCD
  critical point},'' \href{http://dx.doi.org/10.1103/PhysRevD.70.056001}{{\em
  Phys. Rev.} {\bfseries D70} (2004) 056001},
\href{http://arxiv.org/abs/hep-ph/0401052}{{\ttfamily arXiv:hep-ph/0401052
  [hep-ph]}}.

\bibitem{Fujii:2004jt}
H.~Fujii and M.~Ohtani, ``{Sigma and hydrodynamic modes along the critical
  line},'' \href{http://dx.doi.org/10.1103/PhysRevD.70.014016}{{\em Phys. Rev.}
  {\bfseries D70} (2004) 014016},
\href{http://arxiv.org/abs/hep-ph/0402263}{{\ttfamily arXiv:hep-ph/0402263
  [hep-ph]}}.

\bibitem{Stephanov:2008qz}
M.~A. Stephanov, ``{Non-Gaussian fluctuations near the QCD critical point},''
  \href{http://dx.doi.org/10.1103/PhysRevLett.102.032301}{{\em Phys. Rev.
  Lett.} {\bfseries 102} (2009) 032301},
\href{http://arxiv.org/abs/0809.3450}{{\ttfamily arXiv:0809.3450 [hep-ph]}}.

\bibitem{Nouhou:2019nhe}
M.~Agah~Nouhou, M.~Bluhm, A.~Borer, M.~Nahrgang, T.~Sami, and N.~Touroux,
  ``{Finite size effects on cumulants of the critical mode},''
  \href{http://dx.doi.org/10.22323/1.347.0179}{{\em PoS} {\bfseries CORFU2018}
  (2019) 179},
\href{http://arxiv.org/abs/1906.02647}{{\ttfamily arXiv:1906.02647 [nucl-th]}}.

\bibitem{Nahrgang:2011mg}
M.~Nahrgang, S.~Leupold, C.~Herold, and M.~Bleicher, ``{Nonequilibrium chiral
  fluid dynamics including dissipation and noise},''
  \href{http://dx.doi.org/10.1103/PhysRevC.84.024912}{{\em Phys. Rev.}
  {\bfseries C84} (2011) 024912},
\href{http://arxiv.org/abs/1105.0622}{{\ttfamily arXiv:1105.0622 [nucl-th]}}.

\bibitem{Nahrgang:2011mv}
M.~Nahrgang, S.~Leupold, and M.~Bleicher, ``{Equilibration and relaxation times
  at the chiral phase transition including reheating},''
  \href{http://dx.doi.org/10.1016/j.physletb.2012.03.059}{{\em Phys. Lett.}
  {\bfseries B711} (2012) 109--116},
\href{http://arxiv.org/abs/1105.1396}{{\ttfamily arXiv:1105.1396 [nucl-th]}}.

\bibitem{Nahrgang:2011vn}
M.~Nahrgang, C.~Herold, S.~Leupold, I.~Mishustin, and M.~Bleicher, ``{The
  impact of dissipation and noise on fluctuations in chiral fluid dynamics},''
  \href{http://dx.doi.org/10.1088/0954-3899/40/5/055108}{{\em J. Phys.}
  {\bfseries G40} (2013) 055108},
\href{http://arxiv.org/abs/1105.1962}{{\ttfamily arXiv:1105.1962 [nucl-th]}}.

\bibitem{Nahrgang:2013jx}
M.~Nahrgang, C.~Herold, and M.~Bleicher, ``{Influence of an inhomogeneous and
  expanding medium on signals of the QCD phase transition},''
  \href{http://dx.doi.org/10.1016/j.nuclphysa.2013.02.160}{{\em Nucl. Phys.}
  {\bfseries A904-905} (2013) 899c--902c},
\href{http://arxiv.org/abs/1301.2577}{{\ttfamily arXiv:1301.2577 [nucl-th]}}.

\bibitem{Herold:2013bi}
C.~Herold, M.~Nahrgang, I.~Mishustin, and M.~Bleicher, ``{Chiral fluid dynamics
  with explicit propagation of the Polyakov loop},''
  \href{http://dx.doi.org/10.1103/PhysRevC.87.014907}{{\em Phys. Rev.}
  {\bfseries C87} no.~1, (2013) 014907},
\href{http://arxiv.org/abs/1301.1214}{{\ttfamily arXiv:1301.1214 [nucl-th]}}.

\bibitem{Herold:2014zoa}
C.~Herold, M.~Nahrgang, Y.~Yan, and C.~Kobdaj, ``{Net-baryon number variance
  and kurtosis within nonequilibrium chiral fluid dynamics},''
  \href{http://dx.doi.org/10.1088/0954-3899/41/11/115106}{{\em J. Phys.}
  {\bfseries G41} no.~11, (2014) 115106},
\href{http://arxiv.org/abs/1407.8277}{{\ttfamily arXiv:1407.8277 [hep-ph]}}.

\bibitem{Herold:2016uvv}
C.~Herold, M.~Nahrgang, Y.~Yan, and C.~Kobdaj, ``{Dynamical net-proton
  fluctuations near a QCD critical point},''
  \href{http://dx.doi.org/10.1103/PhysRevC.93.021902}{{\em Phys. Rev.}
  {\bfseries C93} no.~2, (2016) 021902},
\href{http://arxiv.org/abs/1601.04839}{{\ttfamily arXiv:1601.04839 [hep-ph]}}.

\bibitem{Herold:2017day}
C.~Herold, M.~Bleicher, M.~Nahrgang, J.~Steinheimer, A.~Limphirat, C.~Kobdaj,
  and Y.~Yan, ``{Broadening of the chiral critical region in a hydrodynamically
  expanding medium},'' \href{http://dx.doi.org/10.1140/epja/i2018-12438-1}{{\em
  Eur. Phys. J.} {\bfseries A54} no.~2, (2018) 19},
\href{http://arxiv.org/abs/1710.03118}{{\ttfamily arXiv:1710.03118 [hep-ph]}}.

\bibitem{Herold:2018ptm}
C.~Herold, A.~Kittiratpattana, C.~Kobdaj, A.~Limphirat, Y.~Yan, M.~Nahrgang,
  J.~Steinheimer, and M.~Bleicher, ``{Entropy production and reheating at the
  chiral phase transition},''
  \href{http://dx.doi.org/10.1016/j.physletb.2019.02.004}{{\em Phys. Lett.}
  {\bfseries B790} (2019) 557--562},
\href{http://arxiv.org/abs/1810.02504}{{\ttfamily arXiv:1810.02504 [hep-ph]}}.

\bibitem{Bluhm:2018qkf}
M.~Bluhm, Y.~Jiang, M.~Nahrgang, J.~M. Pawlowski, F.~Rennecke, and N.~Wink,
  ``{Time-evolution of fluctuations as signal of the phase transition dynamics
  in a QCD-assisted transport approach},''
  \href{http://dx.doi.org/10.1016/j.nuclphysa.2018.09.058}{{\em Nucl. Phys.}
  {\bfseries A982} (2019) 871--874},
\href{http://arxiv.org/abs/1808.01377}{{\ttfamily arXiv:1808.01377 [hep-ph]}}.

\bibitem{Herold:2017omo}
C.~Herold, A.~Limphirat, C.~Kobdaj, Y.~Yan, and M.~Nahrgang, ``{Dynamical
  Fluctuations Near the QCD Critical Point and Their Impact on the Net-proton
  Kurtosis},''
\href{http://dx.doi.org/10.5506/APhysPolBSupp.10.907}{{\em Acta Phys. Polon.
  Supp.} {\bfseries 10} (2017) 907--911}.

\bibitem{Dexheimer:2009hi}
V.~A. Dexheimer and S.~Schramm, ``{A Novel Approach to Model Hybrid Stars},''
  \href{http://dx.doi.org/10.1103/PhysRevC.81.045201}{{\em Phys. Rev.}
  {\bfseries C81} (2010) 045201},
\href{http://arxiv.org/abs/0901.1748}{{\ttfamily arXiv:0901.1748
  [astro-ph.SR]}}.

\bibitem{Nahrgang:2016eou}
M.~Nahrgang and C.~Herold, ``{Phenomena at the QCD phase transition in
  nonequilibrium chiral fluid dynamics (N$\chi$FD)},''
  \href{http://dx.doi.org/10.1140/epja/i2016-16240-9}{{\em Eur. Phys. J.}
  {\bfseries A52} no.~8, (2016) 240},
\href{http://arxiv.org/abs/1602.07223}{{\ttfamily arXiv:1602.07223 [nucl-th]}}.

\bibitem{Stephanov:2017ghc}
M.~Stephanov and Y.~Yin, ``{Hydrodynamics with parametric slowing down and
  fluctuations near the critical point},''
  \href{http://dx.doi.org/10.1103/PhysRevD.98.036006}{{\em Phys. Rev.}
  {\bfseries D98} no.~3, (2018) 036006},
\href{http://arxiv.org/abs/1712.10305}{{\ttfamily arXiv:1712.10305 [nucl-th]}}.

\bibitem{Rajagopal:2019xwg}
K.~Rajagopal, G.~Ridgway, R.~Weller, and Y.~Yin, ``{Hydro+ in Action:
  Understanding the Out-of-Equilibrium Dynamics Near a Critical Point in the
  QCD Phase Diagram},''
\href{http://arxiv.org/abs/1908.08539}{{\ttfamily arXiv:1908.08539 [hep-ph]}}.

\bibitem{Baier:2006gy}
R.~Baier and P.~Romatschke, ``{Causal viscous hydrodynamics for central
  heavy-ion collisions},''
  \href{http://dx.doi.org/10.1140/epjc/s10052-007-0308-5}{{\em Eur. Phys. J.}
  {\bfseries C51} (2007) 677--687},
\href{http://arxiv.org/abs/nucl-th/0610108}{{\ttfamily arXiv:nucl-th/0610108
  [nucl-th]}}.

\bibitem{Akamatsu:2018vjr}
Y.~Akamatsu, D.~Teaney, F.~Yan, and Y.~Yin, ``{Transits of the QCD Critical
  Point},'' \href{http://dx.doi.org/10.1103/PhysRevC.100.044901}{{\em Phys.
  Rev.} {\bfseries C100} no.~4, (2019) 044901},
\href{http://arxiv.org/abs/1811.05081}{{\ttfamily arXiv:1811.05081 [nucl-th]}}.

\bibitem{Donev:2009}
A.~Donev, ``Asynchronous event-driven particle algorithms,'' {\em CoRR}
  {\bfseries abs/cs/0703096} (2007) ,
  \href{http://arxiv.org/abs/cs/0703096}{{\ttfamily arXiv:cs/0703096}}.
  \url{http://arxiv.org/abs/cs/0703096}.

\bibitem{Huovinen:2012is}
P.~Huovinen and H.~Petersen, ``{Particlization in hybrid models},''
  \href{http://dx.doi.org/10.1140/epja/i2012-12171-9}{{\em Eur. Phys. J.}
  {\bfseries A48} (2012) 171},
\href{http://arxiv.org/abs/1206.3371}{{\ttfamily arXiv:1206.3371 [nucl-th]}}.

\bibitem{Steinheimer:2017dpb}
J.~Steinheimer and V.~Koch, ``{Effect of finite particle number sampling on
  baryon number fluctuations},''
  \href{http://dx.doi.org/10.1103/PhysRevC.96.034907}{{\em Phys. Rev.}
  {\bfseries C96} no.~3, (2017) 034907},
\href{http://arxiv.org/abs/1705.08538}{{\ttfamily arXiv:1705.08538 [nucl-th]}}.

\bibitem{Fraga:2003mu}
E.~S. Fraga and R.~Venugopalan, ``{Finite size effects on nucleation in a first
  order phase transition},''
  \href{http://dx.doi.org/10.1016/j.physa.2004.07.045}{{\em Physica} {\bfseries
  A345} (2004) 121--129},
\href{http://arxiv.org/abs/hep-ph/0304094}{{\ttfamily arXiv:hep-ph/0304094
  [hep-ph]}}.

\bibitem{Palhares:2009tf}
L.~F. Palhares, E.~S. Fraga, and T.~Kodama, ``{Chiral transition in a finite
  system and possible use of finite size scaling in relativistic heavy ion
  collisions},'' \href{http://dx.doi.org/10.1088/0954-3899/38/8/085101}{{\em J.
  Phys.} {\bfseries G38} (2011) 085101},
\href{http://arxiv.org/abs/0904.4830}{{\ttfamily arXiv:0904.4830 [nucl-th]}}.

\bibitem{Fraga:2011hi}
E.~S. Fraga, L.~F. Palhares, and P.~Sorensen, ``{Finite-size scaling as a tool
  in the search for the QCD critical point in heavy ion data},''
  \href{http://dx.doi.org/10.1103/PhysRevC.84.011903}{{\em Phys. Rev.}
  {\bfseries C84} (2011) 011903},
\href{http://arxiv.org/abs/1104.3755}{{\ttfamily arXiv:1104.3755 [hep-ph]}}.

\bibitem{Hippert:2015rwa}
M.~Hippert, E.~S. Fraga, and E.~M. Santos, ``{Critical versus spurious
  fluctuations in the search for the QCD critical point},''
  \href{http://dx.doi.org/10.1103/PhysRevD.93.014029}{{\em Phys. Rev.}
  {\bfseries D93} no.~1, (2016) 014029},
  \href{http://arxiv.org/abs/1507.04764}{{\ttfamily arXiv:1507.04764
  [hep-ph]}}.
[Phys. Rev.D93,014029(2016)].

\bibitem{Hippert:2017xoj}
M.~Hippert and E.~S. Fraga, ``{Multiplicity fluctuations near the QCD critical
  point},'' \href{http://dx.doi.org/10.1103/PhysRevD.96.034011}{{\em Phys.
  Rev.} {\bfseries D96} no.~3, (2017) 034011},
\href{http://arxiv.org/abs/1702.02028}{{\ttfamily arXiv:1702.02028 [hep-ph]}}.

\bibitem{Koch:2008ia}
V.~Koch, \href{http://dx.doi.org/10.1007/978-3-642-01539-7_20}{``{Hadronic
  Fluctuations and Correlations},''} in {\em Relativistic Heavy Ion Physics},
  R.~Stock, ed., pp.~626--652.
\newblock 2010.
\newblock \href{http://arxiv.org/abs/0810.2520}{{\ttfamily arXiv:0810.2520
  [nucl-th]}}.
\newblock
\url{http://materials.springer.com/lb/docs/sm_lbs_978-3-642-01539-7_20}.
\newblock

\bibitem{Philipsen:2010gj}
O.~Philipsen, ``{Lattice QCD at non-zero temperature and baryon density},'' in
  {\em {Modern perspectives in lattice QCD: Quantum field theory and high
  performance computing. Proceedings, International School, 93rd Session, Les
  Houches, France, August 3-28, 2009}}, pp.~273--330.
\newblock 2010.
\newblock
\href{http://arxiv.org/abs/1009.4089}{{\ttfamily arXiv:1009.4089 [hep-lat]}}.
\newblock

\bibitem{Karsch:2010ck}
F.~Karsch and K.~Redlich, ``{Probing freeze-out conditions in heavy ion
  collisions with moments of charge fluctuations},''
  \href{http://dx.doi.org/10.1016/j.physletb.2010.10.046}{{\em Phys. Lett.}
  {\bfseries B695} (2011) 136--142},
\href{http://arxiv.org/abs/1007.2581}{{\ttfamily arXiv:1007.2581 [hep-ph]}}.

\bibitem{Karsch:2012wm}
F.~Karsch, ``{Determination of Freeze-out Conditions from Lattice QCD
  Calculations},'' \href{http://dx.doi.org/10.2478/s11534-012-0074-3}{{\em
  Central Eur. J. Phys.} {\bfseries 10} (2012) 1234--1237},
\href{http://arxiv.org/abs/1202.4173}{{\ttfamily arXiv:1202.4173 [hep-lat]}}.

\bibitem{Bzdak:2016sxg}
A.~Bzdak, V.~Koch, and N.~Strodthoff, ``{Cumulants and correlation functions
  versus the QCD phase diagram},''
  \href{http://dx.doi.org/10.1103/PhysRevC.95.054906}{{\em Phys. Rev.}
  {\bfseries C95} no.~5, (2017) 054906},
\href{http://arxiv.org/abs/1607.07375}{{\ttfamily arXiv:1607.07375 [nucl-th]}}.

\bibitem{Abelev:2009ai}
{\bfseries STAR} Collaboration, B.~I. Abelev {\em et~al.}, ``{K/pi Fluctuations
  at Relativistic Energies},''
  \href{http://dx.doi.org/10.1103/PhysRevLett.103.092301}{{\em Phys. Rev.
  Lett.} {\bfseries 103} (2009) 092301},
\href{http://arxiv.org/abs/0901.1795}{{\ttfamily arXiv:0901.1795 [nucl-ex]}}.

\bibitem{Acharya:2017cpf}
{\bfseries ALICE} Collaboration, S.~Acharya {\em et~al.}, ``{Relative particle
  yield fluctuations in $\text{ Pb-Pb }$ collisions at $\sqrt{s_\mathrm{{NN}}}
  =2.76\hbox { TeV}$},''
  \href{http://dx.doi.org/10.1140/epjc/s10052-019-6711-x}{{\em Eur. Phys. J.}
  {\bfseries C79} no.~3, (2019) 236},
\href{http://arxiv.org/abs/1712.07929}{{\ttfamily arXiv:1712.07929 [nucl-ex]}}.

\bibitem{Koch:2009dg}
V.~Koch and T.~Schuster, ``{On the energy dependence of K/pi fluctuations in
  relativistic heavy ion collisions},''
  \href{http://dx.doi.org/10.1103/PhysRevC.81.034910}{{\em Phys. Rev.}
  {\bfseries C81} (2010) 034910},
\href{http://arxiv.org/abs/0911.1160}{{\ttfamily arXiv:0911.1160 [nucl-th]}}.

\bibitem{Pruneau:2019baa}
C.~A. Pruneau, ``{The Role of Baryon Number Conservation in Measurements of
  Fluctuations},''
\href{http://arxiv.org/abs/1903.04591}{{\ttfamily arXiv:1903.04591 [nucl-th]}}.

\bibitem{Gazdzicki:1992ri}
M.~Gazdzicki and S.~Mrowczynski, ``{A Method to study 'equilibration' in
  nucleus-nucleus collisions},''
\href{http://dx.doi.org/10.1007/BF01881715}{{\em Z. Phys.} {\bfseries C54}
  (1992) 127--132}.

\bibitem{Gorenstein:2011vq}
M.~I. Gorenstein and M.~Gazdzicki, ``{Strongly Intensive Quantities},''
  \href{http://dx.doi.org/10.1103/PhysRevC.84.014904}{{\em Phys. Rev.}
  {\bfseries C84} (2011) 014904},
\href{http://arxiv.org/abs/1101.4865}{{\ttfamily arXiv:1101.4865 [nucl-th]}}.

\bibitem{Gazdzicki:2013ana}
M.~Gazdzicki, M.~I. Gorenstein, and M.~Mackowiak-Pawlowska, ``{Normalization of
  strongly intensive quantities},''
  \href{http://dx.doi.org/10.1103/PhysRevC.88.024907}{{\em Phys. Rev.}
  {\bfseries C88} no.~2, (2013) 024907},
\href{http://arxiv.org/abs/1303.0871}{{\ttfamily arXiv:1303.0871 [nucl-th]}}.

\bibitem{Bialas:1976ed}
A.~Bialas, M.~Bleszynski, and W.~Czyz, ``{Multiplicity Distributions in
  Nucleus-Nucleus Collisions at High-Energies},''
\href{http://dx.doi.org/10.1016/0550-3213(76)90329-1}{{\em Nucl. Phys.}
  {\bfseries B111} (1976) 461--476}.

\bibitem{Bialas:1985jb}
A.~Bialas and R.~B. Peschanski, ``{Moments of Rapidity Distributions as a
  Measure of Short Range Fluctuations in High-Energy Collisions},''
\href{http://dx.doi.org/10.1016/0550-3213(86)90386-X}{{\em Nucl. Phys.}
  {\bfseries B273} (1986) 703--718}.

\bibitem{Bialas:1990xd}
A.~Bialas and R.~C. Hwa, ``{Intermittency parameters as a possible signal for
  quark - gluon plasma formation},''
\href{http://dx.doi.org/10.1016/0370-2693(91)91747-J}{{\em Phys. Lett.}
  {\bfseries B253} (1991) 436--438}.

\bibitem{Berdnikov:1999ph}
B.~Berdnikov and K.~Rajagopal, ``{Slowing out-of-equilibrium near the QCD
  critical point},'' \href{http://dx.doi.org/10.1103/PhysRevD.61.105017}{{\em
  Phys. Rev.} {\bfseries D61} (2000) 105017},
\href{http://arxiv.org/abs/hep-ph/9912274}{{\ttfamily arXiv:hep-ph/9912274
  [hep-ph]}}.

\bibitem{Sun:2017xrx}
K.-J. Sun, L.-W. Chen, C.~M. Ko, and Z.~Xu, ``{Probing QCD critical
  fluctuations from light nuclei production in relativistic heavy-ion
  collisions},'' \href{http://dx.doi.org/10.1016/j.physletb.2017.09.056}{{\em
  Phys. Lett.} {\bfseries B774} (2017) 103--107},
\href{http://arxiv.org/abs/1702.07620}{{\ttfamily arXiv:1702.07620 [nucl-th]}}.

\bibitem{Sun:2018jhg}
K.-J. Sun, L.-W. Chen, C.~M. Ko, J.~Pu, and Z.~Xu, ``{Light nuclei production
  as a probe of the QCD phase diagram},''
  \href{http://dx.doi.org/10.1016/j.physletb.2018.04.035}{{\em Phys. Lett.}
  {\bfseries B781} (2018) 499--504},
\href{http://arxiv.org/abs/1801.09382}{{\ttfamily arXiv:1801.09382 [nucl-th]}}.

\bibitem{Anticic:2010mp}
{\bfseries NA49} Collaboration, T.~Anticic {\em et~al.}, ``{Centrality
  dependence of proton and antiproton spectra in Pb+Pb collisions at 40A GeV
  and 158A GeV measured at the CERN SPS},''
  \href{http://dx.doi.org/10.1103/PhysRevC.83.014901}{{\em Phys. Rev.}
  {\bfseries C83} (2011) 014901},
\href{http://arxiv.org/abs/1009.1747}{{\ttfamily arXiv:1009.1747 [nucl-ex]}}.

\bibitem{Blume:2007kw}
{\bfseries Na49} Collaboration, C.~Blume, ``{Centrality and energy dependence
  of proton, light fragment and hyperon production},''
  \href{http://dx.doi.org/10.1088/0954-3899/34/8/S133}{{\em J. Phys.}
  {\bfseries G34} (2007) S951--954},
\href{http://arxiv.org/abs/nucl-ex/0701042}{{\ttfamily arXiv:nucl-ex/0701042
  [nucl-ex]}}.

\bibitem{Anticic:2016ckv}
{\bfseries NA49} Collaboration, T.~Anticic {\em et~al.}, ``{Production of
  deuterium, tritium, and He3 in central Pb + Pb collisions at 20A,30A,40A,80A
  , and 158A GeV at the CERN Super Proton Synchrotron},''
  \href{http://dx.doi.org/10.1103/PhysRevC.94.044906}{{\em Phys. Rev.}
  {\bfseries C94} no.~4, (2016) 044906},
\href{http://arxiv.org/abs/1606.04234}{{\ttfamily arXiv:1606.04234 [nucl-ex]}}.

\bibitem{Adam:2019wnb}
{\bfseries STAR} Collaboration, J.~Adam {\em et~al.}, ``{Beam energy dependence
  of (anti-)deuteron production in Au + Au collisions at the BNL Relativistic
  Heavy Ion Collider},''
  \href{http://dx.doi.org/10.1103/PhysRevC.99.064905}{{\em Phys. Rev.}
  {\bfseries C99} no.~6, (2019) 064905},
\href{http://arxiv.org/abs/1903.11778}{{\ttfamily arXiv:1903.11778 [nucl-ex]}}.

\bibitem{Zhang:2019wun}
{\bfseries STAR} Collaboration, D.~Zhang, ``{Energy Dependence of Light Nuclei
  ($d$, $t$) Production at STAR},''
\href{http://arxiv.org/abs/1909.07028}{{\ttfamily arXiv:1909.07028 [nucl-ex]}}.

\bibitem{Adam:2015vda}
{\bfseries ALICE} Collaboration, J.~Adam {\em et~al.}, ``{Production of light
  nuclei and anti-nuclei in pp and Pb-Pb collisions at energies available at
  the CERN Large Hadron Collider},''
  \href{http://dx.doi.org/10.1103/PhysRevC.93.024917}{{\em Phys. Rev.}
  {\bfseries C93} no.~2, (2016) 024917},
\href{http://arxiv.org/abs/1506.08951}{{\ttfamily arXiv:1506.08951 [nucl-ex]}}.

\bibitem{Shuryak:2019ikv}
E.~Shuryak and J.~M. Torres-Rincon, ``{Baryon preclustering at the freeze-out
  of heavy-ion collisions and light-nuclei production},''
\href{http://arxiv.org/abs/1910.08119}{{\ttfamily arXiv:1910.08119 [nucl-th]}}.

\bibitem{Bazavov:2012jq}
{\bfseries HotQCD} Collaboration, A.~Bazavov {\em et~al.}, ``{Fluctuations and
  Correlations of net baryon number, electric charge, and strangeness: A
  comparison of lattice QCD results with the hadron resonance gas model},''
  \href{http://dx.doi.org/10.1103/PhysRevD.86.034509}{{\em Phys. Rev.}
  {\bfseries D86} (2012) 034509},
\href{http://arxiv.org/abs/1203.0784}{{\ttfamily arXiv:1203.0784 [hep-lat]}}.

\bibitem{Nahrgang:2014fza}
M.~Nahrgang, M.~Bluhm, P.~Alba, R.~Bellwied, and C.~Ratti, ``{Impact of
  resonance regeneration and decay on the net-proton fluctuations in a hadron
  resonance gas},''
  \href{http://dx.doi.org/10.1140/epjc/s10052-015-3775-0}{{\em Eur. Phys. J.}
  {\bfseries C75} no.~12, (2015) 573},
\href{http://arxiv.org/abs/1402.1238}{{\ttfamily arXiv:1402.1238 [hep-ph]}}.

\bibitem{Luo:2013bmi}
X.~Luo, J.~Xu, B.~Mohanty, and N.~Xu, ``{Volume fluctuation and
  auto-correlation effects in the moment analysis of net-proton multiplicity
  distributions in heavy-ion collisions},''
  \href{http://dx.doi.org/10.1088/0954-3899/40/10/105104}{{\em J. Phys.}
  {\bfseries G40} (2013) 105104},
\href{http://arxiv.org/abs/1302.2332}{{\ttfamily arXiv:1302.2332 [nucl-ex]}}.

\bibitem{Braun-Munzinger:2016yjz}
P.~Braun-Munzinger, A.~Rustamov, and J.~Stachel, ``{Bridging the gap between
  event-by-event fluctuation measurements and theory predictions in
  relativistic nuclear collisions},''
  \href{http://dx.doi.org/10.1016/j.nuclphysa.2017.01.011}{{\em Nucl. Phys.}
  {\bfseries A960} (2017) 114--130},
\href{http://arxiv.org/abs/1612.00702}{{\ttfamily arXiv:1612.00702 [nucl-th]}}.

\bibitem{Braun-Munzinger:2019yxj}
P.~Braun-Munzinger, A.~Rustamov, and J.~Stachel, ``{The role of the local
  conservation laws in fluctuations of conserved charges},''
\href{http://arxiv.org/abs/1907.03032}{{\ttfamily arXiv:1907.03032 [nucl-th]}}.

\bibitem{Ling:2015yau}
B.~Ling and M.~A. Stephanov, ``{Acceptance dependence of fluctuation measures
  near the QCD critical point},''
  \href{http://dx.doi.org/10.1103/PhysRevC.93.034915}{{\em Phys. Rev.}
  {\bfseries C93} no.~3, (2016) 034915},
\href{http://arxiv.org/abs/1512.09125}{{\ttfamily arXiv:1512.09125 [nucl-th]}}.

\bibitem{Stephanov:1999zu}
M.~A. Stephanov, K.~Rajagopal, and E.~V. Shuryak, ``{Event-by-event
  fluctuations in heavy ion collisions and the QCD critical point},''
  \href{http://dx.doi.org/10.1103/PhysRevD.60.114028}{{\em Phys. Rev.}
  {\bfseries D60} (1999) 114028},
\href{http://arxiv.org/abs/hep-ph/9903292}{{\ttfamily arXiv:hep-ph/9903292
  [hep-ph]}}.

\bibitem{Kitazawa:2015ira}
M.~Kitazawa, ``{Rapidity window dependences of higher order cumulants and
  diffusion master equation},''
  \href{http://dx.doi.org/10.1016/j.nuclphysa.2015.07.008}{{\em Nucl. Phys.}
  {\bfseries A942} (2015) 65--96},
\href{http://arxiv.org/abs/1505.04349}{{\ttfamily arXiv:1505.04349 [nucl-th]}}.

\bibitem{Bzdak:2017ltv}
A.~Bzdak and V.~Koch, ``{Rapidity dependence of proton cumulants and
  correlation functions},''
  \href{http://dx.doi.org/10.1103/PhysRevC.96.054905}{{\em Phys. Rev.}
  {\bfseries C96} no.~5, (2017) 054905},
\href{http://arxiv.org/abs/1707.02640}{{\ttfamily arXiv:1707.02640 [nucl-th]}}.

\bibitem{Andronic:2017pug}
A.~Andronic, P.~Braun-Munzinger, K.~Redlich, and J.~Stachel, ``{Decoding the
  phase structure of QCD via particle production at high energy},''
  \href{http://dx.doi.org/10.1038/s41586-018-0491-6}{{\em Nature} {\bfseries
  561} no.~7723, (2018) 321--330},
\href{http://arxiv.org/abs/1710.09425}{{\ttfamily arXiv:1710.09425 [nucl-th]}}.

\bibitem{Begun:2006jf}
V.~V. Begun, M.~I. Gorenstein, M.~Hauer, V.~P. Konchakovski, and O.~S. Zozulya,
  ``{Multiplicity Fluctuations in Hadron-Resonance Gas},''
  \href{http://dx.doi.org/10.1103/PhysRevC.74.044903}{{\em Phys. Rev.}
  {\bfseries C74} (2006) 044903},
\href{http://arxiv.org/abs/nucl-th/0606036}{{\ttfamily arXiv:nucl-th/0606036
  [nucl-th]}}.

\bibitem{Fu:2013gga}
J.~Fu, ``{Higher moments of net-proton multiplicity distributions in heavy ion
  collisions at chemical freeze-out},''
\href{http://dx.doi.org/10.1016/j.physletb.2013.04.018}{{\em Phys. Lett.}
  {\bfseries B722} (2013) 144--150}.

\bibitem{Jeon:1999gr}
S.~Jeon and V.~Koch, ``{Fluctuations of particle ratios and the abundance of
  hadronic resonances},''
  \href{http://dx.doi.org/10.1103/PhysRevLett.83.5435}{{\em Phys. Rev. Lett.}
  {\bfseries 83} (1999) 5435--5438},
\href{http://arxiv.org/abs/nucl-th/9906074}{{\ttfamily arXiv:nucl-th/9906074
  [nucl-th]}}.

\bibitem{Ohlson:2019erm}
{\bfseries ALICE} Collaboration, A.~Ohlson, ``{Investigating correlated
  fluctuations of conserved charges with net-$\Lambda$ fluctuations in Pb-Pb
  collisions at ALICE},''
  \href{http://dx.doi.org/10.1016/j.nuclphysa.2018.11.020}{{\em Nucl. Phys.}
  {\bfseries A982} (2019) 299--302},
\href{http://arxiv.org/abs/1901.00744}{{\ttfamily arXiv:1901.00744 [nucl-ex]}}.

\bibitem{Gazdzicki:2011xz}
M.~Gazdzicki, K.~Grebieszkow, M.~Mackowiak, and S.~Mrowczynski, ``{Identity
  method to study chemical fluctuations in relativistic heavy-ion
  collisions},'' \href{http://dx.doi.org/10.1103/PhysRevC.83.054907}{{\em Phys.
  Rev.} {\bfseries C83} (2011) 054907},
\href{http://arxiv.org/abs/1103.2887}{{\ttfamily arXiv:1103.2887 [nucl-th]}}.

\bibitem{Gorenstein:2011hr}
M.~I. Gorenstein, ``{Identity Method for Particle Number Fluctuations and
  Correlations},'' \href{http://dx.doi.org/10.1103/PhysRevC.84.024902,
  10.1103/PhysRevC.97.029903}{{\em Phys. Rev.} {\bfseries C84} (2011) 024902},
  \href{http://arxiv.org/abs/1106.4473}{{\ttfamily arXiv:1106.4473 [nucl-th]}}.
[Erratum: Phys. Rev.C97,no.2,029903(2018)].

\bibitem{Rustamov:2012bx}
A.~Rustamov and M.~I. Gorenstein, ``{Identity Method for Moments of
  Multiplicity Distribution},''
  \href{http://dx.doi.org/10.1103/PhysRevC.86.044906}{{\em Phys. Rev.}
  {\bfseries C86} (2012) 044906},
\href{http://arxiv.org/abs/1204.6632}{{\ttfamily arXiv:1204.6632 [nucl-th]}}.

\bibitem{Behera:2018wqk}
{\bfseries ALICE} Collaboration, N.~K. Behera, ``{Higher moment fluctuations of
  identified particle distributions from ALICE},''
  \href{http://dx.doi.org/10.1016/j.nuclphysa.2018.11.030}{{\em Nucl. Phys.}
  {\bfseries A982} (2019) 851--854},
\href{http://arxiv.org/abs/1807.06780}{{\ttfamily arXiv:1807.06780 [hep-ex]}}.

\bibitem{Abelev:2012pv}
{\bfseries ALICE} Collaboration, B.~Abelev {\em et~al.}, ``{Net-Charge
  Fluctuations in Pb-Pb collisions at $\sqrt{s}_{NN} = 2.76$ TeV},''
  \href{http://dx.doi.org/10.1103/PhysRevLett.110.152301}{{\em Phys. Rev.
  Lett.} {\bfseries 110} no.~15, (2013) 152301},
\href{http://arxiv.org/abs/1207.6068}{{\ttfamily arXiv:1207.6068 [nucl-ex]}}.

\bibitem{Abelev:2013csa}
{\bfseries ALICE} Collaboration, B.~Abelev {\em et~al.}, ``{Charge correlations
  using the balance function in Pb-Pb collisions at $\sqrt{s_{NN}}$ = 2.76
  TeV},'' \href{http://dx.doi.org/10.1016/j.physletb.2013.05.039}{{\em Phys.
  Lett.} {\bfseries B723} (2013) 267--279},
\href{http://arxiv.org/abs/1301.3756}{{\ttfamily arXiv:1301.3756 [nucl-ex]}}.

\bibitem{Adam:2015gda}
{\bfseries ALICE} Collaboration, J.~Adam {\em et~al.}, ``{Multiplicity and
  transverse momentum evolution of charge-dependent correlations in pp, p–Pb,
  and Pb–Pb collisions at the LHC},''
  \href{http://dx.doi.org/10.1140/epjc/s10052-016-3915-1}{{\em Eur. Phys. J.}
  {\bfseries C76} no.~2, (2016) 86},
\href{http://arxiv.org/abs/1509.07255}{{\ttfamily arXiv:1509.07255 [nucl-ex]}}.

\bibitem{Nonaka:2017kko}
T.~Nonaka, M.~Kitazawa, and S.~Esumi, ``{More efficient formulas for efficiency
  correction of cumulants and effect of using averaged efficiency},''
  \href{http://dx.doi.org/10.1103/PhysRevC.95.064912}{{\em Phys. Rev.}
  {\bfseries C95} no.~6, (2017) 064912},
\href{http://arxiv.org/abs/1702.07106}{{\ttfamily arXiv:1702.07106
  [physics.data-an]}}.

\bibitem{Pruneau:2017fim}
C.~A. Pruneau, ``{Identity method reexamined},''
  \href{http://dx.doi.org/10.1103/PhysRevC.96.054902}{{\em Phys. Rev.}
  {\bfseries C96} no.~5, (2017) 054902},
\href{http://arxiv.org/abs/1706.01333}{{\ttfamily arXiv:1706.01333
  [physics.data-an]}}.

\bibitem{Abelev:2008jg}
{\bfseries STAR} Collaboration, B.~I. Abelev {\em et~al.}, ``{Beam-Energy and
  System-Size Dependence of Dynamical Net Charge Fluctuations},''
  \href{http://dx.doi.org/10.1103/PhysRevC.79.024906}{{\em Phys. Rev.}
  {\bfseries C79} (2009) 024906},
\href{http://arxiv.org/abs/0807.3269}{{\ttfamily arXiv:0807.3269 [nucl-ex]}}.

\bibitem{Aggarwal:2010wy}
{\bfseries STAR} Collaboration, M.~M. Aggarwal {\em et~al.}, ``{Higher Moments
  of Net-proton Multiplicity Distributions at RHIC},''
  \href{http://dx.doi.org/10.1103/PhysRevLett.105.022302}{{\em Phys. Rev.
  Lett.} {\bfseries 105} (2010) 022302},
\href{http://arxiv.org/abs/1004.4959}{{\ttfamily arXiv:1004.4959 [nucl-ex]}}.

\bibitem{Adamczyk:2017wsl}
{\bfseries STAR} Collaboration, L.~Adamczyk {\em et~al.}, ``{Collision Energy
  Dependence of Moments of Net-Kaon Multiplicity Distributions at RHIC},''
  \href{http://dx.doi.org/10.1016/j.physletb.2018.07.066}{{\em Phys. Lett.}
  {\bfseries B785} (2018) 551--560},
\href{http://arxiv.org/abs/1709.00773}{{\ttfamily arXiv:1709.00773 [nucl-ex]}}.

\bibitem{Abdelwahab:2014yha}
{\bfseries STAR} Collaboration, N.~M. Abdelwahab {\em et~al.}, ``{Energy
  Dependence of $K/\pi$, $p/\pi$, and $K/p$ Fluctuations in Au+Au Collisions
  from $\rm \sqrt{s_{NN}}$ = 7.7 to 200 GeV},''
  \href{http://dx.doi.org/10.1103/PhysRevC.92.021901}{{\em Phys. Rev.}
  {\bfseries C92} no.~2, (2015) 021901},
\href{http://arxiv.org/abs/1410.5375}{{\ttfamily arXiv:1410.5375 [nucl-ex]}}.

\bibitem{Adam:2019rsf}
{\bfseries STAR} Collaboration, J.~Adam {\em et~al.}, ``{Collision-energy
  dependence of $p_t$ correlations in Au + Au collisions at energies available
  at the BNL Relativistic Heavy Ion Collider},''
  \href{http://dx.doi.org/10.1103/PhysRevC.99.044918}{{\em Phys. Rev.}
  {\bfseries C99} no.~4, (2019) 044918},
\href{http://arxiv.org/abs/1901.00837}{{\ttfamily arXiv:1901.00837 [nucl-ex]}}.

\bibitem{Adamczyk:2015yga}
{\bfseries STAR} Collaboration, L.~Adamczyk {\em et~al.}, ``{Beam-energy
  dependence of charge balance functions from Au + Au collisions at energies
  available at the BNL Relativistic Heavy Ion Collider},''
  \href{http://dx.doi.org/10.1103/PhysRevC.94.024909}{{\em Phys. Rev.}
  {\bfseries C94} no.~2, (2016) 024909},
\href{http://arxiv.org/abs/1507.03539}{{\ttfamily arXiv:1507.03539 [nucl-ex]}}.

\bibitem{Luo:2015ewa}
{\bfseries STAR} Collaboration, X.~Luo, ``{Energy Dependence of Moments of
  Net-Proton and Net-Charge Multiplicity Distributions at STAR},''
  \href{http://dx.doi.org/10.22323/1.217.0019}{{\em PoS} {\bfseries CPOD2014}
  (2015) 019},
\href{http://arxiv.org/abs/1503.02558}{{\ttfamily arXiv:1503.02558 [nucl-ex]}}.

\bibitem{Skokov:2013abc}
V.~Skokov, B.~Friman, and K.~Redlich, ``Volume fluctuations and higher-order
  cumulants of the net baryon number,''
  \href{http://dx.doi.org/10.1103/physrevc.88.034911}{{\em Phys. Rev. C}
  {\bfseries 88} no.~3, (2013) 034911},
  \href{http://arxiv.org/abs/arXiv:1205.4756 [hep-ph]}{{\ttfamily
  arXiv:1205.4756 [hep-ph]}}.

\bibitem{Bzdak:2013pha}
A.~Bzdak and V.~Koch, ``{Local Efficiency Corrections to Higher Order
  Cumulants},'' \href{http://dx.doi.org/10.1103/PhysRevC.91.027901}{{\em Phys.
  Rev.} {\bfseries C91} no.~2, (2015) 027901},
\href{http://arxiv.org/abs/1312.4574}{{\ttfamily arXiv:1312.4574 [nucl-th]}}.

\bibitem{Grebieszkow:2019yjd}
{\bfseries NA61/SHINE} Collaboration, K.~Grebieszkow, ``{New results on spectra
  and fluctuations from NA61/SHINE},''
  \href{http://dx.doi.org/10.22323/1.347.0152}{{\em PoS} {\bfseries CORFU2018}
  (2019) 152},
\href{http://arxiv.org/abs/1904.03165}{{\ttfamily arXiv:1904.03165 [nucl-ex]}}.

\bibitem{Czopowicz:2015mfa}
{\bfseries NA61/SHINE} Collaboration, T.~Czopowicz, ``{Transverse momentum and
  multiplicity fluctuations in Be+Be energy scan from NA61/SHINE},''
  \href{http://dx.doi.org/10.22323/1.217.0054}{{\em PoS} {\bfseries CPOD2014}
  (2015) 054},
\href{http://arxiv.org/abs/1503.01619}{{\ttfamily arXiv:1503.01619 [nucl-ex]}}.

\bibitem{Aduszkiewicz:2015jna}
{\bfseries NA61/SHINE} Collaboration, A.~Aduszkiewicz {\em et~al.},
  ``{Multiplicity and transverse momentum fluctuations in inelastic
  proton–proton interactions at the CERN Super Proton Synchrotron},''
  \href{http://dx.doi.org/10.1140/epjc/s10052-016-4450-9}{{\em Eur. Phys. J.}
  {\bfseries C76} no.~11, (2016) 635},
\href{http://arxiv.org/abs/1510.00163}{{\ttfamily arXiv:1510.00163 [hep-ex]}}.

\bibitem{Seryakov:2017sss}
{\bfseries NA61/SHINE} Collaboration, A.~Seryakov, ``{Multiplicity fluctuations
  in Ar+Sc collisions at the CERN SPS from NA61/SHINE},''
  \href{http://dx.doi.org/10.5506/APhysPolBSupp.10.723}{{\em Acta Phys. Polon.
  Supp.} {\bfseries 10} (2017) 723},
\href{http://arxiv.org/abs/1704.00751}{{\ttfamily arXiv:1704.00751 [hep-ex]}}.

\bibitem{Begun:2006uu}
V.~V. Begun, M.~Gazdzicki, M.~I. Gorenstein, M.~Hauer, V.~P. Konchakovski, and
  B.~Lungwitz, ``{Multiplicity fluctuations in relativistic nuclear collisions:
  Statistical model versus experimental data},''
  \href{http://dx.doi.org/10.1103/PhysRevC.76.024902}{{\em Phys. Rev.}
  {\bfseries C76} (2007) 024902},
\href{http://arxiv.org/abs/nucl-th/0611075}{{\ttfamily arXiv:nucl-th/0611075
  [nucl-th]}}.

\bibitem{Aduszkiewicz:2287091}
{\bfseries NA61/SHINE} Collaboration, A.~Aduszkiewicz, ``{Report from the
  NA61/SHINE experiment at the CERN SPS},'' Tech. Rep. CERN-SPSC-2017-038.
  SPSC-SR-221, CERN, Geneva, Oct, 2017.
\newblock \url{https://cds.cern.ch/record/2287091}.

\bibitem{Andronov:2016ddd}
{\bfseries NA61/SHINE} Collaboration, E.~Andronov, ``{Transverse momentum and
  multiplicity fluctuations in Ar+Sc collisions at the CERN SPS from
  NA61/SHINE},'' \href{http://dx.doi.org/10.5506/APhysPolBSupp.10.449}{{\em
  Acta Phys. Polon. Supp.} {\bfseries 10} (2017) 449--453},
\href{http://arxiv.org/abs/1710.06197}{{\ttfamily arXiv:1710.06197 [nucl-ex]}}.

\bibitem{Davis:2017puj}
{\bfseries NA61/SHINE} Collaboration, N.~Davis, N.~Antoniou, and F.~Diakonos,
  ``{Search for the critical point of strongly interacting matter through
  power-law fluctuations of the proton density in NA61/SHINE},''
\href{http://dx.doi.org/10.22323/1.311.0054}{{\em PoS} {\bfseries CPOD2017}
  (2018) 054}.

\bibitem{MMP:QM2019}
{\bfseries NA61/SHINE} Collaboration, M.~Mackowiak-Pawlowska, ``{NA61/SHINE
  results on fluctuations and correlations at CERN SPS energies},'' {\em to
  appear in the proceedings of QM2019} .

\bibitem{Anticic:2012xb}
{\bfseries NA49} Collaboration, T.~Anticic {\em et~al.}, ``{Critical
  fluctuations of the proton density in A+A collisions at 158$A$ GeV},''
  \href{http://dx.doi.org/10.1140/epjc/s10052-015-3738-5}{{\em Eur. Phys. J.}
  {\bfseries C75} no.~12, (2015) 587},
\href{http://arxiv.org/abs/1208.5292}{{\ttfamily arXiv:1208.5292 [nucl-ex]}}.

\bibitem{OHara:2002pqs}
K.~M. O'Hara, S.~L. Hemmer, M.~E. Gehm, S.~R. Granade, and J.~E. Thomas,
  ``{Observation of a Strongly Interacting Degenerate Fermi Gas of Atoms},''
  \href{http://dx.doi.org/10.1126/science.1079107}{{\em Science} {\bfseries
  298} (2002) 2179--2182},
\href{http://arxiv.org/abs/cond-mat/0212463}{{\ttfamily arXiv:cond-mat/0212463
  [cond-mat.supr-con]}}.

\bibitem{Cao:2010wa}
C.~Cao, E.~Elliott, J.~Joseph, H.~Wu, J.~Petricka, T.~Schäfer, and J.~E.
  Thomas, ``{Universal Quantum Viscosity in a Unitary Fermi Gas},''
  \href{http://dx.doi.org/10.1126/science.1195219}{{\em Science} {\bfseries
  331} (2011) 58},
\href{http://arxiv.org/abs/1007.2625}{{\ttfamily arXiv:1007.2625
  [cond-mat.quant-gas]}}.

\bibitem{Gaebler_2010}
J.~P. Gaebler, J.~T. Stewart, T.~E. Drake, D.~S. Jin, A.~Perali, P.~Pieri, and
  G.~C. Strinati, ``Observation of pseudogap behaviour in a strongly
  interacting fermi gas,'' \href{http://dx.doi.org/10.1038/nphys1709}{{\em
  Nature Physics} {\bfseries 6} no.~8, (Jul, 2010) 569–573}.
  \url{http://dx.doi.org/10.1038/nphys1709}.

\bibitem{Tan_2008}
S.~Tan, ``Generalized virial theorem and pressure relation for a strongly
  correlated fermi gas,''
  \href{http://dx.doi.org/10.1016/j.aop.2008.03.003}{{\em Annals of Physics}
  {\bfseries 323} no.~12, (Dec, 2008) 2987–2990}.
  \url{http://dx.doi.org/10.1016/j.aop.2008.03.003}.

\bibitem{baird:2019}
L.~Baird, X.~Wang, S.~Roof, and J.~E. Thomas, ``Measuring the hydrodynamic
  linear response of a unitary fermi gas,'' 2019.

\bibitem{patel:2019}
P.~B. Patel, Z.~Yan, B.~Mukherjee, R.~J. Fletcher, J.~Struck, and M.~W.
  Zwierlein, ``Universal sound diffusion in a strongly interacting fermi gas,''
  2019.

\bibitem{Sanner:2010}
C.~Sanner, E.~J. Su, A.~Keshet, R.~Gommers, Y.-i. Shin, W.~Huang, and
  W.~Ketterle, ``Suppression of density fluctuations in a quantum degenerate
  fermi gas,'' \href{http://dx.doi.org/10.1103/physrevlett.105.040402}{{\em
  Physical Review Letters} {\bfseries 105} no.~4, (Jul, 2010) }.
  \url{http://dx.doi.org/10.1103/PhysRevLett.105.040402}.

\bibitem{Ku_2012}
M.~J.~H. Ku, A.~T. Sommer, L.~W. Cheuk, and M.~W. Zwierlein, ``Revealing the
  superfluid lambda transition in the universal thermodynamics of a unitary
  fermi gas,'' \href{http://dx.doi.org/10.1126/science.1214987}{{\em Science}
  {\bfseries 335} no.~6068, (Jan, 2012) 563–567}.
  \url{http://dx.doi.org/10.1126/science.1214987}.

\bibitem{onuki_2002}
A.~Onuki, \href{http://dx.doi.org/10.1017/CBO9780511534874}{{\em Phase
  Transition Dynamics}}.
\newblock Cambridge University Press, 2002.

\bibitem{Regal_2005}
C.~A. Regal, M.~Greiner, S.~Giorgini, M.~Holland, and D.~S. Jin, ``Momentum
  distribution of a fermi gas of atoms in the bcs-bec crossover,''
  \href{http://dx.doi.org/10.1103/physrevlett.95.250404}{{\em Physical Review
  Letters} {\bfseries 95} no.~25, (Dec, 2005) }.
  \url{http://dx.doi.org/10.1103/PhysRevLett.95.250404}.

\bibitem{Martinez:2019bsn}
M.~Martinez, T.~Schäfer, and V.~Skokov, ``{Critical behavior of the bulk
  viscosity in QCD},''
  \href{http://dx.doi.org/10.1103/PhysRevD.100.074017}{{\em Phys. Rev.}
  {\bfseries D100} no.~7, (2019) 074017},
\href{http://arxiv.org/abs/1906.11306}{{\ttfamily arXiv:1906.11306 [hep-ph]}}.

\bibitem{Bluhm:2017rnf}
M.~Bluhm, J.~Hou, and T.~Schäfer, ``{Determination of the density and
  temperature dependence of the shear viscosity of a unitary Fermi gas based on
  hydrodynamic flow},''
  \href{http://dx.doi.org/10.1103/PhysRevLett.119.065302}{{\em Phys. Rev.
  Lett.} {\bfseries 119} no.~6, (2017) 065302},
\href{http://arxiv.org/abs/1704.03720}{{\ttfamily arXiv:1704.03720
  [cond-mat.quant-gas]}}.

\bibitem{Nascimbene_2009}
S.~Nascimbène, N.~Navon, K.~J. Jiang, L.~Tarruell, M.~Teichmann, J.~McKeever,
  F.~Chevy, and C.~Salomon, ``Collective oscillations of an imbalanced fermi
  gas: Axial compression modes and polaron effective mass,''
  \href{http://dx.doi.org/10.1103/physrevlett.103.170402}{{\em Physical Review
  Letters} {\bfseries 103} no.~17, (Oct, 2009) }.
  \url{http://dx.doi.org/10.1103/PhysRevLett.103.170402}.

\bibitem{Elliott:2013}
J.~B. Elliott, P.~T. Lake, L.~G. Moretto, and L.~Phair, ``Determination of the
  coexistence curve, critical temperature, density, and pressure of bulk
  nuclear matter from fragment emission data,''
  \href{http://dx.doi.org/10.1103/PhysRevC.87.054622}{{\em Phys. Rev. C}
  {\bfseries 87} (May, 2013) 054622}.
  \url{https://link.aps.org/doi/10.1103/PhysRevC.87.054622}.

\bibitem{Vovchenko:2015pya}
V.~Vovchenko, D.~V. Anchishkin, M.~I. Gorenstein, and R.~V. Poberezhnyuk,
  ``{Scaled variance, skewness, and kurtosis near the critical point of nuclear
  matter},'' \href{http://dx.doi.org/10.1103/PhysRevC.92.054901}{{\em Phys.
  Rev.} {\bfseries C92} no.~5, (2015) 054901},
\href{http://arxiv.org/abs/1506.05763}{{\ttfamily arXiv:1506.05763 [nucl-th]}}.

\bibitem{Vovchenko:2017ayq}
V.~Vovchenko, L.~Jiang, M.~I. Gorenstein, and H.~Stoecker, ``{Critical point of
  nuclear matter and beam energy dependence of net proton number
  fluctuations},'' \href{http://dx.doi.org/10.1103/PhysRevC.98.024910}{{\em
  Phys. Rev.} {\bfseries C98} no.~2, (2018) 024910},
\href{http://arxiv.org/abs/1711.07260}{{\ttfamily arXiv:1711.07260 [nucl-th]}}.

\bibitem{Danielewicz:2019mvp}
P.~Danielewicz, H.~Lin, J.~R. Stone, and Y.~Iwata, ``{Spinodal Instability at
  the Onset of Collective Expansion in Nuclear Collisions},''
\href{http://arxiv.org/abs/1910.10500}{{\ttfamily arXiv:1910.10500 [nucl-th]}}.

\bibitem{Hanauske:2019qgs}
M.~Hanauske, J.~Steinheimer, A.~Motornenko, V.~Vovchenko, L.~Bovard, E.~R.
  Most, L.~J. Papenfort, S.~Schramm, and H.~Stöcker, ``{Neutron Star Mergers:
  Probing the EoS of Hot, Dense Matter by Gravitational Waves},''
\href{http://dx.doi.org/10.3390/particles2010004}{{\em Particles} {\bfseries 2}
  no.~1, (2019) 44--56}.

\bibitem{Perego:2019adq}
A.~Perego, S.~Bernuzzi, and D.~Radice, ``{Thermodynamics conditions of matter
  in neutron star mergers},''
  \href{http://dx.doi.org/10.1140/epja/i2019-12810-7}{{\em Eur. Phys. J.}
  {\bfseries A55} no.~8, (2019) 124},
\href{http://arxiv.org/abs/1903.07898}{{\ttfamily arXiv:1903.07898 [gr-qc]}}.

\bibitem{Most:2018eaw}
E.~R. Most, L.~J. Papenfort, V.~Dexheimer, M.~Hanauske, S.~Schramm,
  H.~Stöcker, and L.~Rezzolla, ``{Signatures of quark-hadron phase transitions
  in general-relativistic neutron-star mergers},''
  \href{http://dx.doi.org/10.1103/PhysRevLett.122.061101}{{\em Phys. Rev.
  Lett.} {\bfseries 122} no.~6, (2019) 061101},
\href{http://arxiv.org/abs/1807.03684}{{\ttfamily arXiv:1807.03684
  [astro-ph.HE]}}.

\bibitem{Bauswein:2018bma}
A.~Bauswein, N.-U.~F. Bastian, D.~B. Blaschke, K.~Chatziioannou, J.~A. Clark,
  T.~Fischer, and M.~Oertel, ``{Identifying a first-order phase transition in
  neutron star mergers through gravitational waves},''
  \href{http://dx.doi.org/10.1103/PhysRevLett.122.061102}{{\em Phys. Rev.
  Lett.} {\bfseries 122} no.~6, (2019) 061102},
\href{http://arxiv.org/abs/1809.01116}{{\ttfamily arXiv:1809.01116
  [astro-ph.HE]}}.

\bibitem{Floerchinger:2012xd}
S.~Floerchinger and C.~Wetterich, ``{Chemical freeze-out in heavy ion
  collisions at large baryon densities},''
  \href{http://dx.doi.org/10.1016/j.nuclphysa.2012.07.009}{{\em Nucl. Phys.}
  {\bfseries A890-891} (2012) 11--24},
\href{http://arxiv.org/abs/1202.1671}{{\ttfamily arXiv:1202.1671 [nucl-th]}}.

\bibitem{Mazeliauskas:2017wyz}
A.~Mazeliauskas, Y.~Akamatsu, and D.~Teaney, ``{Out-of-equilibrium hydrodynamic
  fluctuations in the expanding QGP},''
\href{http://dx.doi.org/10.22323/1.311.0038}{{\em PoS} {\bfseries CPOD2017}
  (2018) 038}.

\bibitem{Tsypin:1994nh}
M.~M. Tsypin, ``{Universal effective potential for scalar field theory in
  three-dimensions by Monte Carlo computation},''
\href{http://dx.doi.org/10.1103/PhysRevLett.73.2015}{{\em Phys. Rev. Lett.}
  {\bfseries 73} (1994) 2015--2018}.

\bibitem{Tsypin:1997zz}
M.~M. Tsypin, ``{Effective potential for a scalar field in three dimensions:
  Ising model in the ferromagnetic phase},''
\href{http://dx.doi.org/10.1103/PhysRevB.55.8911}{{\em Phys. Rev.} {\bfseries
  B55} (1997) 8911--8917}.

\bibitem{Bluhm:2016trm}
M.~Bluhm, M.~Nahrgang, S.~A. Bass, and T.~Schäfer, ``{Behavior of universal
  critical parameters in the QCD phase diagram},''
  \href{http://dx.doi.org/10.1088/1742-6596/779/1/012074}{{\em J. Phys. Conf.
  Ser.} {\bfseries 779} no.~1, (2017) 012074},
\href{http://arxiv.org/abs/1612.04564}{{\ttfamily arXiv:1612.04564 [nucl-th]}}.

\bibitem{Guida:1996ep}
R.~Guida and J.~Zinn-Justin, ``{3-D Ising model: The Scaling equation of
  state},'' \href{http://dx.doi.org/10.1016/S0550-3213(96)00704-3}{{\em Nucl.
  Phys.} {\bfseries B489} (1997) 626--652},
\href{http://arxiv.org/abs/hep-th/9610223}{{\ttfamily arXiv:hep-th/9610223
  [hep-th]}}.

\bibitem{Bluhm:2016byc}
M.~Bluhm, M.~Nahrgang, S.~A. Bass, and T.~Schaefer, ``{Impact of resonance
  decays on critical point signals in net-proton fluctuations},''
  \href{http://dx.doi.org/10.1140/epjc/s10052-017-4771-3}{{\em Eur. Phys. J.}
  {\bfseries C77} no.~4, (2017) 210},
\href{http://arxiv.org/abs/1612.03889}{{\ttfamily arXiv:1612.03889 [nucl-th]}}.

\bibitem{Alkofer:2018guy}
R.~Alkofer, A.~Maas, W.~A. Mian, M.~Mitter, J.~París-López, J.~M. Pawlowski,
  and N.~Wink, ``{Bound state properties from the functional renormalization
  group},'' \href{http://dx.doi.org/10.1103/PhysRevD.99.054029}{{\em Phys.
  Rev.} {\bfseries D99} no.~5, (2019) 054029},
\href{http://arxiv.org/abs/1810.07955}{{\ttfamily arXiv:1810.07955 [hep-ph]}}.

\bibitem{Cyrol:2017ewj}
A.~K. Cyrol, M.~Mitter, J.~M. Pawlowski, and N.~Strodthoff, ``{Nonperturbative
  quark, gluon, and meson correlators of unquenched QCD},''
  \href{http://dx.doi.org/10.1103/PhysRevD.97.054006}{{\em Phys. Rev.}
  {\bfseries D97} no.~5, (2018) 054006},
\href{http://arxiv.org/abs/1706.06326}{{\ttfamily arXiv:1706.06326 [hep-ph]}}.

\bibitem{Herbst:2010rf}
T.~K. Herbst, J.~M. Pawlowski, and B.-J. Schaefer, ``{The phase structure of
  the Polyakov–quark–meson model beyond mean field},''
  \href{http://dx.doi.org/10.1016/j.physletb.2010.12.003}{{\em Phys. Lett.}
  {\bfseries B696} (2011) 58--67},
\href{http://arxiv.org/abs/1008.0081}{{\ttfamily arXiv:1008.0081 [hep-ph]}}.

\bibitem{Rennecke:2016tkm}
F.~Rennecke and B.-J. Schaefer, ``{Fluctuation-induced modifications of the
  phase structure in (2+1)-flavor QCD},''
  \href{http://dx.doi.org/10.1103/PhysRevD.96.016009}{{\em Phys. Rev.}
  {\bfseries D96} no.~1, (2017) 016009},
\href{http://arxiv.org/abs/1610.08748}{{\ttfamily arXiv:1610.08748 [hep-ph]}}.

\bibitem{Cyrol:2018xeq}
A.~K. Cyrol, J.~M. Pawlowski, A.~Rothkopf, and N.~Wink, ``{Reconstructing the
  gluon},'' \href{http://dx.doi.org/10.21468/SciPostPhys.5.6.065}{{\em SciPost
  Phys.} {\bfseries 5} (2018) 065},
\href{http://arxiv.org/abs/1804.00945}{{\ttfamily arXiv:1804.00945 [hep-ph]}}.

\bibitem{Floerchinger:2011sc}
S.~Floerchinger, ``{Analytic Continuation of Functional Renormalization Group
  Equations},'' \href{http://dx.doi.org/10.1007/JHEP05(2012)021}{{\em JHEP}
  {\bfseries 05} (2012) 021},
\href{http://arxiv.org/abs/1112.4374}{{\ttfamily arXiv:1112.4374 [hep-th]}}.

\bibitem{Kamikado:2013sia}
K.~Kamikado, N.~Strodthoff, L.~von Smekal, and J.~Wambach, ``{Real-time
  correlation functions in the $O(N)$ model from the functional renormalization
  group},'' \href{http://dx.doi.org/10.1140/epjc/s10052-014-2806-6}{{\em Eur.
  Phys. J.} {\bfseries C74} no.~3, (2014) 2806},
\href{http://arxiv.org/abs/1302.6199}{{\ttfamily arXiv:1302.6199 [hep-ph]}}.

\bibitem{Tripolt:2013jra}
R.-A. Tripolt, N.~Strodthoff, L.~von Smekal, and J.~Wambach, ``{Spectral
  Functions for the Quark-Meson Model Phase Diagram from the Functional
  Renormalization Group},''
  \href{http://dx.doi.org/10.1103/PhysRevD.89.034010}{{\em Phys. Rev.}
  {\bfseries D89} no.~3, (2014) 034010},
\href{http://arxiv.org/abs/1311.0630}{{\ttfamily arXiv:1311.0630 [hep-ph]}}.

\bibitem{Pawlowski:2017gxj}
J.~M. Pawlowski, N.~Strodthoff, and N.~Wink, ``{Finite temperature spectral
  functions in the O(N)-model},''
  \href{http://dx.doi.org/10.1103/PhysRevD.98.074008}{{\em Phys. Rev.}
  {\bfseries D98} no.~7, (2018) 074008},
\href{http://arxiv.org/abs/1711.07444}{{\ttfamily arXiv:1711.07444 [hep-th]}}.

\bibitem{Cardy:1996xt}
J.~L. Cardy, {\em {Scaling and renormalization in statistical physics}}.
\newblock
1996.
\newblock

\bibitem{Kapusta:2012zb}
J.~I. Kapusta and J.~M. Torres-Rincon, ``{Thermal Conductivity and Chiral
  Critical Point in Heavy Ion Collisions},''
  \href{http://dx.doi.org/10.1103/PhysRevC.86.054911}{{\em Phys. Rev.}
  {\bfseries C86} (2012) 054911},
\href{http://arxiv.org/abs/1209.0675}{{\ttfamily arXiv:1209.0675 [nucl-th]}}.

\bibitem{Plumberg:2017tvu}
C.~Plumberg and J.~I. Kapusta, ``{Hydrodynamic fluctuations near a critical
  endpoint and Hanbury-Brown–Twiss interferometry},''
  \href{http://dx.doi.org/10.1103/PhysRevC.95.044910}{{\em Phys. Rev.}
  {\bfseries C95} no.~4, (2017) 044910},
\href{http://arxiv.org/abs/1702.01368}{{\ttfamily arXiv:1702.01368 [nucl-th]}}.

\bibitem{Ling:2013ksb}
B.~Ling, T.~Springer, and M.~Stephanov, ``{Hydrodynamics of charge fluctuations
  and balance functions},''
  \href{http://dx.doi.org/10.1103/PhysRevC.89.064901}{{\em Phys. Rev.}
  {\bfseries C89} no.~6, (2014) 064901},
\href{http://arxiv.org/abs/1310.6036}{{\ttfamily arXiv:1310.6036 [nucl-th]}}.

\end{thebibliography}\endgroup

\end{document}